%% file: scalar3.tex
\documentclass[review]{elsarticle}
 \biboptions{comma,sort&compress}
\usepackage{graphicx}
\usepackage{amsmath}
\usepackage{here}
\usepackage{cuted}
\def\beq{\begin{equation}}
\def\eeq{\end{equation}}
\def\beqn{\begin{equation*}}
\def\eeqn{\end{equation*}}
\def\bea{\begin{eqnarray}}
\def\eea{\end{eqnarray}}
\def\bq{\begin{quote}}
\def\eq{\end{quote}}

\def\nnb{\nonumber}
\def\ga{\left(}
\def\dr{\right)}

\def\rar{\rightarrow}
\def\lrar{\Longrightarrow}
																
\def\nnb{\nonumber}

\def\la{\langle}
\def\ra{\rangle}

\def\ba{\vspace*{-0.2cm}\begin{array}}
\def\ea{\end{array}\vspace*{-0.2cm}}

\def\b{$\bullet~$}

\def\d{$\diamond~$}

\def\als{\alpha_s}

\def\gg{\la\alpha_s G^2 \ra}
\def\ggg{\la g^3f_{abc} G^aG^bG^c \ra}
\def\gggg{\la\als^2 G^4\ra}

\usepackage{amsmath}
\usepackage{slashed}
\usepackage{color}

\newcommand{\pipi}{\mbox{$\pi\pi$}}

\begin{document}

%\markboth{Stephan Narison, Montpellier (FR)}{ }
\begin{frontmatter}

%%
%%%%%%%%%%%%%%%%%%%%%%%%%%%%%%%%%%%%%%%%%%%%%%%%%
%\begin{document}
%\title{Fully heavy scalar molecules and  tetraquarks states from QCD at NLO}
%\title{On the Tetraquarks / Molecules Nature of X(2900) from QCD Laplace sum rule}
%\begin{center}
\title{ \large Di-Gluonium Sum Rules, $I=0$ Scalar Mesons and Conformal Anomaly
%Puzzling Sigma and Scalar Gluonia
}
\author{Stephan Narison\corref{cor1}
}
\cortext[cor1]{ICTP-Trieste consultant for Madagascar.}
\address{Laboratoire
Univers et Particules de Montpellier (LUPM), CNRS-IN2P3, \\
Case 070, Place Eug\`ene
Bataillon, 34095 - Montpellier, France\\
and\\
Institute of High-Energy Physics of Madagascar (iHEPMAD)\\
University of Ankatso, Antananarivo 101, Madagascar}
\ead{snarison@yahoo.fr}

\begin{abstract}
\noindent

We revisit, scrutinize, improve, confirm and complete our previous results\,\cite{VENEZIA,SNG,SNGREV,BN} from the scalar di-gluonium sum rules within the standard SVZ-expansion at N2LO  {\it without instantons} and {\it beyond the minimal duality ansatz : ``one resonance $\oplus$ QCD continuum"} parametrization of the spectral function which is necessary for a better understanding of the complex spectra of the $I=0$ scalar mesons. We select different (un)subtracted  sum rules (USR) moments of degree $\leq$ 4 for extracting the two lowest gluonia masses and couplings. 
%We notice the novel feature tha the $\tau$-sum rule scale behaviour of the QCD moment is mainly due to the re-organization of the huge radiative corrections in the PT series rather than to the QCD condensates in the OPE. 
We obtain:  $[M_{\sigma_B},f_{\sigma_B}]=[1.07(13),0.46(16)],~[M_{G_1},f_{G_1}][1.55(12),0.37(11)]$ GeV and the corresponding masses of the radial excitations : $M_{\sigma'_B}$= 1.11(12)  and $M_{G'_1}=1.56(14)$ GeV which are (unexpectedly) almost degenerated with the ground states.  The 2nd radial excitation is found to have a much heavier  mass: $M_{G_2}\simeq$ 2.99(22) GeV. Combining these results with some Low-Energy Vertex Sum Rules (LEV-SR), we predict some  hadronic widths and classify them into two groups : -- The $\sigma$-like ($\sigma_B,\sigma'_B$) which decay copiously to $\pi\pi$ from OZI-violating process and the $\sigma'_B$ to $2(\pi\pi)S$ through $\sigma\sigma$. -- The $G$-like $(G_1,~G'_1$ and eventually $G_2$) which decay into $\eta'\eta, ~\eta\eta$ through the $U(1)_A$ gluonic vertex. Besides some eventual mixings with quarkonia states, we may expect that the observed $\sigma/f_0(500)$ and $f_0(137)$ are  $\sigma$-like while the $f_0(1.5)$ and $f_0(1.7)$ are $G$-like gluonia. The high mass $G_2(2.99)$ can also mix with the $G_1,~G'_1$ to bring the gluon component of the gluonia candidates above 2 GeV. We also estimate the conformal charge $\psi_G(0)=2.09(29)$ GeV$^4$ % [which we compare with the low-energy theorem (LET) one] 
and  its slope  $10^2\times\psi'_G(0)=-22(29)$ GeV$^2$. Our results are summarized in Table\,\ref{tab:res}.   
\end{abstract}
%\end{center}

%% keywords
%\keywords{
%}
%\ccode{Pac numbers: 11.55.Hx, 12.38.Lg, 13.20-Gd, 14.65.Dw, 14.65.Fy, 14.70.Dj}  
\end{frontmatter}
%%%%%%%%%%%%%%%%%%%%%%%%%%%%%%%%%%
%\vspace*{-1.5cm}
\vfill\eject
\pagestyle{plain}
 %%%%%%%%%%%%%%%%%%%%%%%%%%%%%%%%%%%
% \section*{References}
%\end{document}

 \section{Introduction}
 %%%%%%%%%%%%%%%%%%%%%%%%%%%%%%%%%%%
Gluonia / Glueballs bound states are expected to be a consequence of QCD\,\cite{MIN}. However, despite considerable theoretical and experimental efforts, there are not (at present) any clear indication of their signature. The difficulty is also  due to the fact that the observed candidates can be a strong mixing of the gluonia with the $\bar qq$ light mesons or some other exotic mesons (four-quark, hybrid states). In particular, the $(I=0)$ Isoscalar  Scalar Channel channel is overpopulated beyond the conventional scalar nonet such that one suspects that some of these states can be of exotic origins and may (for instance) contain some large gluon component in their wave functions. 

In some previous works \,\cite{VENEZIA,SNG,SNGREV,SNU1,SNG0,PAK,PAVER,SNA0,BN,SN06,PABAN,SHORE} based on QCD spectral sum rules (QSSR) \`a la SVZ\,\cite{SVZ,ZA,SNB} combined with some low-energy theorems (LET), we have tried to understand the (pseudo)scalar gluonia channels. Since then, some progresses have been accomplished from experiments and from some other approaches which we shall briefly remind below. 

After this short reminder, we update and improve our previous QSSR analysis.  The  impact of these novel results on our present understanding of the peculiar $(I=0)$ isoscalar scalar channel is discussed. 
 
 %%%%%%%%%%%%%%%%%%%%%%%%%%%%%%
% \section{Experimental facts and Phenomenlogical analysis}
 %%%%%%%%%%%%%%%%%%%%%%%%%%%%%%
 
 %%%%%%%%%%%%%%%%%%%%%%%%%%%%%%%%%%%%%
\section{$\sigma/f_0(500)$ and $f_0(980)$ from scatterings data}
 %%%%%%%%%%%%%%%%%%%%%%%%%%%%%%%%%%%%%
To quantify the observables of the low-lying scalar states, we have phenomenologically studied the $\pi\pi, \bar KK, \gamma\gamma$\,\cite{MNO,KMN}, $K_{e4}$\cite{WANG1}, $ J/\psi,\phi$ radiative and $D_s$ semileptonic decays\,\cite{WANG2,DOSCH} data with the aim to understand the internal substructure of the $\sigma / f_0(500)$ and $f_0(980)$ mesons\,\footnote{A more complete and comprehensive discussion on production processes and decays of scalar mesons can be found in the recent reviews of\,\cite{OCHS,GASTALDI,AMSLER,PDG}.}.  
%%%%%%%%%%%%%%%%%%%%%%%%%%%%%%%%%%%%%%%%%%
\subsection*{\b Masses and hadronic widths from $\pi\pi\to\pi\pi,KK$ and $Ke_4$ data}
%%%%%%%%%%%%%%%%%%%%%%%%%%%%%%%%%%%%%%%%%%
\d Using a K-matrix \,\cite{MENES,PEAN} analysis of the $\pi\pi$ elastic scattering data below 0.7 GeV, we found for the complex pole mass and the residue using one bare resonance $\oplus$ one channel \,\cite{MNO} :
\beq
M^{pole}_\sigma[{\rm MeV}]\simeq 422-i290. %~~~~  \Gamma^{dir}_\sigma|_{pole}=0.13(5){\rm keV},~~~\Gamma^{resc}_\sigma =2.7(4)~{\rm keV}
\label{eq:one res}
\eeq

\d We extend the previous analysis until 900 MeV and use two bare resonances [$\sigma(500)$, $f_0(980)$] $\oplus$ two open channels ($\pi\pi,\bar KK$) parametrization of the data \,\cite{KMN,WANG1}. To the data used previously, we add  the new precise measurement from NA48/2\,\cite{NA48}  on $Ke_4~(K\to \pi\pi e\nu_e)$ data  for the $\pi\pi$ phase shift  below 390 MeV.
% and use from 400 to 900 MeV the CERN-Munich\,\cite{MUNICH} and
%Hyams et al.\,\cite{HYAMS}  $\pi\pi$-phase from  $\pi\pi \to  \pi\pi$ which agree
%each others above 400 MeV. 
Then, averaging this result with the previous one in Eq.\ref{eq:one res}, one obtains the final estimate :
\beq
M^{pole}_\sigma[{\rm MeV}]\simeq 452(12)-i260(15) ~,~~~\vert g_{\sigma\pi^+\pi^-}\vert=1.12(31)~{\rm GeV},~~~\frac{\vert g_{f_0 K^+K^-}\vert} { \vert g_{f_0\pi^+\pi^-}\vert}=2.58(1.34)~,
\eeq
which agrees with the ones based on the  analytic continuation and unitarity properties of the amplitude to the deep imaginary region\cite{LEUT,YND}\,:
\beq
M^{pole}_\sigma[{\rm MeV}]\simeq 444-i272~, ~~~{\rm and}~~~M^{pole}_\sigma[{\rm MeV}]\simeq 489-i264~.
\eeq 
\d However, in order to compare these results with the QCD spectral sum rules ones where the analysis is done in the real axis, one has to introduce the On-shell or Breit-Wigner (os) mass and width where the amplitude is purely imaginary at the phase $90^0$\,:
\beq
Re{\cal D}((M_{\sigma}^{os})^2)=0~~~~~~~\lrar ~~~~~~~(M^{os}_\sigma, \Gamma^{os}_\sigma) = (920, 700)~{\rm MeV}~, ~~~~~~~
\label{eq:sigdata}
\eeq
where ${\cal D}$ is the propagator appearing in the unitary $\pi\pi$ amplitude.  Similar result has been obtained in\,\cite{OCHSMIN,AU} where the wide resonance centred at 1 GeV has been interpreted in\,\cite{OCHSMIN} as a resonance  interfering destructively with the $f_0(980)$ and $f_0(1500)$. 

\d For the $f_0(980)$, one obtains \,\cite{KMN,WANG1}:
\beq
M^{pole}_{f_0}[{\rm MeV}]\simeq 981(34)-i18(11) ~,~~~\vert g_{f_0\pi^+\pi^-}\vert= 2.65(10)~{\rm GeV},~~~\frac{\vert g_{\sigma K^+K^-}\vert}{ \vert g_{\sigma\pi^+\pi^-}\vert}=0.37(6)~,
\eeq
and
\beq
M^{os}_{f_0}\simeq M^{pole}_{f_0},~~~~~~~~~~~~~~~~~~~~\Gamma^{os}_{f_0}\simeq  \Gamma^{pole}_{f_0}~,
\label{eq:f0data}
\eeq
where Eq.\,\ref{eq:f0data} comes from the fact that $f_0(980)$ is narrow. 
%In this way, we obtain :
%%%%%%%%%%%%%%%%%%%%%%%%%%%%%%%%%%%%%%%%%%%%%%%
\subsection*{\b $\sigma/f_0(500)$ and $f_0(980)$ $\gamma\gamma$ widths from $\gamma\gamma\to\pi\pi,KK$ scatterings}
%%%%%%%%%%%%%%%%%%%%%%%%%%%%%%%%%%%%%%%%%%%%%%%
\d We proceed as above and deduce the average (in units of keV):
\beq
\Gamma^{dir}_{\sigma_{pole}}=0.16(4)~,~~~~ \Gamma^{resc}_{\sigma_{pole}} =1.89(81)~,~~~~  \Gamma^{tot}_{\sigma_{pole}} =3.08(82)
\eeq
for the direct, rescattering and total (direct $\oplus$ rescattering) $\gamma\gamma$ widths.  For the on-shell mass, these lead to :
\beq
\Gamma^{dir}_{\sigma_{os}}=1.2(3)~{\rm keV},~~~~~~~~\Gamma^{resc}_{\sigma_{pole}} =\Gamma^{resc}_{\sigma_{os}}~.
%~~~~  \Gamma^{tot}_\sigma|_{os} =3.08(82)
\eeq
\d For  the $f_0(980)$, one obtains in units of keV :
\beq
\Gamma^{dir}_{f_0}|_{pole}\simeq \Gamma^{dir}_{f_0}|_{os}=0.28(1),~~~~~~\Gamma^{res}_{f_0}=0.85(5), ~~~~~~\Gamma^{tot}_{f_0}=0.16(1)~.
\eeq
%%%%%%%%%%%%%%%%%%%%%%%%%%%
\subsection*{\b Production of $\sigma/f_0(500)$ and $f_0(980)$ from some other channels}
%%%%%%%%%%%%%%%%%%%%%%%%%%%%
Isoscalar Scalar production in some other channels has been reviewed in\,\cite{OCHS,GASTALDI,AMSLER,PDG}, where the structures:
\beq
\sigma/f_0(500)~, ~~~~~f_0(980)~,~~~~~ f_0(1370)~, ~~~~~f_0(1500)
\eeq 
have been also found in different processes ($J/\psi$-decays, central production in $pp$ and $e^+e^-$, $\bar pp$ and $\bar pn$ annihilation at rest, $\pi p$ and $K p$). 

\d We  notice the production of the $\sigma/f_0(500)$ from  $J/\psi\to \omega \pi^+\pi^-$\,\cite{BES05}
with the extracted complex pole mass and width:
\beq
M^{pole}_\sigma=541(39)-i252(42)~,
\eeq
and of the 
$f_0(980)$ from $J/\psi\to\gamma\pi^0\pi^0$ by BESIII\,\cite{BES16} 
and from $J/\psi\to \phi \pi^+\pi^-$ and $J/\psi\to \phi K^+K^-$\,\cite{BES05} with a ratio of coupling :
\beq
 g_{f_0K^+K^-} / g_{f_0\pi^+\pi^-} = 4.21\pm 0.32~.
\eeq

\d The  $f_0(980)$ is also produced from $\phi\to (\pi^+\pi^-, ~K^+K^-)+\gamma$, a  glue filter channel at KLOE\,\cite{KLOE} with the couplings : 
 \beq
 g_{f_0K^+K^-}=(3.97\sim 4.74)~{\rm GeV}~,~~~~~~~~~~g_{f_0\pi^+\pi^-}=-(2.22\sim 1.82) ~{\rm GeV}~,
\eeq
where in these last two experiments the presence of the $\sigma/f_0(500)$  improves the quality of the fit. 
The non-vanishing of $g_{f_0\pi^+\pi^-}$ indicates that the $f_0(980)$ cannot be a pure $\bar ss$ state. 

%%%%%%%%%%%%%%%%%%%%%%%%%%%%%%%%%%%%%%%%%%%%%%%%%
\section{Production of some other $(I=0$) isoscalar scalar mesons}
%%%%%%%%%%%%%%%%%%%%%%%%%%%%%%%%%%%%%%%%%%%%%%%%%
 To these data, we add the production of the new structures which can have gluon component\,\cite{PDG} :
\beq
f_0(1765)~, ~~~~~~~~~~~~f_0(2020)~, ~~~~~~~~~~~~f_0(2100), ~~~~~~~~~~~~f_0(2200)~.
\eeq
observed from $J/\psi\to \gamma \pi^+\pi^-,K^+k^-,4\pi$ radiative decays,  $\pi p,~\bar pp,~pp \to \eta\eta', \eta\eta,4\pi$\,\cite{PDG}  and
recently from $\eta_c\to [f_0(1.7)\to K^+K^-], f_0(2.1) \to\pi^+\pi^-]\eta'$ by BABAR\,\cite{PALANO}. 
%by BESIII\,\cite{BES05,BES16} and compiled in PDG. 
%\end{document}
%%%%%%%%%%%%%%%%%%%%%%%%%%%%%%%%%%
\section{Impacts of the previous phenomenological analysis and data}
%%%%%%%%%%%%%%%%%%%%%%%%%%%%%%%%%%
One can notice that the $I=0$ scalar channel is overpopulated and goes beyond the usual nonet
expectations. From the previous phenomenological analysis, one may expect that:
%%%%%%%%%%%%%%%%%%%%%%
\subsubsection*{\b $\sigma/f_0(500)$ meson}
%%%%%%%%%%%%%%%%%%%%%%
\d It cannot be a pure four-quark $(\bar uu+\bar dd)$ or / and $\pi\pi$ molecule state as it has non-vanishing coupling to $\bar KK$. 

\d However, from the large value of its $\gamma\gamma$ rescattering width (see e.g.\,\cite{MNO}), one may be tempted to interpret it as  $\pi\pi$ or/and $\bar KK$
molecule state but the $\bar KK$ molecule mass is expected to be around 1 GeV due to $SU(3)$ breakings. 

\d From the size of its direct on-shell width of about 1 keV to  $\gamma\gamma$\,\cite{MNO}, it cannot be a pure $(\bar uu+\bar dd)$ state which has a $\gamma\gamma$-width expected to be about 4 keV from QCD spectral sum rules\,\cite{SN06} and quark model\,\cite{ROSNER} but larger than for a pure gluonium state of about (0.2-0.6) KeV\,\cite{SNG,MNO}. It cannot also be a four-quark state as the model predicts much smaller (0.00..keV) direct $\gamma\gamma$ widths\,\cite{ACHASOV,SNA0}.

%%%%%%%%%%%%%%%%%%%%%%
\subsubsection*{\b $f_0(980)$ meson}
%%%%%%%%%%%%%%%%%%%%%%
\d A large $\bar ss$ component would lead to a  direct $\gamma\gamma$ width of 0.4 keV which are comparable with the data but its non-zero coupling to $\pi\pi$ rules out this possibility. An estimate of a mass of the $\bar ss$ state from QSSR leads to a value around 1.4 GeV\,\cite{SN06} which is (relatively) too high for the $f_-0(980)$. 
%%%%%%%%%%%%%%%%%%%%%%%%%%%%%%%%
\subsubsection*{\b $f_0(1.37)$  meson}
%%%%%%%%%%%%%%%%%%%%%%%%%%%%%%%%
The large decay width of this  meson to $2(\pi\pi)$ in a S-wave may signal a large gluon component in its wave functions\,\cite{VENEZIA,SNG}.
%%%%%%%%%%%%%%%%%%%%%%%%%%%%%%%%
\subsubsection*{\b $f_0(1.37)$ and $f_0(1.5)$ mesons}
%%%%%%%%%%%%%%%%%%%%%%%%%%%%%%%%
Its decay into $\eta'\eta$ and $\eta\eta$ and the presence of a $2(\pi\pi)_S$ indicate a large gluon component in its wave function. 
%%%%%%%%%%%%%%%%%%%%%
\section{Theoretical expectations}
%%%%%%%%%%%%%%%%%%%%%
\b In the $I=0$  isoscalar scalar channel, we have \,\cite{VENEZIA,SNG} explored the possibility that the $\sigma/f_0(500)$ meson can be the lowest scalar gluonium state $\sigma_B$ (hereafter, the subindex $B$ refers to unmixed gluonium state) having a mass around 1 GeV. From our approach the $\sigma$ is composed mainly with gluons and it  couples strongly and universally to $\pi\pi$ and $\bar KK$ explaining its large width while its ``direct" decay to $\gamma\gamma$ is expected to be about (0.2-0.6) keV .  

\b A maximal mixing of this gluonium with a ($\bar uu+\bar dd$) state may explain the feature of the observed $\sigma/f_0(500)$ and $f_0(980)$ : narrow width in $\pi\pi$ but strong coupling to $\bar KK$\,\cite{BN}\,\footnote{An estimate of this mixing angle using Gaussian sum rules favours  a maximal mixing\,\cite{STEELE2}.}.

\b We have also argued\,\cite{VENEZIA,SNG} that the contribution of the $\sigma_B$ to the subtracted (SSR) and unsubtracted (USR) sum rules is necessary for resolving the apparent inconsistency between these two sum rules where, in addition,  a higher mass gluonium G(1.5) is needed which we have identified\,\cite{VENEZIA} with the G(1.6) found from the GAMS data\,\cite{GAMS}. The contribution of $\sigma_B$  to the SSR compensates the large contribution of the two-point subtraction constant to the SSR without appealing to more speculative direct instanton non-perturbative contributions.

%\d  Instantons are expected to have high dimensions $(d\geq 5)$ and enter into the Wilson coefficients\,\cite{NOVIKOV,NSVZ,SHURYAK} in contrast to the standard SVZ Operator Product Expansion (OPE) where the perturbative Wilson coeffiicients can be unambiguously separated from the non-perturbative condensates\,\cite{SVZ,ZA}. These instanton contributions have been argued to break the OPE\,\cite{FORKEL} (see however\,\cite{STEELE}).  Unfortunately, its size and density are not under a good control (no quoted error) and are model dependent such that its contribution remains very qualitative. 

\b  Another striking feature of the $\sigma_B$ is its analogy with its chiral partner the $\eta'$ which plays a crucial role for the $U(1)_A$ anomaly by its contribution to the topological charge (subtraction constant of the $U(1)_A$ two-point correlator)\,\cite{SNU1,WITTEN, VENEZIANO,DIVECCHIA,GIACOMO}. The $\sigma_B$, being the dilaton particle of the conformal anomaly\,\cite{NOVIKOV,NSVZ,CHANO,LANIK,VENEZIA}, is expected to be associated to the trace of the energy-momentum tensor $\theta_\mu^\mu$:
\bea
\theta^\mu_\mu&=&\frac{1}{4} \beta(\alpha_s) G^{\mu\nu}_aG^a_{\mu\nu}+\ga1+\gamma_m(\alpha_s)\dr\sum_{u,d,s}m_i\bar\psi_i\psi_i ,
\label{eq:theta}
\eea
with :  $\gamma_m=2a_s+\cdots$ is the quark mass anomalous dimension and $a_s\equiv \alpha_s/\pi$.  $\beta(\alpha_s)$ is the $\beta$-function normalized as :
\bea
\beta(\alpha_s)&=&\beta_1a_s+\beta_2a_s^2+\beta_3a_s^3+\cdots
\eea
with:
\bea
 \beta_1&=&-\frac{11}{2} + \frac{n_f}{3}~,~~\beta_2=-\frac{51}{4} +\frac{19}{12}n_f~,~~~
\beta_3 = \frac{1}{64}\ga -2857 + \frac{5033}{9}n_f - \frac{325}{27}n_f^2\dr
\eea
for $n_f$ number of quark flavours. For $n_f=3$ used in this paper, one has :
\beq
 \beta_1=-9/2,~~~~~~~~  \beta_2=-8,~~~~~~~~ \beta_3=-20.1198~.
\eeq
%%%%%%%%%%%%%%%%%%%%%%%%%%%%%%%%%%%%%%%%%%
\section{Prospects}
%%%%%%%%%%%%%%%%%%%%%%%%%%%%%%%%%%%%%%%%%%
In this paper, we shall :

\b Use selected low and high degree moments of QCD spectral sum rules (QSSR) within the {\it standard SVZ expansion without instantons} and parametrize the spectral function beyond the minimal duality ansatz {\it ``one resonance $\oplus$ QCD continuum"} in order to extract the masses and decay constants of the scalar gluonia.

\b Extract the value of the conformal charge $\psi_G(0)$ (value of the two-point correlator at zero momentum) in order to test the Low Energy Theorem (LET)\,\cite{NSVZ} estimate. 

\b Discuss some phenomenological implications of our results for an attempt to understand the complex  spectrum of the $I=0$  observed scalar mesons.
%\d Test the influence of  $\sigma_B$ on these new high moments and compare  the result  with the ``one resonance" parametrization used in the sum rule literature and lattice calculations\,\cite{HART,MCNEILE,RAGO, MATHIEU}.

%\d Try to build a gluonuim-quarkonium mixing -scheme for explaining some of the complex isoscalar scalar meson spectroscopy.

%%%%%%%%%%%%%%%%%%%%%%%%%%%%%%%%%%%%%%%%%%%
\section{The QCD anatomy of the two-point correlator}
%%%%%%%%%%%%%%%%%%%%%%%%%%%%%%%%%%%%%%%%%%%
We shall work with the gluonium two-point correlator :
\beq
\psi_G(q^2)=16i\int d^4x \,e^{iqx}\la 0 \vert \ga\theta^\mu_\mu\dr_G(x) \ga\theta^\mu_\mu\dr_G^\dagger(0)\vert 0\ra
\eeq
built from the gluon component of the trace of the energy-momentum tensor in Eq.\,\ref{eq:theta}.  
%%%%%%%%%%%%%%%%%%%%%%%%%%%%%%%%%%%%%%%%%%%%%%%
\subsection*{\b The standard SVZ-expansion}
%%%%%%%%%%%%%%%%%%%%%%%%%%%%%%%%%%%%%%%%%%%%%%%
Using the Operator Product Expansion (OPE) \`a la SVZ , its QCD expression can be written as :

\beq
\psi_G(q^2)=\beta^2(\alpha_s)\ga\frac{2}{\pi^2}\dr\sum_{0,1,2,...}C_{2n}\la  {\cal O}_{2n}\ra~.
\eeq
where $C_{2n}$ is the Wilson coefficients calculable perturbatively while $ \la  {\cal O}_{2n}\ra$ is a short-hand notation for the non-perturbative vacuum condensates $\la 0 \vert {\cal O}_{2n}\vert 0\ra$  of dimension $2n$.
%%%%%%%%%%%%%%%%%%%%%%%%%%%%%%%%%%%
\subsubsection*{\d The unit perturbative operator ($n=0$) }
%%%%%%%%%%%%%%%%%%%%%%%%%%%%%%%%%%%
Its contribution reads :
\bea
C_0&\equiv&- Q^4 L_\mu\Big{[} C_{00} +C_{01} L_\mu+C_{02} L_\mu^2\Big{]}~~{\rm with} :\nnb\\
C_{00} &=& 1+\frac{659}{36}a_s+247.480a_s^2~,~~~C_{01}=-a_s\ga \frac{9}{4}+65.781a_s\dr~,~~C_{02}=5.0625a_s^2,
\label{eq:pert}
\eea
where the NLO (resp. N2LO) contributions have been obtained in\,\cite{KATAEV} (resp.\,\cite{CHETG}) and have been adapted 
for a sum rule use in\,\cite{STEELE}. $L_\mu\equiv {\rm Log}(Q^2/\mu^2)$ where $\mu$ is the subtraction point. We shall use for 3 flavours :
\beq
\Lambda = 340(28)~{\rm MeV}
\eeq
deduced from $\alpha_s(M_Z)= 0.1182(19)$ from $M_{\chi_{c0,b0}}-M_{\eta_c,\eta_b}$mass-splittings\,\cite{SNparam,SNparama},\,$\tau$-decays\,\cite{PICHTAU,SNTAU} and the world average\,\cite{PDG,BETHKE}. We shall use the running QCD coupling to order $\alpha_s^2$:
\beq
a_s(\mu)= a_s^{(0)}\Bigg{\{} 1-a_s^{(0)}\frac{\beta_2}{\beta_1}LL_\mu+\ga a_s^{(0)}\dr^2\Bigg{[}\ga\frac{\beta_2}{\beta_1}\dr^2\ga LL_\mu^2-LL_\mu-1\dr+\frac{\beta_3}{\beta_1}\Bigg{]}\Bigg{\}}~,
\label{eq:alphas}
\eeq
where :  
\beq
 a_s^{(0)}\equiv \frac{1}{-\beta_1{\rm Log} \ga{\mu}/{\Lambda}\dr}~~~~~~~{\rm and}~~~~~~~LL_\mu\equiv {\rm Log} \, \Big{[} {\rm 2\,Log}\ga{\mu}/{\Lambda}\dr\Big{]}. 
 \eeq
%%%%%%%%%%%%%%%%%%%%%%%%%%%%%%%%%%% 
\subsubsection*{\d The dimension-four gluon condensate ($n=2$) }
%%%%%%%%%%%%%%%%%%%%%%%%%%%%%%%%%%%
Its contribution reads :
\beq
C_{4}\la  {\cal O}_{4}\ra\equiv (C_{40}+L_\mu C_{41})\la  \alpha_s G^2 \ra~~:~~ 
C_{40} = 2\pi a_s\ga 1 + \frac{175}{36} a_s\dr,~~C_{41} = -\frac{9}{2}\pi a_s^2~.
\eeq
We shall use the value :
\beq
\gg=(6.35\pm 0.35)\times 10^{-2}\,{\rm GeV}^4
\eeq
determined from light and heavy quark systems\,\cite{SNparam,SNparama,SNB8,SNREV15}. 
%%%%%%%%%%%%%%%%%%%%%%%%%%%%%%%%%%%%%%%%%%%%%%%%%%
\subsection*{\d The dimension-six ($n=3$) gluon condensate}
%%%%%%%%%%%%%%%%%%%%%%%%%%%%%%%%%%%%%%%%%%%%%%%%%%
Its  contribution reads (see \cite{BAGAND6} for the $\alpha_s$ correction for $n_f=0$ and where a $1/\pi$ misprint is corrected):
\beq
C_{6}\la  {\cal O}_{6}\ra= \la C_{60}+L_\mu C_{61}\ra\ggg /Q^2~~:~~~~~~C_{60}=a_s~,~~C_{61}=-\frac{29}{4}a_s^2
\eeq
with\,\cite{SNB8} :
\beq
\ggg=(8.2\pm 1.0)\,{\rm GeV}^2\gg~,
\eeq
which notably differs from the instanton liquid model estimate $\ggg\approx (1.5\pm 0.5)\,{\rm GeV}^2\gg$\,\cite{SVZ,NOVIKOV,NSVZ,SHURYAK} used in\,\cite{SNG}. 
%%%%%%%%%%%%%%%%%%%%%%%%%%%%%%%%%%%%%%%%%%%%%%%%%%
\subsection*{\d The dimension-eight ($n=4$) gluon condensate}
%%%%%%%%%%%%%%%%%%%%%%%%%%%%%%%%%%%%%%%%%%%%%%%%%%
Its contribution reads : 
\beq
C_{8}\la  {\cal O}_{8}\ra= C_{80}\gggg/Q^4~~ :~~~~~~C_{80}=4\pi \alpha_s
\eeq
with :
\beq
\gggg\equiv \Big{[}14\la \ga \alpha_s f_{abc}G^a_{\mu\rho}G^b_{\nu\rho}\dr^2\ra -\la \ga \alpha_s f_{abc}G^a_{\mu\nu}G^b_{\rho\lambda}\dr^2\ra\Big{]}\simeq (0.55\pm 0.01)  \gg^2
\eeq
from a modified factorization with $1/N_c^2$ corrections (13/24) \,\cite{BAGAND8} and factorization (9/16)\,\cite{NOVIKOV}. $(0.55\pm 0.01)$  measures the deviation from factorization which has been found to be largely violated for the four-quark condensates\,\,\cite{SNTAU,LNT,LAUNER}. 
%We have multiplied the average of these results by a factor 2 in order to take into account  an eventual violation of factorization as observed in the case of the four-quark condensate\,\cite{SNTAU,LNT,LAUNER}. 
%%%%%%%%%%%%%%%%%%%%%%%%%%%%%%%%%%%%%%%%%%%%%%%
\subsection*{\b Beyond the standard SVZ-expansion}
%%%%%%%%%%%%%%%%%%%%%%%%%%%%%%%%%%%%%%%%%%%%%%%
\subsubsection*{\d The tachyonic gluon mass ($n=1$)}
To these standard contributions in the OPE, we can consider the one from a dimension-two tachyonic gluon mass contribution introduced by\,\footnote{For reviews, see e.g.  \,\cite{ZAKa,ZAKb}.} which phenomenologically parametrizes the large order terms of the perturbative QCD series\,\cite{CNZb}.
Its existence is supported by some AdS approaches\,\cite{ADS1,ADS2,ADS3}. 
This effect has been calculated explicitly in\,\cite{CNZa}\,:
\beq
C_2\la O_2\ra= C_{21} L_\mu \lambda^2Q^2~ : ~~~~~~ C_{21}= \frac{3}{a_s}
\eeq
where $\lambda^2$ is the tachyonic gluon mass determined from $e^+e^-\to$ hadrons data and the pion channel\,\,\cite{SND21,CNZa,TERAYEV}:
\beq
a_s\lambda^2\simeq -(6.0\pm 0.5)\times 10^{-2}~{\rm GeV}^2~.
\eeq
%%%%%%%%%%%%%%%%%%%%%
\subsubsection*{\d The direct instanton $(n\geq 5/2)$ } 
%%%%%%%%%%%%%%%%%%%%%%
In an instanton liquid model\,\cite{NSVZ,SHURYAK}, the direct instanton contribution is assumed to be dominated by the single instanton-anti-instanton contribution via a non-perturbative contribution to the perturbative Wilson coefficient\,\cite{FORKEL,STEELE3}:
\beq
\psi(Q^2)\vert_{\bar I-I}=32{\beta}^2 \,Q^4\int \rho^4 \Big{[} K_2\ga \rho\sqrt{Q^2}\dr\Big{]}^2 dn(\rho)
\eeq
where $K_2(x)$ is the modified Bessel function of the second kind. At this stage this classical field effect is beyond the SVZ expansion where the later assumes that one can separate without any ambiguity the perturbative Wilson coefficients from the non-perturbative condensate contributions. 

Besides the fact that its contributes in the OPE as $1/Q^5$, i.e.  it acts like other high-dimension condensates not taken into account in the OPE, the above instanton effect depends crucially on the (model-dependent) overall density $\bar n=\int_0^\infty d\rho\,n(\rho)$ and on its average size $\bar\rho=(1/\bar n)\int_0^\infty d\rho\,\rho\,n(\rho)$ which (unfortunately) are not quantitatively under a good control  ($\bar\rho$ ranges from 5 \,\cite{NSVZ} to 1.94,\cite{SNB8} and 1.65 GeV$^{-1}$\cite{SHURYAK}, while $\bar n \approx (0.5\sim 1.2)$\,fm$^{-4}$). They
contribute with a high power in $\rho$ to the spectral function ${\rm Im}\psi(t)$, which can be found explicitly in\,\cite{FORKEL,STEELE3},  and behave as :
\bea
{\rm Im}\psi_G (t)\vert_{\bar I-I}&& \buildrel t\to\infty\over \sim  n\ga \rho\sqrt{t}\dr^{-5}\nnb\\
&& \buildrel t\to 0\over \sim n\ga \rho\sqrt{t}\dr^{4}~.
\eea
%However,  $\bar\rho$ and $\bar n$ are not (unfortunately) quantitatively under a good control  ($\bar\rho$ ranges from 5 \,\cite{NSVZ} to 1.94,\cite{SNcb2,SNcb3} and 1.65 GeV$^{-1}$\cite{SHURYAK}, while $\bar n \approx (0.5\sim 1.2)$\,fm$^{-4}$).

As  such effects are quite inaccurate and model-dependent,  we shall not consider them explicitly  in the analysis.  Instead, 
%we shall test (a posteriori) its effect  by comparing 
an eventual deviation of our results within the standard SVZ expansion 
from some experimental data or/and or some other alternative estimates (Low-Energy Theorems (LET), Lattice calculations,...) may signal the need of such (beyond the standard OPE) effects in the analysis.   One should mention that the approach within the standard SVZ OPE and without a direct instanton effect used in the :

\hspace*{0.5cm} --  $U(1)_A$ channel has predicted successfully the value of the topological charge, its slope and the $\eta'$-mass and decay constant\,\cite{SNU1,SHORE}. 

\hspace*{0.5cm} -- Pseudoscalar pion and kaon channels have reproduced successfully the value of the light quark masses where we have also explicitly shown\,\cite{SNL} that  the direct instanton effect induces a relatively small correction  contrary to some other strong claims\,\cite{IOFFE} in the literature. 

%%%%%%%%%%%%%%%%%%%%%%%%%%%%%%%%%%%%%%%%%%%
\section{The Inverse Laplace transform sum rules}
%%%%%%%%%%%%%%%%%%%%%%%%%%%%%%%%%%%%%%%%%%%
From its QCD asymptotic behaviour $\sim (-q^2)^2{\rm Log}(-q^2/\mu^2)$ one can write a twice subtracted dispersion relation:
\beq
\psi_G(q^2)=\psi_G(0) +q^2\psi'_G(0)+\frac{q^4}{\pi} \int_0^\infty \frac{dt}{t^2}\,\frac{{\rm Im}\psi_G(t)} {(t-q^2-i\epsilon)}~.
\eeq

\b Following standard QSSR techniques\,\cite{SVZ,SNB}, one can derive from it different form of the sum rules.  In this paper, we shall work with the Exponential or Borel \,\cite{SVZ,BELLa,BERTa} or Inverse Laplace transform\,\cite{SNR} Finite Energy sum rule(LSR)\footnote{The name Inverse Laplace transform has been attributed due to the fact that perturbative radiative corrections have this property.}\,:
\beq
{\cal L}^c_n(\tau)=\int_0^{t_c} \hspace*{-0.25cm}dt\,t^n\, e^{-t\tau}\frac{1}{\pi} {\rm Im}\psi_G(t)~: n=-1,0,1,2,3
\eeq
and the corresponding ratios of sum rules :
\beq
{\cal R}^c_{n+l~n}(\tau)\equiv \frac{{\cal L}^c_{n+l}(\tau)}{{\cal L}^c_n(\tau)}~,
\eeq
where $\tau$ is the Laplace sum rule variable. In the duality ansatz :
\beq
\frac{1}{\pi} {\rm Im}\psi_G(t)=2\sum_G f_G^2M_G^4\,\delta(t-M_G^2)+\theta (t-t_c) {\rm "QCD\, continuum"}~, 
\eeq
where the $f_G$ are the lowest resonances couplings normalized as $f_\pi=93$ MeV  and $M_G$ their masses while the "QCD continuum" comes from the discontinuity $ {\rm Im}\psi_G(t)\vert_{QCD}$ of the QCD diagrams from the continuum threshold $t_c$.  In the {\it ``one narrow resonance $\oplus$ QCD continuum"} parametrization of the spectral function:
\beq
{\cal R}^c_{n+l~n}(\tau)\simeq M_G^2~.
\eeq

\b To get ${\cal L}_{-1}$, we find convenient to take the Inverse Laplace transform of the  first superconvergent  2nd derivative of the correlator.
%while for  ${\cal L}_0$, we take the Inverse Laplace transform of the  first superconvergent  3rd derivative. 
It reads explicitly:
\bea
{\cal L}^c_{-1}(\tau)&=&\beta^2(\alpha_s)\ga\frac{2}{\pi^2}\dr\tau^{-2}\sum_{n=0,2,\cdots}\hspace*{-0.25cm}D^{-1}_n +\psi_G(0)~,
\eea
with:
\bea
D^{-1}_0&=&\Big{[}C_{00}+2C_{01}(1-\gamma_E-L_\tau)+3C_{02}\big{[}1-\pi^2/6+(-1+\gamma_E+L_\tau)^2\big{]}\Big{]}(1-\rho_1)\nnb\\
D^{-1}_2&=&C_{21}\lambda^2\tau(1-\rho_0),\nnb\\
D^{-1}_4&=&-\Big{[}C_{40}-\frac{C_{41}}{36}\big{[} 55+6(\gamma_E+L_\tau)\big{]}\Big{]}\gg\tau^2,\nnb\\
D^{-1}_6&=&-\big{[}C_{60}-C_{61}(-1+\gamma_E+L_\tau)\big{]}\ggg\tau^3,\nnb\\
D^{-1}_8&=&\frac{C_{80}}{2}(1.1\pm 0.5)\gg^2\tau^4~,
\eea
with :
\beq
L_\tau\equiv -{\rm Log}(\tau\mu^2)~,~~~~~~~~\rho_n\equiv e^{-t_c\tau}\ga 1+t_c\tau+\cdots \frac{(t_c\tau)^n}{ n !}\dr
\eeq
is the QCD continuum contribution from the discontinuity of the QCD diagrams to the sum rule. We omit some eventual continuum contributions from the non-perturbative condensates which are numerically tiny.  

$\psi_G(0)$ is the value of the two-point correlator at zero momentum and can be fixed by a LET as\,\cite{NOVIKOV,NSVZ}:
\beq
\psi_G(0)\simeq -\frac{16}{\pi}\beta_1 \gg= (1.46\pm 0.08)~{\rm GeV}^4,
\label{eq:let}
\eeq
which we shall test later on. 

\b To get ${\cal L}_0$, we take the Inverse Laplace transform of the 1st superconvergent 3rd derivative of the two-point correlator. In this way, we obtain :
\bea
{\cal L}^c_{0}(\tau)&=&\beta^2(\alpha_s)\ga\frac{2}{\pi^2}\dr\tau^{-2}\sum_{n=0,2,\cdots}\hspace*{-0.25cm}D^{0}_n ~,
\eea
with: 
\bea
D^{0}_0&=&\Big{[}2C_{00}-2C_{01}(3-2\gamma_E-2L_\tau)-6C_{02}\big{[}1-3\gamma_E+\gamma^2-\pi^2/6+(-3+2\gamma_E)L_\tau+L_\tau^2\big{]}\Big{]}(1-\rho_2)\nnb\\
D^{0}_2&=&-\frac{C_{21}}{2}\lambda^2\tau(1-\rho_1),\nnb\\
D^{0}_4&=&-{C_{41}}\gg\tau^2,\nnb\\
%D^{0}_4&=&-\Big{[}C_{40}-\frac{C_{41}}{36}\big{[} 55+6(\gamma_E+L)\big{]}\Big{]}\gg\tau^3,\nnb\\
D^{0}_6&=&\big{[}C_{60}-C_{61}(\gamma_E-L_\tau)\big{]}\ggg\tau^3,\nnb\\
D^{0}_8&=&C_{80}(0.55\pm 0.01)\gg^2\tau^4~,
\eea

\b The other higher degrees sum rules ${\cal L}^c_n(\tau)$ for $n\geq 1$ can be deduced from the $n^{th}$ $\tau$-derivative of ${\cal L}^c_0(\tau)$:
\beq
{\cal L}^c_n(\tau)=(-1)^n\frac{d^n}{d\tau^n} {\cal L}^c_0(\tau)~.
\eeq

\b These superconvergent sum rules obey the homogeneous renormalization group equation (RGE):
\beq
\Big{\{}-\frac{\partial }{\partial t} +\beta(\alpha_s)\alpha_s \frac{\partial }{\partial \alpha_s}\Big{\}}{\cal L}^c_n(e^t \tau,\alpha_s)=0~,
\eeq
with $t\equiv (1/2)L_\tau$ with the renormalization group improved (RGI) solution :
\beq
{\cal L}^c_n(e^t \tau,\alpha_s)= {\cal L}^c_n(t=0,\bar\alpha_s(\tau))~,
\eeq
where $\bar\alpha_s$ is the QCD running coupling. 

\b In the following analysis, we shall work with the family of sum rules having degrees less or equal to 4. Then, we shall select the sum rules which present stability (minimum or inflexion point) in the sum rule variable $\tau$ and in the continuum threshold $t_c$ such that we can extract optimal information from the analysis. 
%%%%%%%%%%%%%%%%%%%%%%%%%%%%%%%%%%%%%%%%%%%
\section{$\tau$ and $t_c$ behaviours of the QCD side of the LSR ${\cal L}^c_n$} 
\begin{figure}[hbt]
%\vspace*{-0.25cm}
\begin{center}
\includegraphics[width=10cm]{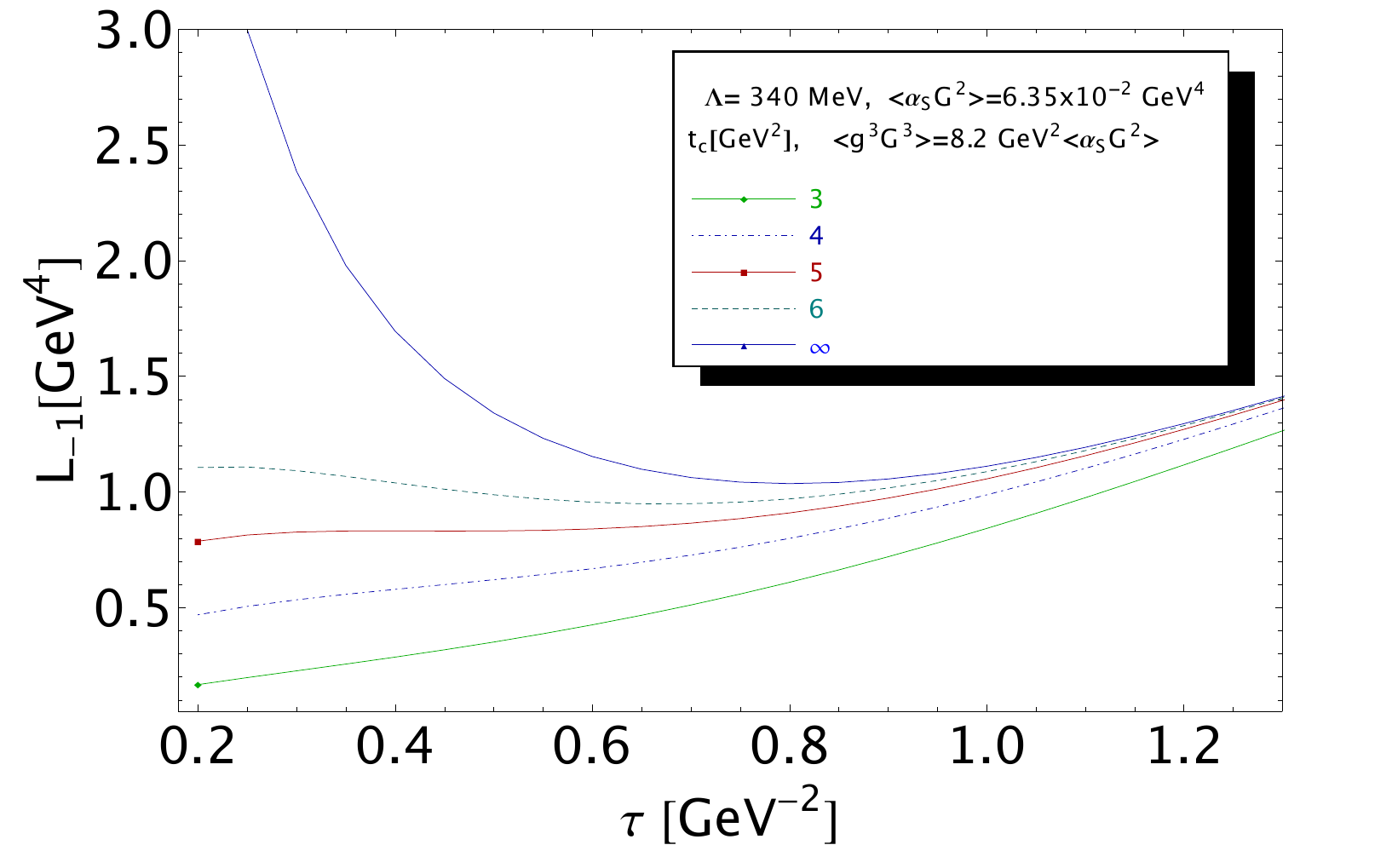} \\
\vspace*{-0.5cm}
\caption{\footnotesize  ${\cal L}^c_{-1}$  as a function of $\tau$ at N2LO for different values of $t_c$.} 
\label{fig:lm1}
\end{center}
%\vspace*{-0.75cm}
\end{figure}

\begin{figure}[hbt]
%\vspace*{-0.25cm}
\begin{center}
%\centerline {\hspace*{-6.cm} \bf a)\hspace*{10cm} }
\vspace{-0.25cm}
%\includegraphics[width=10cm]{L-1_OPE.pdf} \\
%\centerline {\hspace*{-6.cm} \bf b)\hspace*{10cm}}
\includegraphics[width=9cm]{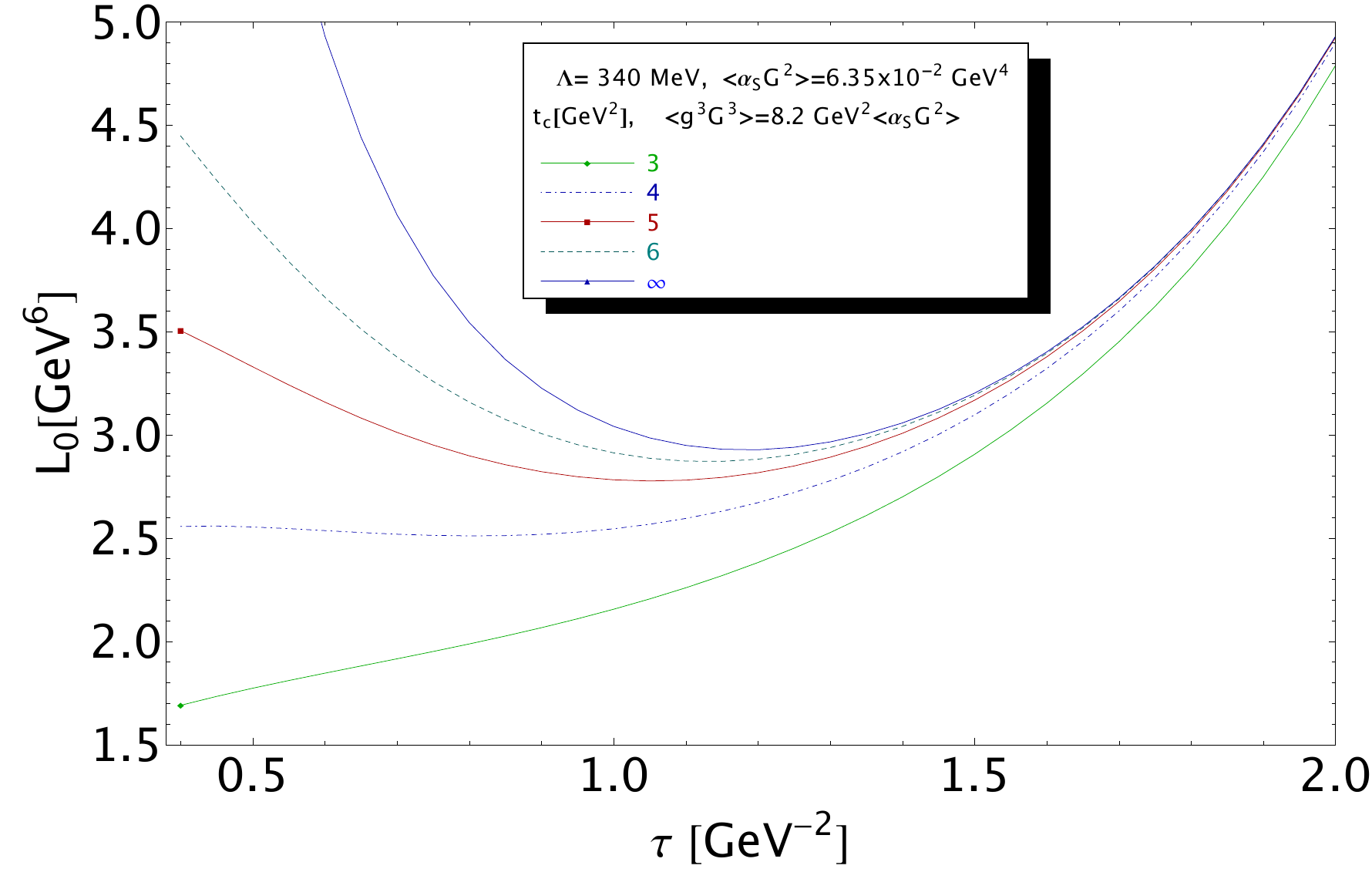}
\vspace*{-0.5cm}
\caption{\footnotesize  ${\cal L}^c_{0}$ as a function of $\tau$ at N2LO for different values of $t_c$.} 
\label{fig:l0}
\end{center}
%\vspace*{-0.75cm}
\end{figure}

%%%%%%%%%%%%%%%%%%%%%%%%%%%%%%%%%%%%%%%%%%%
Before doing the phenomenological analysis of the LSR, we study the $\tau$ and $t_c$ behaviour of their QCD expressions. For instance, we show the subtracted (SSR) ${\cal L}^c_{-1}$ and unsubtracted (USR) ${\cal L}^c_{0}$ sum rules
in Figs\,\ref{fig:lm1} and \,\ref{fig:l0}, where we notice that the subtraction constant shifts the USR minimum (optimization point) at lower values of $\tau$.  This feature has leaded to the inconsistencies of the gluonium mass from these sum rules within a{\it  ``one resonance $\oplus$ QCD continuum"} parametrization of the spectral function as noted in\,\cite{VENEZIA,SNG} which can be cured by working instead with {\it ``two resonances $\oplus$ QCD continuum}"

\b  The other  USR  moments ${\cal L}^c_{2}$ and ${\cal L}^c_{4}$ stabilize for $\tau$ of about 1.5 GeV$^{-2}$ while ${\cal L}^c_{1}$ and ${\cal L}^c_{3}$ do not present $\tau$-stability. To understand these $\tau$-behaviours :

\hspace*{0.5cm} \d We write explicitly the OPE expression of ${\cal L}_0$ and ${\cal L}_1$ to lowest order but includes the $\alpha_s$ correction of the $D=4$ contribution as this contribution vanishes to LO. Normalized to $(2/\pi^2)\beta^2$, they read:
\bea
{\cal L}_0  &=&\tau^{-3}\Big{[} 1+\frac{9}{2}\pi a_s^2\gg\tau^2 +a_s \ggg\tau^3+4\pi \alpha_s\gggg\tau^4\Big{]},\nnb\\
{\cal L}_1&=& 3\tau^{-4}\Big{[} 1-\frac{9}{6}\pi a_s^2\gg\tau^2 -4\pi \alpha_s\gggg\tau^4\Big{]}~,
\eea
where the changes of sign of the condensate contributions destabilize ${\cal L}_1$. 

\hspace*{0.5cm}\d The inclusion of the huge PT radiative corrections shows that  they also participate to the $\tau$-behaviour of the QCD expression 
due to the non-trivial  reorganization of these PT terms  in each Laplace transformed moment.
% induced by the truncated expression of $\alpha_s$ given in Eq.\,\ref{.  

%after the such that the decay constant of the gluonium is mainly due to the PT contributions. These radiative corrections are less important in the ratio of moments used to determine the gluonium mass. 

\begin{figure}[H]
%\vspace*{-0.25cm}
\begin{center}
%\centerline {\hspace*{-6.cm} \bf a)\hspace*{10cm} }
\vspace{-0.25cm}
\includegraphics[width=9cm]{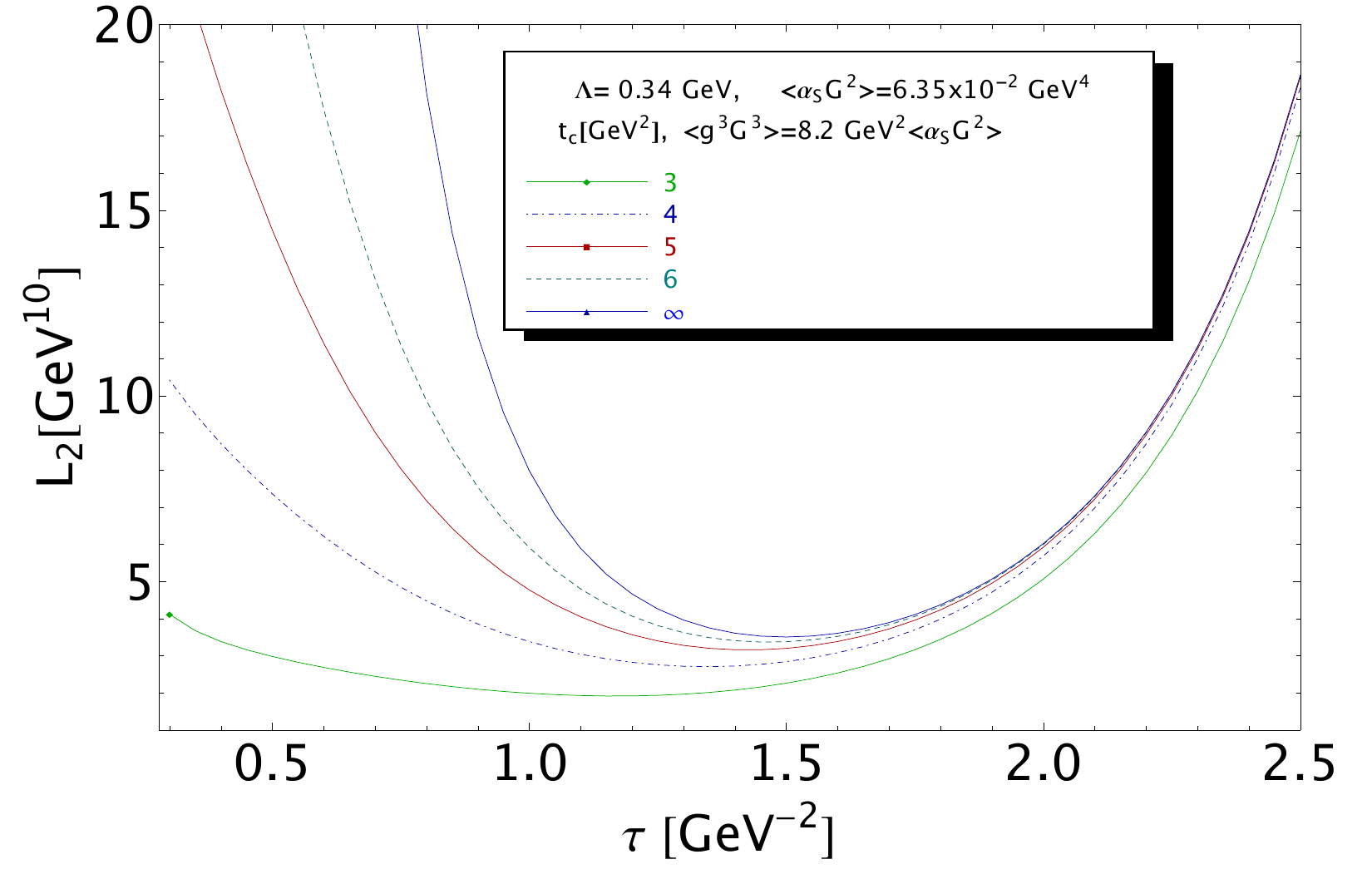}
\vspace*{-0.5cm}
\caption{\footnotesize  ${\cal L}^c_{2}$ as a function of $\tau$ at N2LO for different values of $t_c$.} 
\label{fig:l2}
\end{center}
%\vspace*{-0.75cm}
\end{figure}

\begin{figure}[H]
\begin{center}
\includegraphics[width=9.5cm]{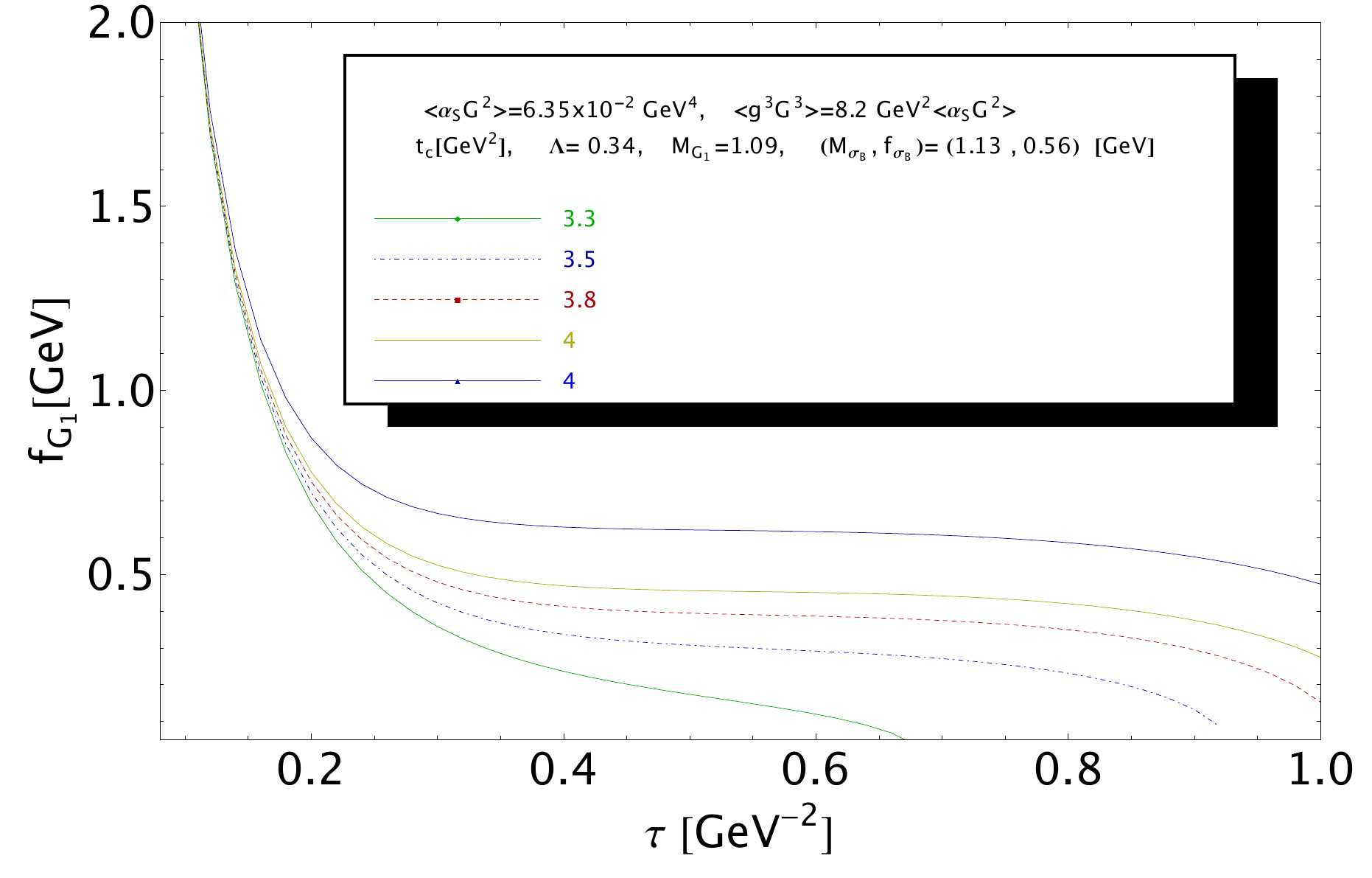} 
\vspace*{-0.5cm}
\caption{\footnotesize  $f_{G_1}$  from ${\cal L}^c_3$  as a function of $\tau$ at N2LO for different values of $t_c$ from a two resonances parametrization and for given values of $(M_{\sigma_B}, f_{\sigma_B})$. }
\label{fig:fg12}
\end{center}
%\vspace*{-0.75cm}
\end{figure}

\b From this explicit analysis, one expects that, a priori:

\hspace*{0.5cm} \d One may extract reliably the decay constant of the gluonium from ${\cal L}^c_{2}$ and ${\cal L}^c_{4}$. However, we shall see later on  that the extraction of the decay constant from ${\cal L}^c_{3}$ can present $\tau$-stability due to the exponential weight $e^{M_G^2\tau}$ brought by the resonance contributions which compensates the decrease of the QCD expression $[{\cal L}^c_{n}\sim \tau^{-(n+3)}]$ for increasing $\tau$.  In the specific example of ${\cal L}^c_{2}$, we shall that the exponential factor shifts the $\tau$-minimum to lower $\tau$-values (Fig.\,\ref{fig:fg11})  at which 
the PT corrections are much smaller than in Fig.\,\ref{fig:l2}, while for ${\cal L}^c_{3}$,
there appears a plateau inflexion point  (Fig.\,\ref{fig:fg12}) which render reliable the extraction of $f_{G_1}$. 

\hspace*{0.5cm} \d The mass of the gluonium can be reliably obtained from the ratios ${\cal R}^c_{nl}$ of the USR ${\cal L}^c_{n,l}$ which optimize at about the same value of $\tau$ and where the PT corrections tend to compensate. 

 %%%%%%%%%%%%%%%%%%%%%%%%%%%%%%%%%%%%%%%
%%%%%%%%%%%%%%%%%%%%%%%%%%%%%%%%%%%%%%%
 %%%%%%%%%%%%%%%%%%%%%%%%%%%%%%%%%%%%%%%
%%%%%%%%%%%%%%%%%%%%%%%%%%%%%%%%%%%%%%%
 %%%%%%%%%%%%%%%%%%%%%%%%%%%%%%%%%%%%%%%
\begin{figure}[hbt]
%\vspace*{-0.25cm}
\begin{center}
%\centerline {\hspace*{-6.cm} \bf a)\hspace*{10cm} }
\vspace{-0.25cm}
\includegraphics[width=10cm]{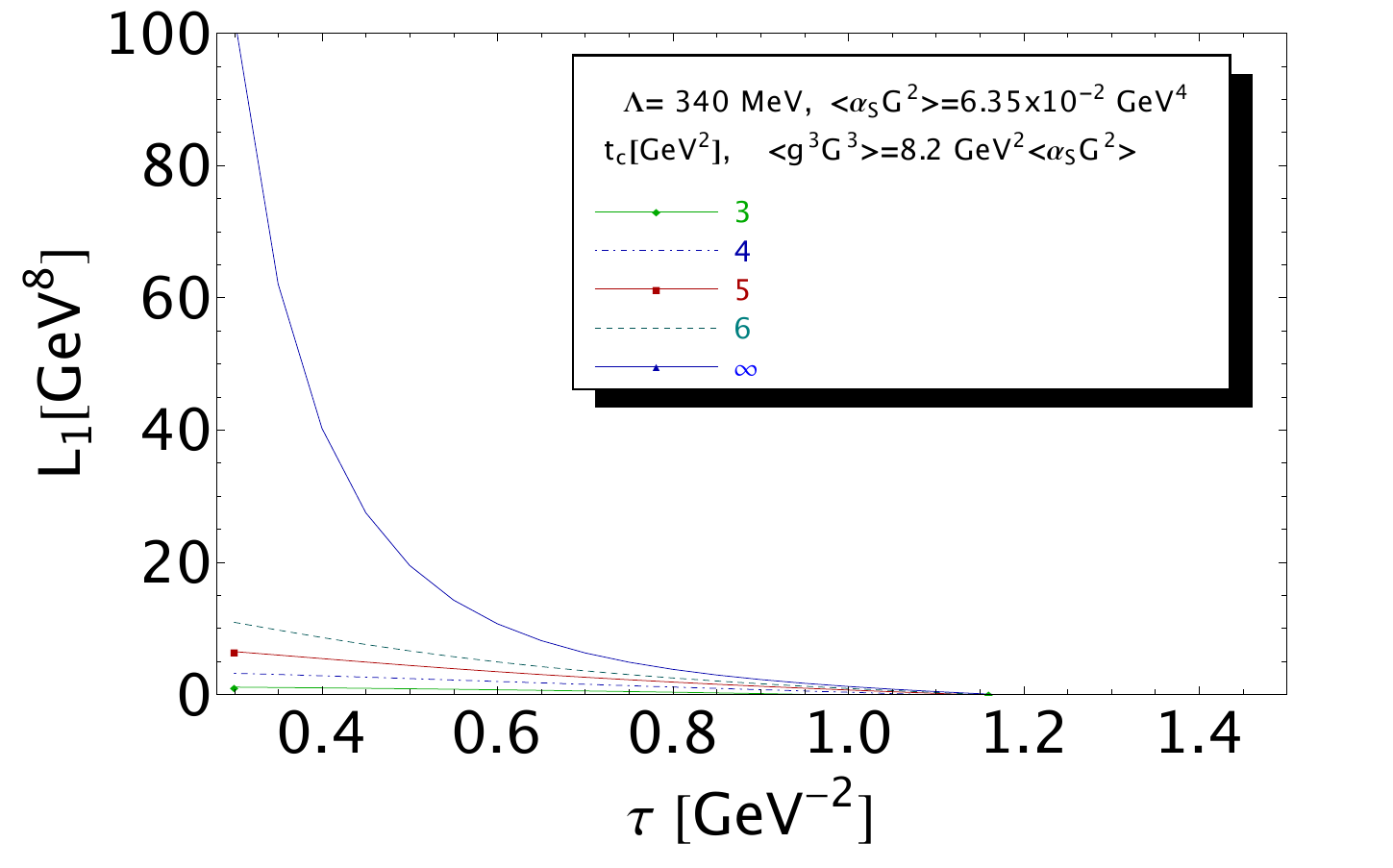}
\vspace*{-0.5cm}
\caption{\footnotesize  ${\cal L}^c_{1}$ as a function of $\tau$ at N2LO for different values of $t_c$.} 
\label{fig:l1}
\end{center}
%\vspace*{-0.75cm}
\end{figure} 
%%%%%%%%%%%%%%%%%%%%%%%%%%%%%%%%%%%%%%%
 %%%%%%%%%%%%%%%%%%%%%%%%%%%%%%%%%%%%%%%
%%%%%%%%%%%%%%%%%%%%%%%%%%%%%%%%%%%%%%%
 %%%%%%%%%%%%%%%%%%%%%%%%%%%%%%%%%%%%%%%
\begin{figure}[H]
%\vspace*{-0.25cm}
\begin{center}
%\centerline {\hspace*{-6.cm} \bf a)\hspace*{10cm} }
\vspace{-0.25cm}
\includegraphics[width=10cm]{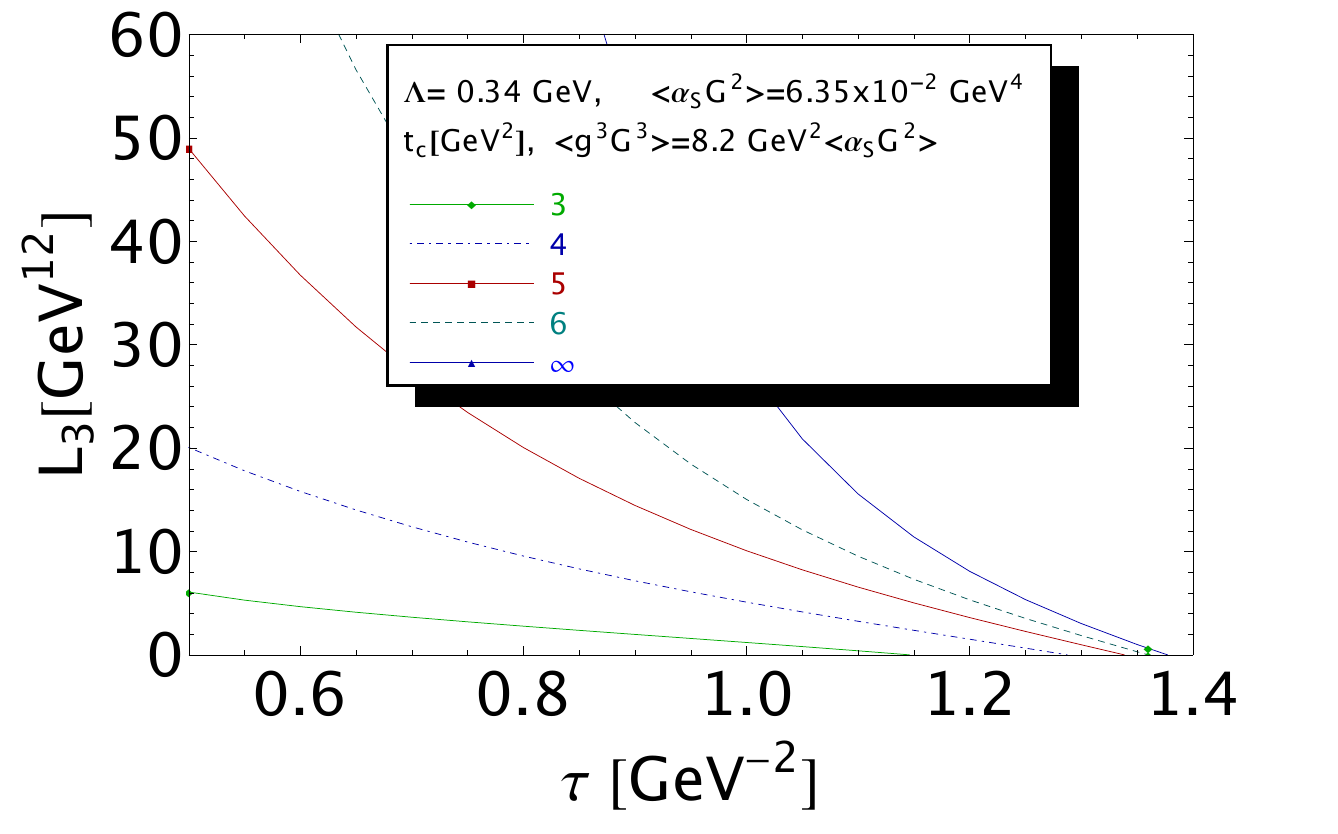} \\
%\centerline {\hspace*{-6.cm} \bf b)\hspace*{10cm}}
%\includegraphics[width=9cm]{L4_OPE.pdf}
\vspace*{-0.5cm}
\caption{\footnotesize  ${\cal L}^c_{3}$ as a function of $\tau$ at N2LO for different values of $t_c$.} 
\label{fig:l3}
\end{center}
%\vspace*{-0.75cm}
\end{figure} 
%%%%%%%%%%%%%%%%%%%%%%%%%%%%%%%%%%%%%%%
% ore eventually with the ratios involving ${\cal L}^c_{1}$ or/and ${\cal L}^c_{3}$ if the corresponding ratio present $\tau$ minimas. 

%%%%%%%%%%%%%%%%%%%%%%%%%%%%%%%%%%%%%%%
\begin{figure}[H]
%\vspace*{-0.25cm}
\begin{center}
%\centerline {\hspace*{-6.cm} \bf a)\hspace*{10cm} }
\vspace{-0.25cm}
%\includegraphics[width=10cm]{L3_OPE.pdf} \\
%\centerline {\hspace*{-6.cm} \bf b)\hspace*{10cm}}
\includegraphics[width=9cm]{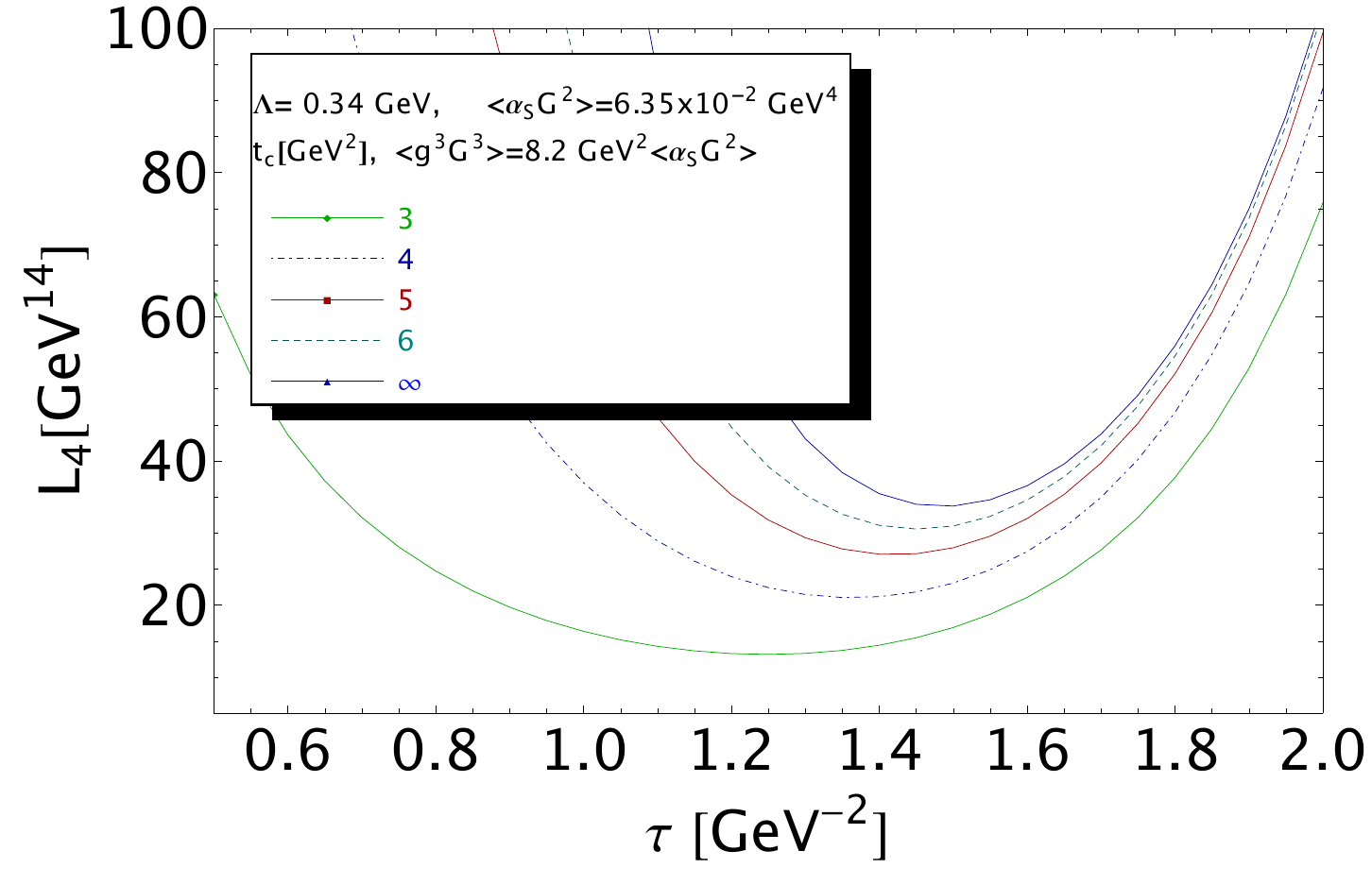}
\vspace*{-0.5cm}
\caption{\footnotesize ${\cal L}^c_{4}$ as a function of $\tau$ at N2LO for different values of $t_c$.} 
\label{fig:l4}
\end{center}
%\vspace*{-0.75cm}
\end{figure} 
%%%%%%%%%%%%%%%%%%%%%%%%%%%%%%%%%%%%%%%
 %%%%%%%%%%%%%%%%%%%%%%%%%%%%%%%%%%%%%%%
\begin{figure}[H]
%\vspace*{-0.25cm}
\begin{center}
%\centerline {\hspace*{-6.cm} \bf a) }
%\vspace{0.25cm}
\includegraphics[width=10cm]{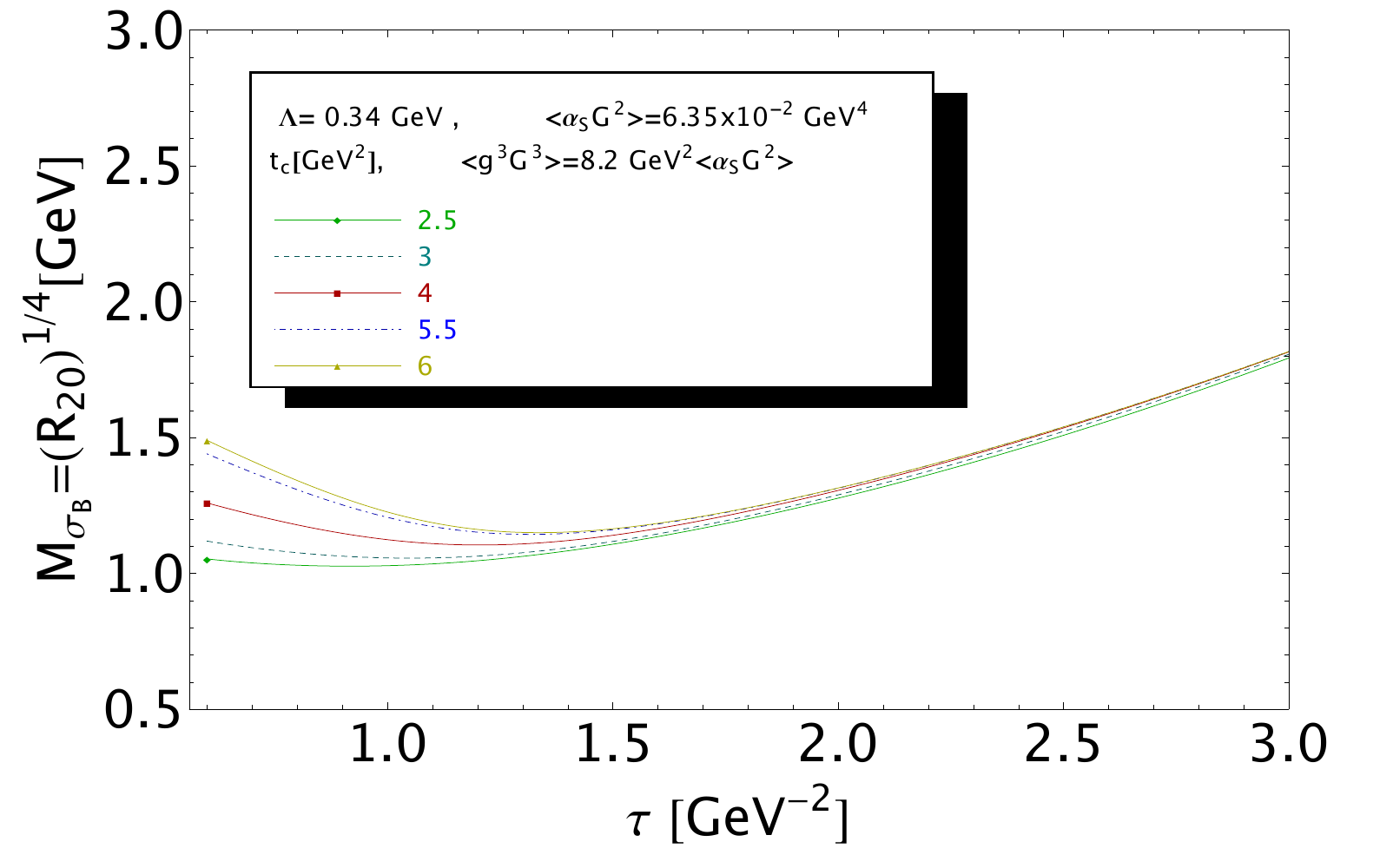} \\
%\vspace{0.25cm}
%\centerline {\hspace*{-6cm} \bf b) }
%\includegraphics[width=10cm]{L20-G.pdf}
\vspace*{-0.5cm}
\caption{\footnotesize  $M_{\sigma_B}$  from ${\cal R}^c_{20}$  as a function of $\tau$ at N2LO for different values of $t_c$ within a one resonance parametrization.} 
\label{fig:r201}
\end{center}
%\vspace*{-0.75cm}
\end{figure} 
%%%%%%%%%%%%%%%%%%%%%%%%%%%%%%%%%%%%%%%

%%%%%%%%%%%%%%%%%%%%%%%%%%%%%%%%%%%%%%%%%%%
\section{The masses of the two lightest gluonia $\sigma_B$ and $G_1$}
%degree sum rules  ${\cal R}^c_{nl}$,  $(n,l=-1,0,1)$  and ${\cal L}^c_{-1}$}
%%%%%%%%%%%%%%%%%%%%%%%%%%%%%%%%%%%%%%%%%%%
In order to get the lowest ground state gluonia masses, we shall work with different forms of USR. We shall not use for this case the observables involving ${\cal L}_{-1}$ in order to avoid the dependence of the result on the input value of $\psi_G(0)$ which we shall test later on.  Among   the possible ratios of sum rules, we have selected ${\cal R}^c_{20}$ and ${\cal R}^c_{42}$ which appear to give the most stable results versus the variation of different parameters. 
%%%%%%%%%%%%%%%%%%%%%%%%%%%%%%%%%%%%%%
\subsection*{\b Mass of the lightest ground state (hereafter named $\sigma_B$) from ${\cal R}^c_{20}$}
\begin{figure}[hbt]
%\vspace*{-0.25cm}
\begin{center}
%\centerline {\hspace*{-6.cm} \bf a) }
%\vspace{0.25cm}
%\includegraphics[width=10cm]{L20.pdf} \\
%\vspace{0.25cm}
%\centerline {\hspace*{-6cm} \bf b) }
\includegraphics[width=10cm]{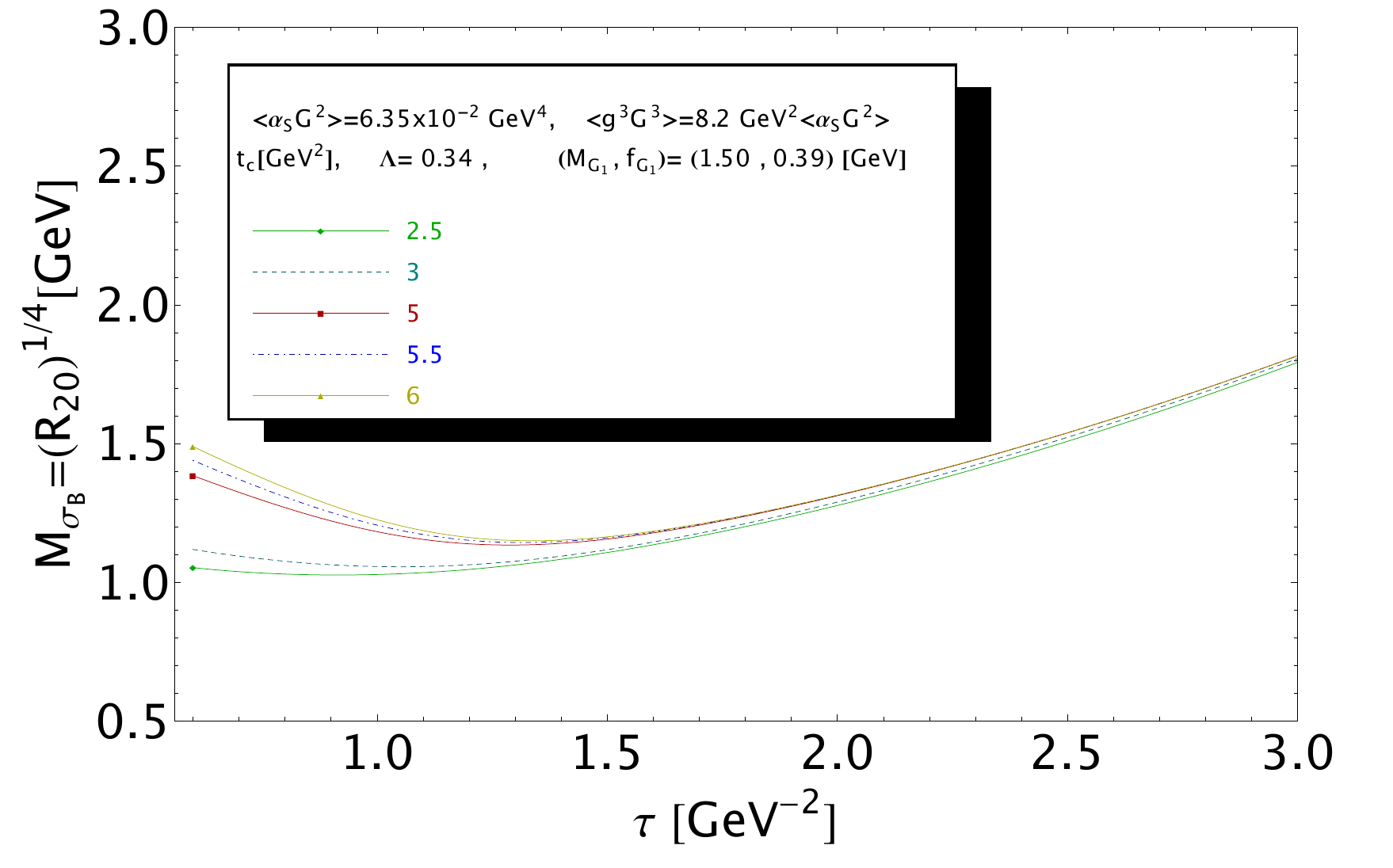}
\vspace*{-0.5cm}
\caption{\footnotesize  $M_{\sigma_B}$  from ${\cal R}^c_{20}$  as a function of $\tau$ at N2LO for different values of $t_c$ including the effect of the 2nd resonance with $(M_{G_1}\,,\,f_{G_1})=(1.50,0.39)$ GeV\,\cite{SNG}.} 
\label{fig:r20}
\end{center}
%\vspace*{-0.75cm}
\end{figure}

%%%%%%%%%%%%%%%%%%%%%%%%%%%%%%%%%%%%%%
In so doing, we work with the ratio of sum rules ${\cal R}^c_{20}$. We show its $\tau$-behaviour for different values of $t_c$ in Fig.\,\ref{fig:r20}.
%%%%%%%%%%%%%%%%%%%%%%%%%%%%%%%%%%%%%%%
\subsubsection*{\d  The case : one resonance $\oplus$ QCD continuum} 
%%%%%%%%%%%%%%%%%%%%%%%%%%%%%%%%%%%%%%%
We show in Fig.\,\ref{fig:r201} the  result with one resonance $\oplus$ QCD continuum parametrization of the spectral function.
 %%%%%%%%%%%%%%%%%%%%%%%%%%%%%%%%%%%%%%%
%%%%%%%%%%%%%%%%%%%%%%%%%%%%%%%%%%%%%%%

The $\tau$-stability starts from $t_c=2.5$ GeV$^2$ while $t_c$-stability is reached around 5.5 GeV$^2$.
We deduce in this range:
\beq
M_{\sigma_B}= 1088(2)_{\tau}(54)_{t_c}(57)_\Lambda(1)_{G^2}(2)_{G^3}=1088(79)~{\rm MeV},
\label{eq:msigma0}
\eeq
where the errors are mainly due to $\Lambda$ and $t_c$. 
 %%%%%%%%%%%%%%%%%%%%%%%%%%%%%%%%%%%%%%%
\subsubsection*{\d  The case : two resonances $\oplus$ QCD continuum} 
%%%%%%%%%%%%%%%%%%%%%%%%%%%%%%%%%%%%%%%
 We show in Fig.\,\ref{fig:r20}  the effect of a 2nd resonance with the  parameters in\,\cite{SNG} :
 \beq
 M_{G_1}= 1.5~{\rm GeV} ~~~~~{\rm and}~~~~~ f_{G_1}= 0.39~{\rm GeV}.
 \label{eq:G1}
 \eeq
 % and :
% \beq
 %M_{G_1}\simeq (1.54\pm 0.12)~{\rm  GeV}~~~~~~{\rm and} ~~~~~~ f_{G_1}\simeq (0.65\pm 0.25)~{\rm GeV}~.
% \eeq
It affects the $\sigma_B$ mass result by a negligible amount of about 22 MeV.
We deduce the conservative result from $t_c=3$ to 5.5 GeV$^2$ and use an input error of 0.1 GeV$^{-2}$ for the localization of $\tau$:
\beq
M_{\sigma_B}=  1085(5)_{\tau}(59)_{t_c}(89)_\Lambda(1)_{G^2}(2)_{G^3}(1)_{\lambda^2}(39)_{M_G}(53)_{f_G}=1085(126)~{\rm MeV},
\label{eq:msigmaf}
\eeq
where the errors due to the non-perturbative condensates are negligible. 
%The effect of the 2nd resonance on the value of $M_{\sigma_B}$ is tiny (22 MeV). 

%%%%%%%%%%%%%%%%%%%%%%%%%%%%%%%%%%%%%%%%%%%
\subsection*{\b Mass of the medium ground state gluonium $G_1$ from ${\cal R}^c_{42}$}
\begin{figure}[H]
%\vspace*{-0.25cm}
\begin{center}
%\centerline {\hspace*{-6.cm} \bf a) }
%\vspace{0.25cm}
\includegraphics[width=10cm]{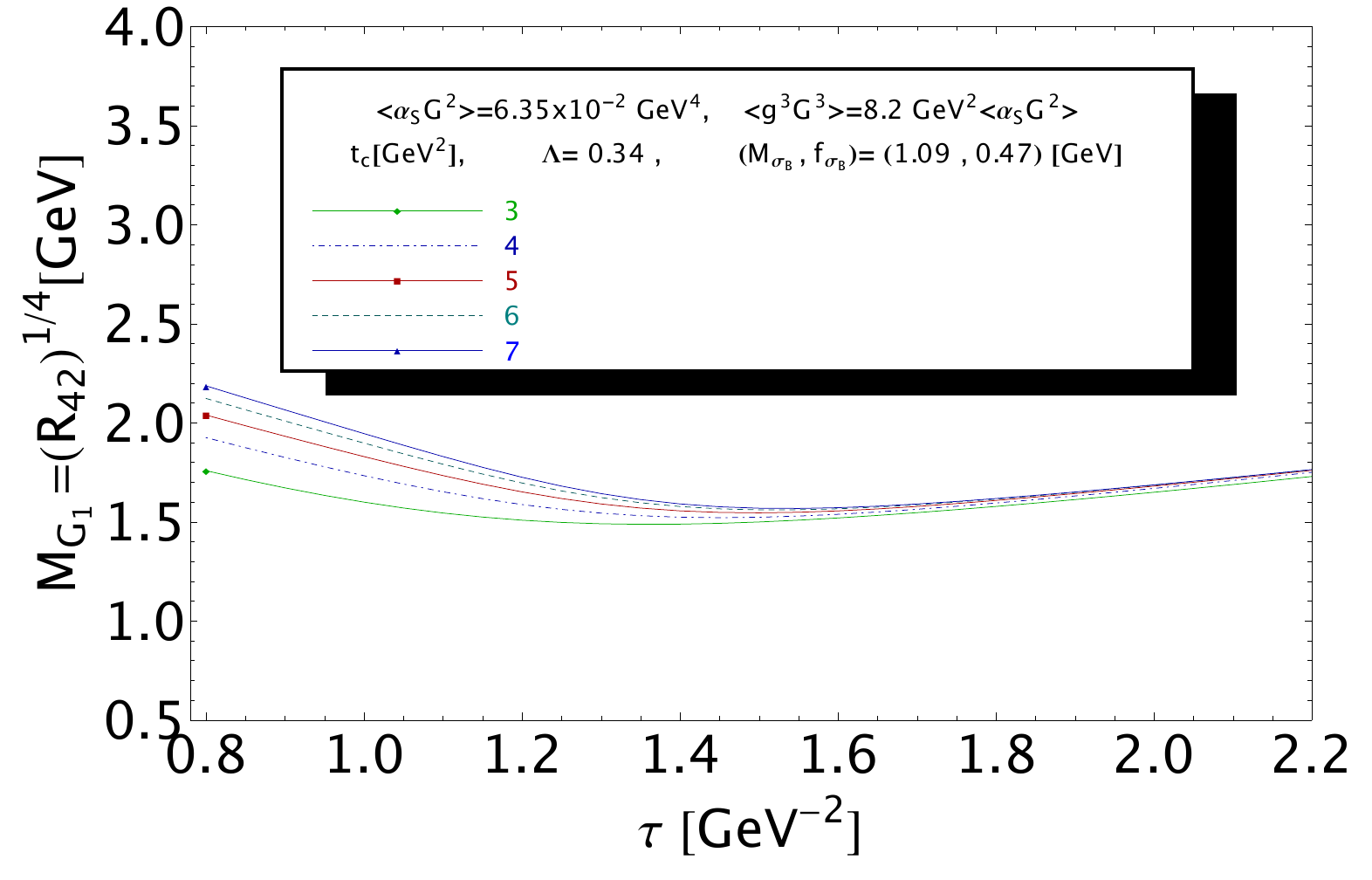} 
%\vspace{0.25cm}
%\centerline {\hspace*{-6cm} \bf b) }
%\includegraphics[width=10cm]{L31.pdf}
\vspace*{-0.5cm}
\caption{\footnotesize  $M_{G_1}$   as a function of $\tau$ at N2LO for different values of $t_c$ from a two resonances parametrization. }
% from : a) ${\cal R}^c_{21}$; b) ${\cal R}^c_{31}$} 
\label{fig:r42}
\end{center}
%\vspace*{-0.75cm}
\end{figure}

%%%%%%%%%%%%%%%%%%%%%%%%%%%%%%%%%%%%%%%%%%5
Among the different combinations of sum rules, we find that ${\cal R}^c_{42}$ can give a reliable prediction of the medium ground state  $G_1$ because  ${\cal L}^c_2$ and ${\cal L}^c_4$ stabilize at about the same value of $\tau\simeq  (1.2\sim 1.5)$ GeV$^{-2}$ (see Figs\,\ref{fig:l2} and \,\ref{fig:l4}).
 We show in Fig.\,\ref{fig:r42} the $\tau$ behaviour of  ${\cal R}_{42}$ for different values of $t_c$. 

%%%%%%%%%%%%%%%%%%%%%%%%%%%%%%%%%%%%%%%
%%%%%%%%%%%%%%%%%%%%%%%%%%%%%%%%%%%%%%%

%%%%%%%%%%%%%%%%%%%%%%%%%%%%%%%%%%%%%%%
\subsubsection*{\d  The case :  two resonances  $\oplus$ QCD continuum} 
%%%%%%%%%%%%%%%%%%%%%%%%%%%%%%%%%%%%%%%
We obtain for $t_c=3$ to 6 GeV$^2$:
\beq
M_{G_1}= 1524(36)_{t_c}(7)_\tau(115)_\Lambda(2)_{G^2}(2)_{G^3}(0)_{\lambda^2}(6)_{M_\sigma}(11)_{f_\sigma}= 1524(121)~{\rm MeV}~,
\label{eq:mg1}
\eeq
where the error comes mainly from the one of $\Lambda$. We have used $f_{\sigma_B}=0.47$ as anticipated from the next section.
%%%%%%%%%%%%%%%%%%%%%%%%%%%%%%%%%%%%%%%
\subsubsection*{\d  The case : one resonance $\oplus$ QCD continuum} 
%%%%%%%%%%%%%%%%%%%%%%%%%%%%%%%%%%%%%%%
A similar analysis leads to:
\beq
M_{G_1}=1515(123)~{\rm MeV}
\label{eq:mg11}
\eeq
indicating that the $\sigma_B$ effect to the value of the mass is quite small (about 9 MeV).
%%%%%%%%%%%%%%%%%%%%%%%%%%%%%%%%%%%%%%%%%%%
\section{Decay constants $f_{\sigma_B}$ and $f_{G_1}$}
%%%%%%%%%%%%%%%%%%%%%%%%%%%%%%%%%%%%%%%%%%%
To extract these decay constants, we work simultaneously with ${\cal L}^c_2$, ${\cal L}^c_3$ and ${\cal L}^c_{-1}$ and use an iteration procedure. The two former are expected to be more sensitive to $f_{G_1}$ while the third to $f_{\sigma_B}$. 

\begin{figure}[H]
\begin{center}
\includegraphics[width=9.5cm]{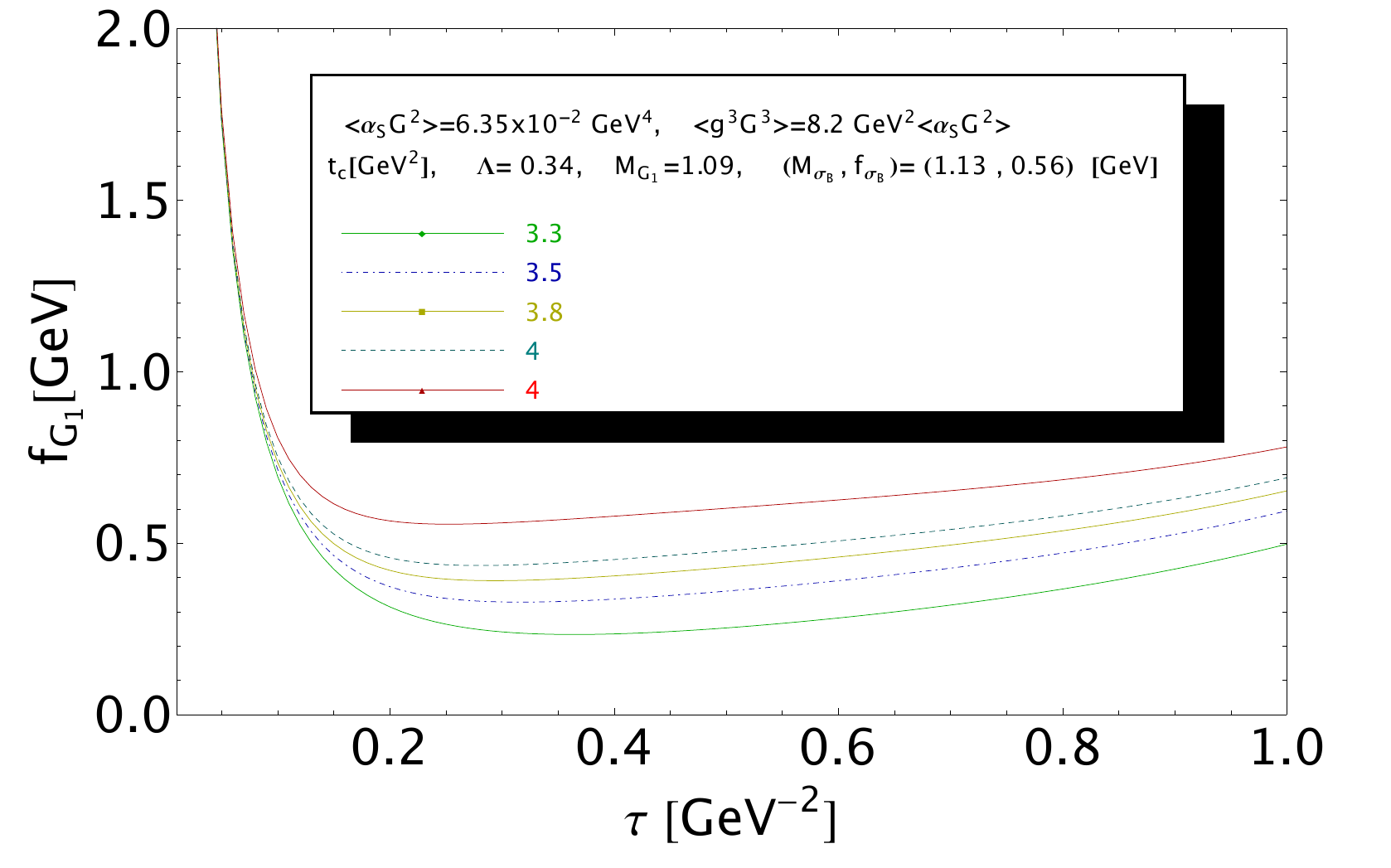} 
\vspace*{-0.5cm}
\caption{\footnotesize  $f_{G_1}$  from ${\cal L}^c_2$  as a function of $\tau$ at N2LO for different values of $t_c$ from a two resonances parametrization and for given values of $(M_{\sigma_B}, (f_{\sigma_B})$. }
\label{fig:fg11}
\end{center}
%\vspace*{-0.75cm}
\end{figure}

\begin{figure}[H]
\begin{center}
\includegraphics[width=9.cm]{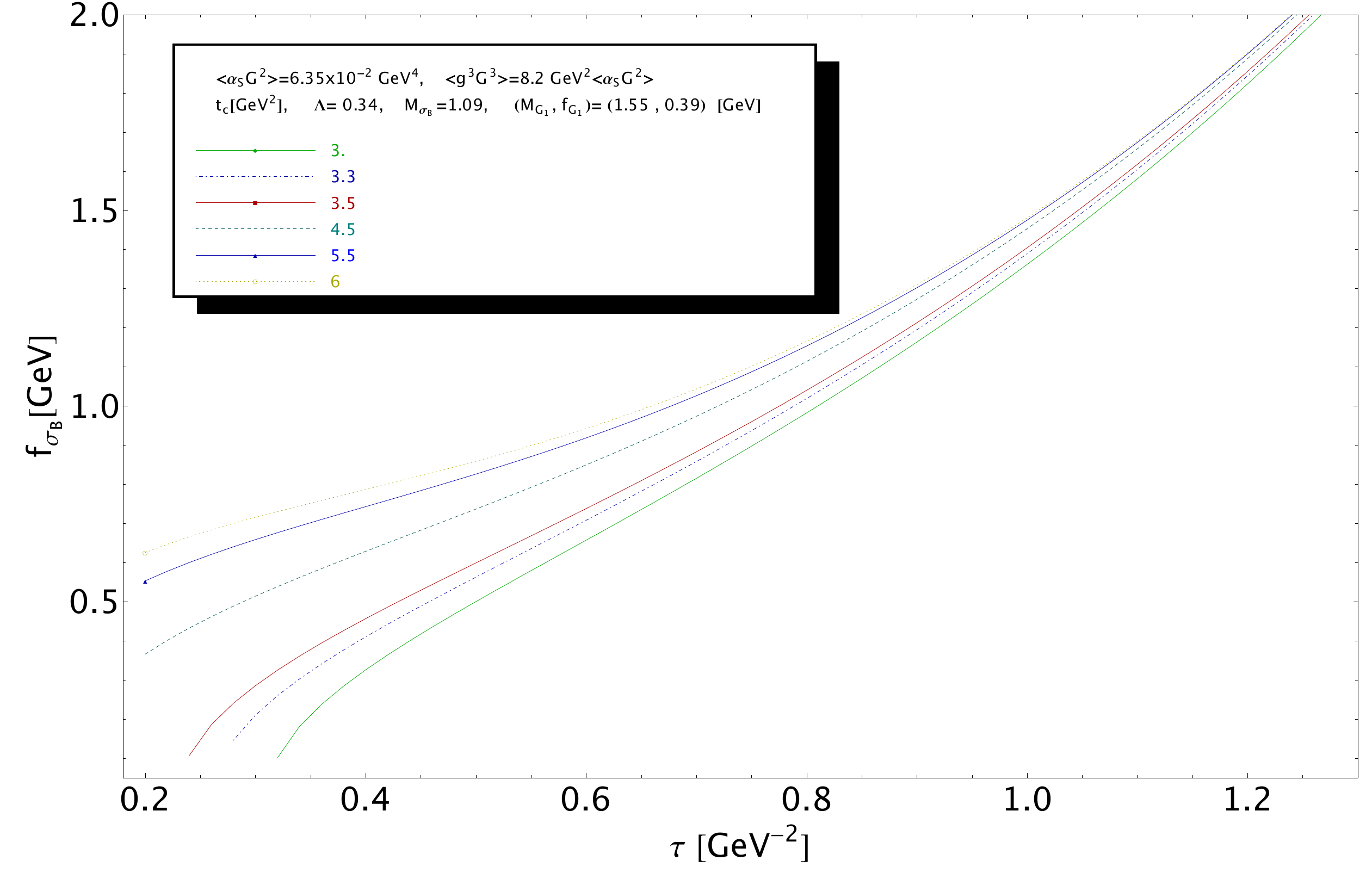} 
\vspace*{-0.5cm}
\caption{\footnotesize  $f_{\sigma_B}$   as a function of $\tau$ at N2LO for different values of $t_c$ from a two resonances parametrization and for given values of $(M_{G_1}, f_{G_1})$. }
\label{fig:fsigma}
\end{center}
%\vspace*{-0.75cm}
\end{figure}

 \d We shall take $M_{\sigma_B}$ from Eq.\,\ref{eq:msigmaf} and $M_{G_1}=1.548$ GeV anticipated from Eq.\,\ref{eq:mg1}. We start the iteration by using  $f_{\sigma_B}$ =800 MeV from\,\cite{VENEZIA,SNG}. Then, we compare the value of $f_{G_1}$   from ${\cal L}^c_2$, ${\cal L}^c_3$. Then, we use the common solution of $f_{G_1}$   from the two sum rules into ${\cal L}^c_{-1}$ to extract $f_{\sigma_B}$. We continue the iterations until we reach convergent solutions. The  analysis corresponding to the final value of the decay constants is shown in Figs.\,\ref{fig:fg11} to \,\ref{fig:fsigma}. 
 
 -- In Figs.\,\ref{fig:fg11} and \,ref{fig:fg12}, we show the $\tau$ and $t_c$ behaviours of $f_{G_1}$ given the value of $f_{\sigma_B}$. 
--  In Fig.\,\ref{fig:fg11}, we show the results from ${\cal L}^c_2$, ${\cal L}^c_3$ at the $\tau$ minimum or plateau from which we deduce the common solution for $f_{G_1}$ for fixed $f_{\sigma_B}$ which allows to fix :
\beq
t_c\simeq (3.3\pm0.3)\,{\rm GeV}^2~.
\eeq

--  In Fig.\,\ref{fig:fsigma}, we show the $\tau$ and $t_c$ behaviours of $f_{\sigma_B}$ where we take the value $\tau=0.5$ GeV$^{-2}$ at the inflexion point which is more pronounced for higher values of $t_c$. 

 Then, we obtain :
 \bea
f_{G_1}&=&394(45)_{t_c}(2)_{\tau}(22)_\Lambda(0)_{G^2}(1)_{G^3}(23)_{M_{\sigma_B}}(9)_{f_{\sigma_B}}(94)_{M_{G_1}}= 394(109)~{\rm MeV}~~~{\rm from}~~~{\cal L}_3^c\nnb\\
f_{\sigma_B}&=&563(36)_{t_c}(64)_{\tau}(81)_\Lambda(9)_{G^2}(11)_{G^3}(10)_{M_\sigma}(3)_{M_{G_1}}
(107)_{f_{G_1}}= 563(154)~{\rm MeV}~~~{\rm from}~~~{\cal L}_{-1}^c
\label{eq:decay-const}
\eea
One can note that value of $f_{G_1}$  agrees perfectly  with the one in\,\cite{SNG} while $f_{\sigma_B}$ is in the range of the one obtained in\,\cite{VENEZIA,SNG}. 
 
 %%%%%%%%%%%%%%%%%%%%%%%%%%%%%%%%%%%%%%%
%%%%%%%%%%%%%%%%%%%%%%%%%%%%%%%%%%%%%%%

%%%%%%%%%%%%%%%%%%%%%%%%%%%%%%%%%%%%%%%
%%%%%%%%%%%%%%%%%%%%%%%%%%%%%%%%%%%%%%%

%%%%%%%%%%%%%%%%%%%%%%%%%%%%%%%%%%%%%%%
\begin{figure}[H]
\begin{center}
\includegraphics[width=10cm]{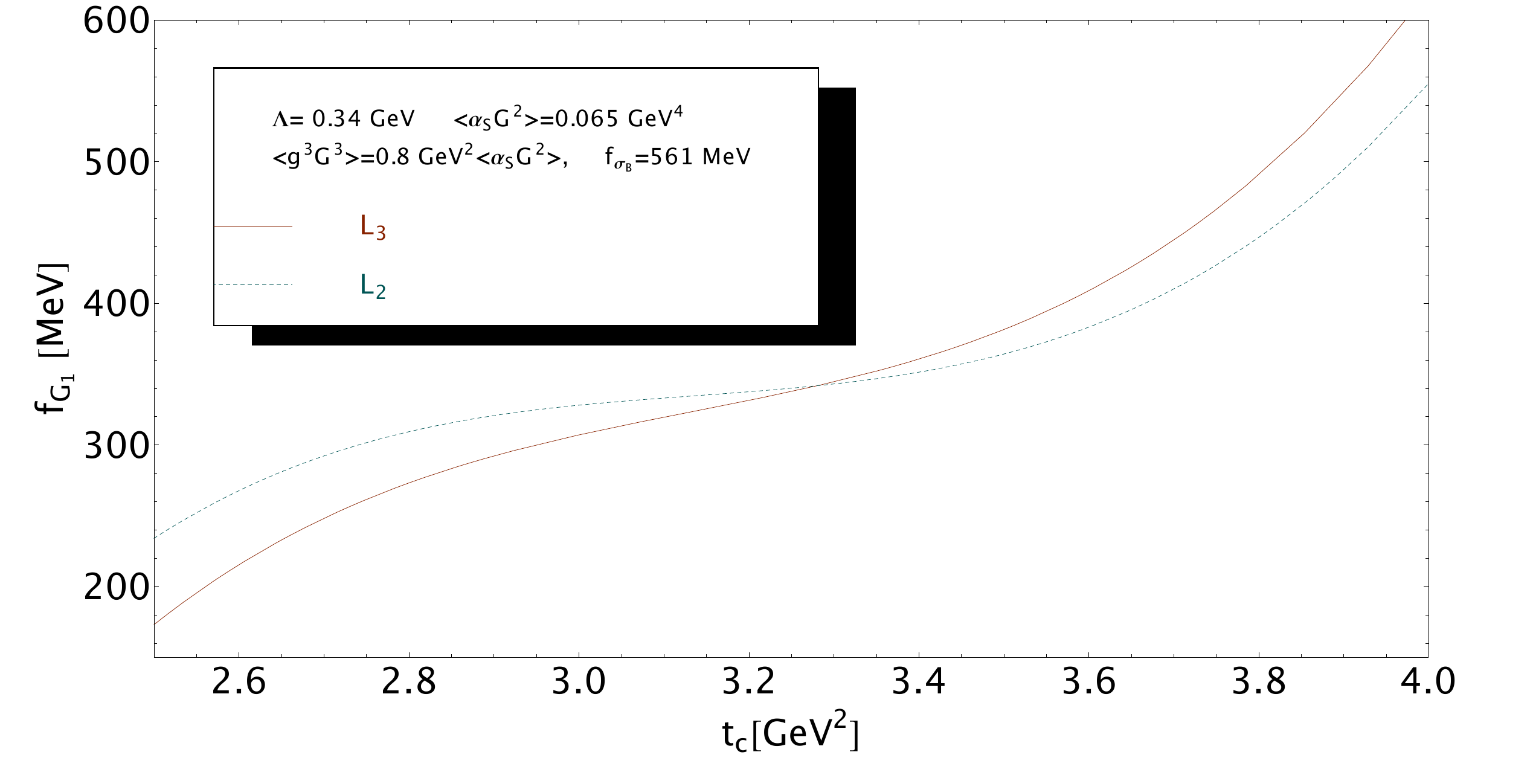} 
\vspace*{-0.5cm}
\caption{\footnotesize  $f_{G_1}$  from ${\cal L}^c_2$ and ${\cal L}^c_3$  from the $\tau$-minimum $\tau\simeq 0.5$ GeV$^{-2}$  from previous figures as a function of  $t_c$ for given values of $(M_{\sigma_B}, f_{\sigma_B})$. }
\label{fig:fg13}
\end{center}
%\vspace*{-0.75cm}
\end{figure} 
%%%%%%%%%%%%%%%%%%%%%%%%%%%%%%%%%%%%%%%
%%%%%%%%%%%%%%%%%%%%%%%%%%%%%%%%%%%%%%%
%%%%%%%%%%%%%%%%%%%%%%%%%%%%%%%%%%%%%%%

%%%%%%%%%%%%%%%%%%%%%%%%%%%%%%%%%%%%%%%%%%%
%\section{The first radial excitation}
 %%%%%%%%%%%%%%%%%%%%%%%%%%%%%%%%%%%%%%%%%%%
\section{First radial excitation $\sigma'_B$ of $\sigma_B$}
%%%%%%%%%%%%%%%%%%%%%%%%%%%%%%%%%%%%%%%%%%%
%%%%%%%%%%%%%%%%%%%%%%%%%%%%%%%%%%%%%%%
\subsubsection*{\d The $\sigma'_B$ mass}
\begin{figure}[H]
%\vspace*{-0.25cm}
\begin{center}
\includegraphics[width=10cm]{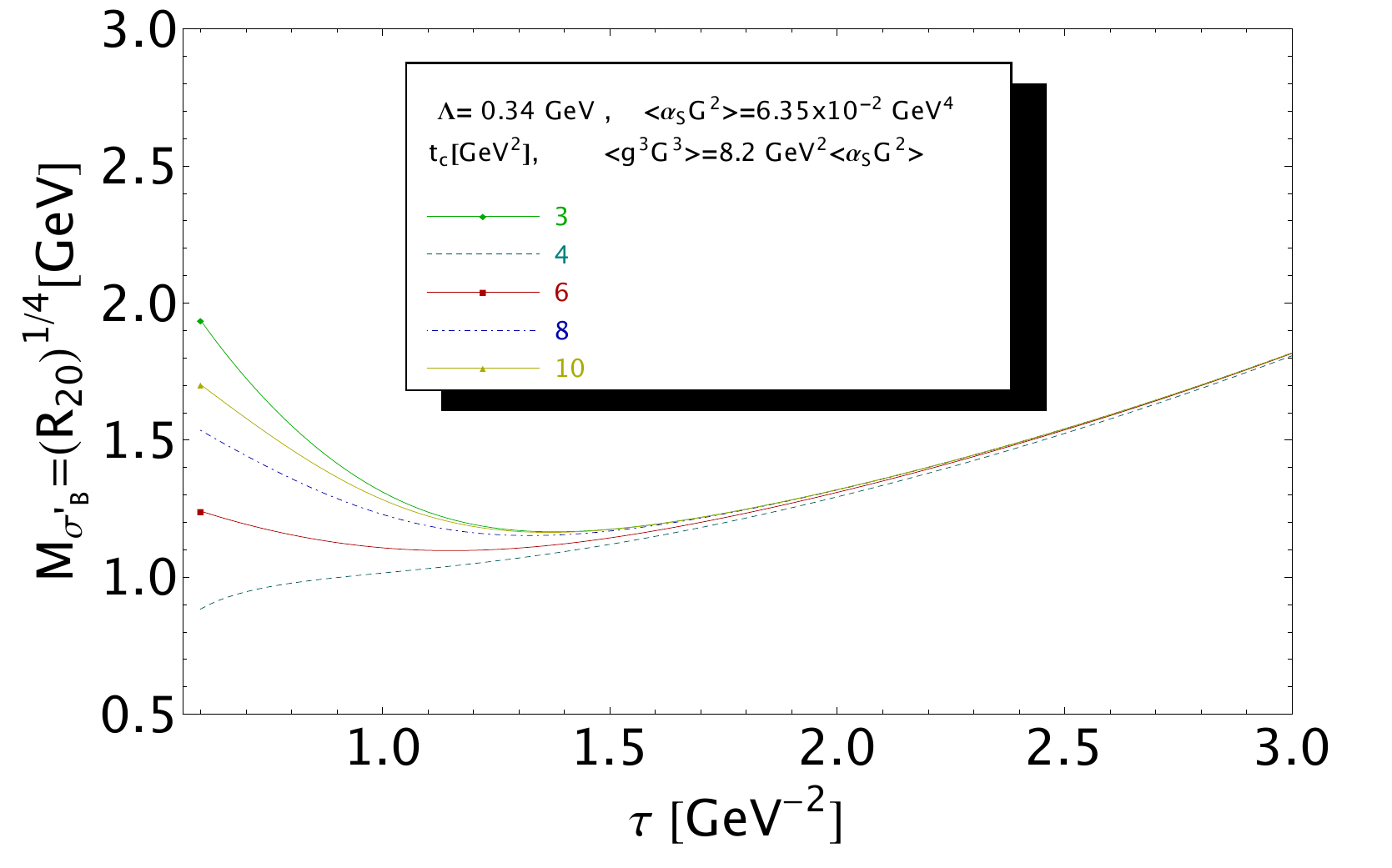}
\vspace*{-0.5cm}
\caption{\footnotesize  $M_{\sigma'_B}$   as a function of $\tau$ at N2LO for different values of $t_c$ within a three resonances parametrization
from  ${\cal R}^c_{20}$.} 
\label{fig:l20-radial}
\end{center}
%\vspace*{-0.75cm}
\end{figure}

%%%%%%%%%%%%%%%%%%%%%%%%%%%%%%%%%%%%%%%%
We attempt to extract its mass from ${\cal R}^c_{20}$ which we have used to estimate $M_{\sigma_B}$. In so doing,  we subtract the contributions
of the $\sigma_B$ and $G_1$ at take the result for very large values of $t_c$ i.e. by  replacing the QCD continuum contribution by the $\sigma'_B$.
In this way, we obtain the result in Fig.\,\ref{fig:l20-radial} from ${\cal R}^c_{20}$. %and in Fig.\,\ref{fig:l42-radial} from ${\cal L}^2_{42}$. 
%%%%%%%%%%%%%%%%%%%%%%%%%%%%%%%%%%%%%%%
%%%%%%%%%%%%%%%%%%%%%%%%%%%%%%%%%%%%%%%

We obtain for $t_c\simeq (6\sim 8) \,$ GeV$^2$: at the $\tau$-minimum :
\beq
M_{\sigma'_B}= 1110(25)_{t_c}(10)_\tau(110)_\Lambda(0)_{G^2}(3)_{G^3}(1)_{M_{\sigma_B}}(12)_{f_{\sigma_B}}
(7){M_{G_1}}(23)_{f_{G_1}}=1110(117)~{\rm MeV},
\label{eq:msigmap}
\eeq
where the contributions of $\sigma_B$ and $G_1$ have been included after iteration from 2  to 3 resonances resonances parametrization of the spectral function and using the values of the $\sigma _{B}$ and $G_{1}$ masses and decay constants from 2 resonances parametrization. 
%from which we deduce the average:
%\beq
%M_{\sigma'_B}=1283(93)~{\rm MeV}~.
%\label{eq:msigmap}
%\eeq
%%%%%%%%%%%%%%%%%%%%%%%%%%%%%%%%%%%%%%%
\subsubsection*{\d The $\sigma'_B$ decay constant}
%%%%%%%%%%%%%%%%%%%%%%%%%%%%%%%%%%%%%%%%
%%%%%%%%%%%%%%%%%%%%%%%%%%%%%%%%%%%%%%%
\begin{figure}[hbt]
\vspace*{-0.25cm}
\begin{center}
\includegraphics[width=10cm]{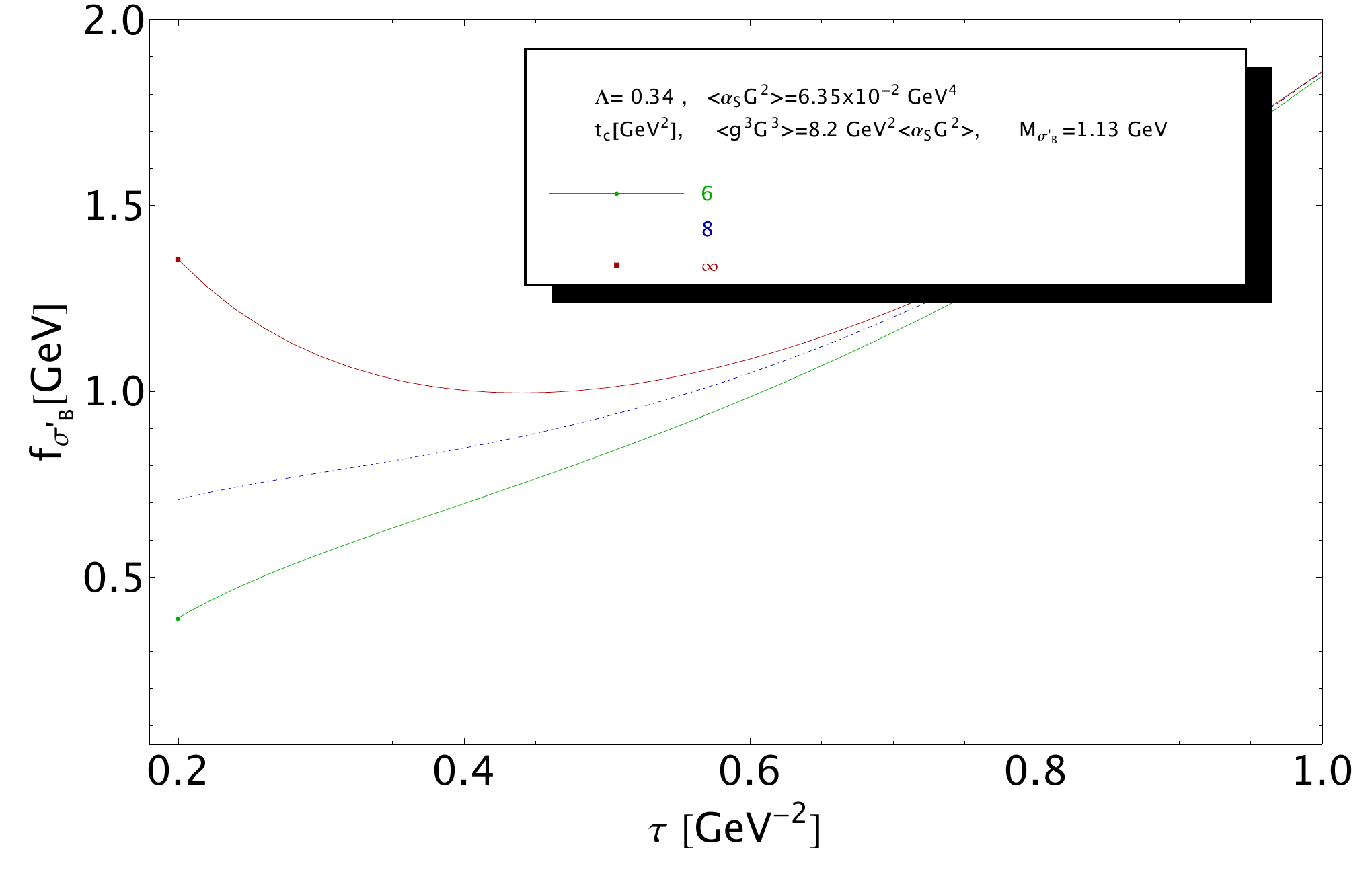}
\vspace*{-0.5cm}
\caption{\footnotesize  $f_{\sigma'_B}$   as a function of $\tau$ at N2LO for different values of $t_c$ within a three resonances parametrization
from  ${\cal L}^c_{-1}$.} 
\label{fig:l42-radial}
\end{center}
%\vspace*{-0.75cm}
\end{figure} 
%%%%%%%%%%%%%%%%%%%%%%%%%%%%%%%%%%%%%%%
We extract the decay constant from  ${\cal L}^c_{-1}$  using the same range of $t_c$ as for $M_{\sigma'_B}$ and taking $\tau\simeq 0.4$ GeV$^{-2}$ (Fig.\,\ref{fig:l42-radial}). We obtain:
\bea
f_{\sigma'_B}&=& 648(75)_{t_c}(43)_\tau(124)_\Lambda(20)_{G^2}(7)_{G^3}(24)_{M_{\sigma_B}}(137)_{f_{\sigma_B}}
(0){M_{G_1}}(56)_{f_{G_1}} (33)_{M_{\sigma'_B}}\nnb\\
&=& 648(216)~{\rm MeV}~~~{\rm from}~~~{\cal L}^c_{-1}.
\label{eq:fsigmap}
\eea
Anticipating the estimate in Section\,\ref{sec:sigmap} by extracting the
decay constant from $f_0(1.37)\to 2(\pi\pi)_S$ data through a LEV-SR, we shall work instead with the more precise value:
\beq
f_{\sigma'_B}=329(30)~{\rm MeV}.
\eeq
%%%%%%%%%%%%%%%%%%%%%%%%%%%%%%%%%%%%%%%%%%%%%%%%%%
\section{Iterative improvements of the previous estimates}
%%%%%%%%%%%%%%%%%%%%%%%%%%%%%%%%%%%%%%%%%%%%%%%%%%
In the following, we study the effect of the ${\sigma'_B}$ which is below $M_{G_1}$ on the previous estimates
%%%%%%%%%%%%%%%%%%%%
\subsection*{\b The mass $M_{\sigma_B}$}
%%%%%%%%%%%%%%%%%%%%
We include the contribution of the $\sigma'_B$ which is below the $G_1$ mass. Using ${\cal R}_{20}^c$ as in previous section, the curves are similar to the ones in  Fig.\,\ref{fig:r20}. The mass in Eq.\,\ref{eq:msigmaf} becomes :
\beq
M_{\sigma_B}= 1070(126)(1)_{f_{\sigma'_B}}(1)_{M_{\sigma'_B}}~{\rm MeV}~,
\label{eq:msigmaf1}
\eeq
which is our final estimate. 
%%%%%%%%%%%%%%%%%%%%
\subsection*{\b The mass $M_{G_1}$}
%%%%%%%%%%%%%%%%%%%%
Including the contribution of the $\sigma'_B$  into ${\cal R}_{42}^c$, the mass in Eq.\,\ref{eq:mg1} becomes :
\beq
M_{G_1}= 1548(118)(15)_{f_{\sigma'_B}}(13)_{M_{\sigma'_B}}=1548(120)~{\rm MeV}~,
\label{eq:mgf1}
\eeq
which is our final estimate.
%%%%%%%%%%%%%%%%%%%%%%%%%%%%%%%%%%
\subsection*{\b The decay constants  $f_{\sigma_B}$ and $f_{G_1}$}
%%%%%%%%%%%%%%%%%%%%%%%%%%%%%%%%%%
We repeat the procedure used previously by extracting simultaneously  $f_{\sigma_B}$ and $f_{G_1}$ from ${\cal L}^c_{-1}$ and from ${\cal L}_2^c,~ {\cal L}_3^c$ by including the $\sigma'_B$ contribution.  The value of $t_c$ at which the solutions from ${\cal L}_2^c,~ {\cal L}_3^c$ meet is slightly shifted from 3.3 to 3.2 GeV$^2$. We obtain the final estimate:
\beq
f_{\sigma_B}=456(154)(8)_{M_{\sigma'_B}}(12)_{f_{\sigma'_B}}=456(157)~{\rm MeV},
\label{eq:fsigmaf1}
\eeq
and:
\beq
f_{G_1}=365(109)(17)_{M_{\sigma'_B}}(1)_{f_{\sigma'_B}}=365(110)~{\rm MeV}.
\label{eq:fgf1}
\eeq
 %%%%%%%%%%%%%%%%%%%%%%%%%%%%%%%%%%%%%%%%%%%
\section{The first radial excitation $G'_1$ of $G_1$}
%%%%%%%%%%%%%%%%%%%%%%%%%%%%%%%%%%%%%%%%%%%
\subsubsection*{\b The $G'_1$ mass}
\begin{figure}[H]
%\vspace*{-0.25cm}
\begin{center}
\includegraphics[width=10cm]{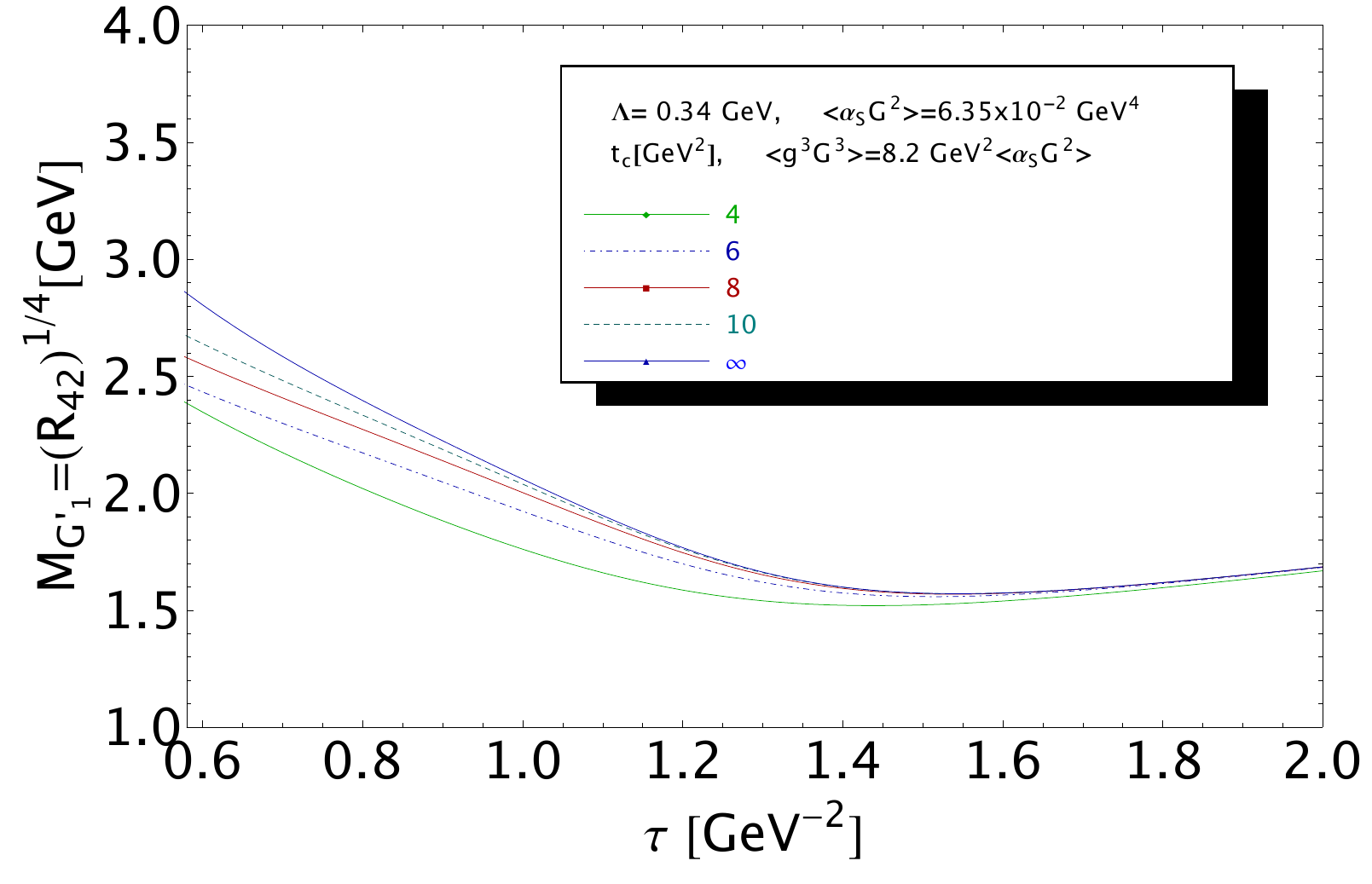} 
\vspace*{-0.5cm}
\caption{\footnotesize  $M_{G'_1}$   as a function of $\tau$ at N2LO for  from a four resonances parametrization. }
\label{fig:mGP1}
\end{center}
\end{figure}

%%%%%%%%%%%%%%%%%%%%%%%%%%%%%%%%%%%%%%%%
\d We attempt to estimate the mass of the first radial excitation by replacing the QCD continuum by a 4th resonance in the sum rule ${\cal R}^c_{42}$ which we have used to extract $M_{G_1}$. 
The analysis is shown in Fig.\ref{fig:mGP1}
%%%%%%%%%%%%%%%%%%%%%%%%%%%%%%%%%%%%%%%
%%%%%%%%%%%%%%%%%%%%%%%%%%%%%%%%%%%%%%%

We deduce for $t_c\simeq (4\sim 8)$ GeV$^2$ at the $\tau$-minimum:
\beq
M_{G'_1}=1563(23)_{t_c}(2)_\tau(138)_\Lambda(3)_{G^2}(3)_{G^3}(6)_{f_{\sigma_B}}(2)_{M_{\sigma_B}}(3)_{f_{G_1}}(9)_{M_{G_1}}(11)_{f_{\sigma'_B}}(1)_{M_{\sigma'_B}}=1563(141)~{\rm MeV}
\label{eq:mg1p}
\eeq
where the error comes mainly from $\Lambda$. We have included the contributions of the $\sigma_B, \sigma'_B$ and $G_1$. 
%%%%%%%%%%%%%%%%%%%%%%%%%%%%%%%%%%%%%%%
\begin{figure}[H]
%\vspace*{-0.25cm}
\begin{center}
\includegraphics[width=10cm]{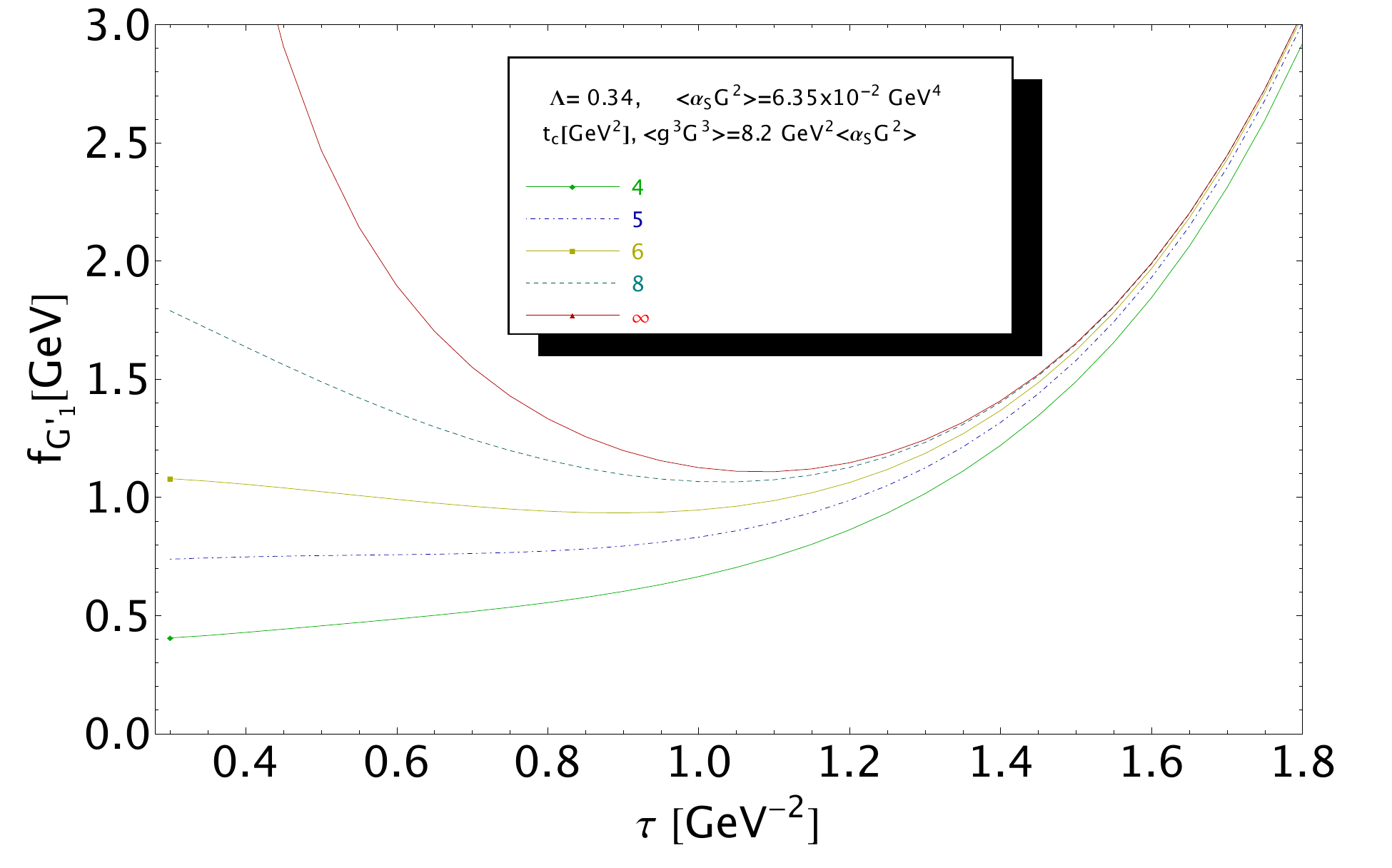} 
\vspace*{-0.5cm}
\caption{\footnotesize  $f_{G'_1}$   as a function of $\tau$ at N2LO  from a three resonances parametrization. }
\label{fig:fGP1}
\end{center}
\end{figure} 
%%%%%%%%%%%%%%%%%%%%%%%%%%%%%%%%%%%%%%%%%%%
%%%%%%%%%%%%%%%%%%%%%%%%%%%%%%%%%%%%%%%%%%%
\subsubsection*{\b The decay constant $f_{G'_1}$}
%%%%%%%%%%%%%%%%%%%%%%%%%%%%%%%%%%%%%%%%
To extract the decay constant we shall work with the sum rules ${\cal L}_2^c$ or/and  ${\cal L}_3^c$ used previously to get $f_{G_1}$ where ${\cal L}_2^c$
presents a $\tau$-minimum (Fig.\,\ref{fig:fGP1}) while  ${\cal L}_3^c$ an inflexion point. From ${\cal L}_2^c$, one deduces:
\bea
f_{G'_1}&=&1000(65)_{t_c}(11)_{\tau}(150)_\Lambda(1)_{G^2}(2)_{G^3}(15)_{f_{\sigma_B}}(58)_{M_{\sigma_B}}(53)_{f_{G_1}}(63)_{M_{G_1}}(1)_{f_{\sigma'_B}}(8)_{M_{\sigma'_B}}(125)_{M_{G'_1}}\nnb\\
&=&1000(230)~{\rm MeV}.
\label{eq:fg1p}
\eea
%%%%%%%%%%%%%%%%%%
%%%%%%%%%%%%%%%%%%%%%%%%%%%%%%%%%%%%%%%%%%%
\section{The second radial excitation $G_2$}
 %%%%%%%%%%%%%%%%%%%%%%%%%%%%%%%%%%%%%%%%%%%
\subsection*{\b The mass $M_{G_2}$}
\begin{figure}[H]
%\vspace*{-0.25cm}
\begin{center}
\includegraphics[width=10cm]{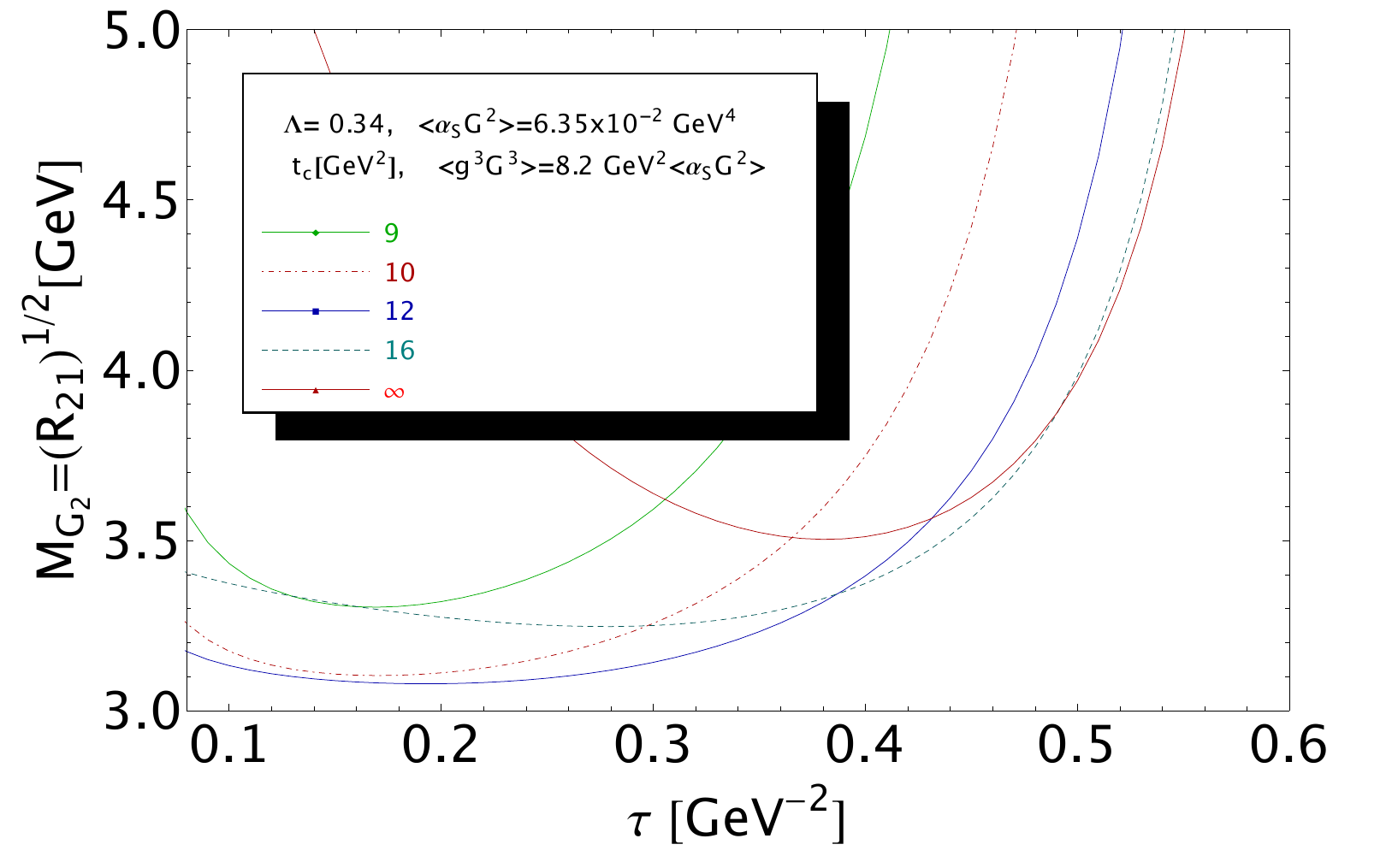} \\
\vspace*{-0.5cm}
\caption{\footnotesize  $M_{G_2}$   as a function of $\tau$ at N2LO  for a four resonances parametrization
from ${\cal R}^c_{21}$.} 
\label{fig:r21}
\end{center}
%\vspace*{-0.75cm}
\end{figure}

\begin{figure}[H]
%\vspace*{-0.25cm}
\begin{center}
\includegraphics[width=10cm]{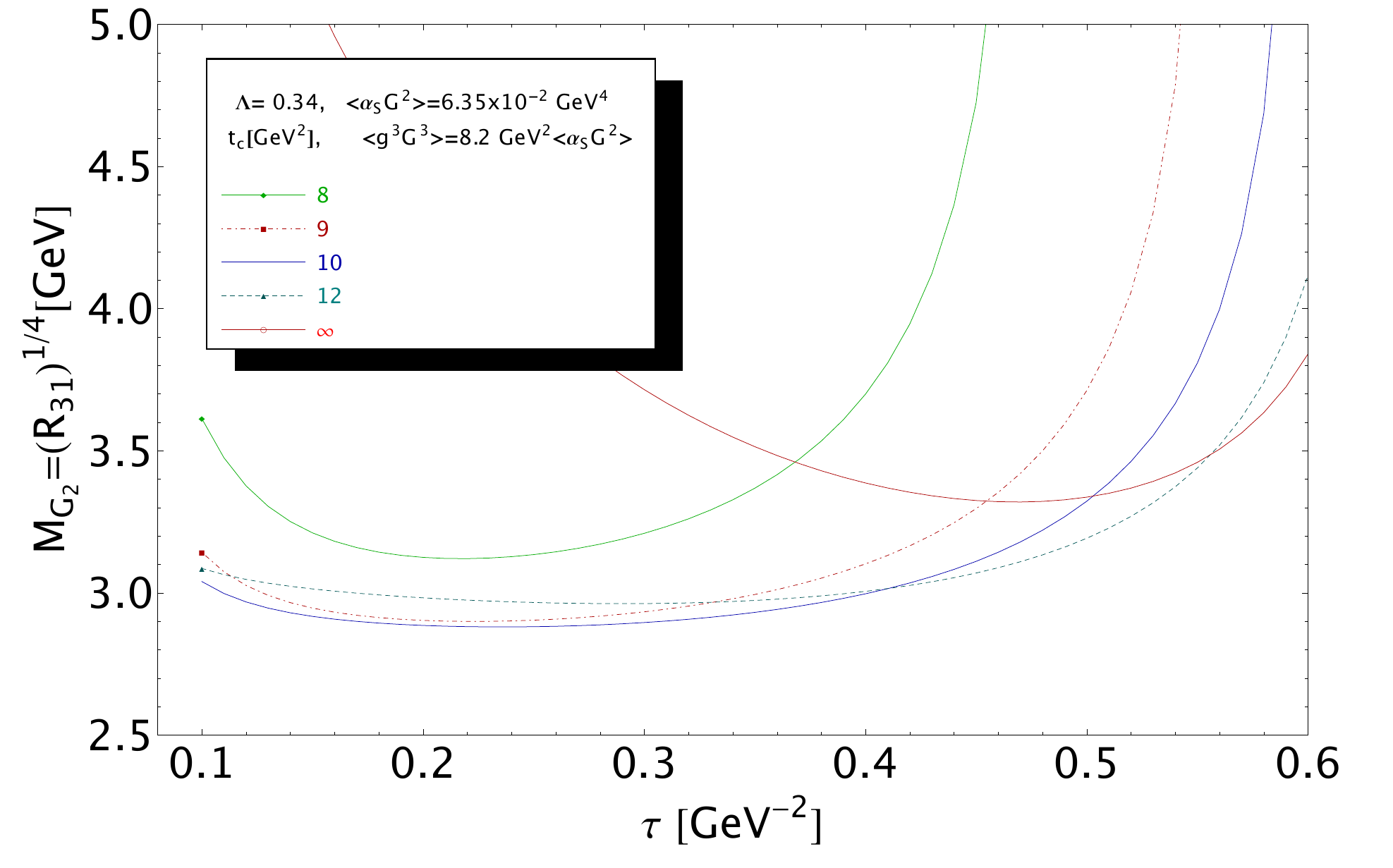}
\vspace*{-0.5cm}
\caption{\footnotesize  $M_{G_2}$   as a function of $\tau$ at N2LO  for a four resonances parametrization
from  ${\cal R}^c_{31}$} 
\label{fig:r31}
\end{center}
%\vspace*{-0.75cm}
\end{figure}

%%%%%%%%%%%%%%%%%%%%%%%%%%%%%%%%%%%%%%%%%%%%%%%%%%%%%%%%%%%%%
In so doing, we work with
some judicious choice of sum rules which optimize at smaller values of $\tau$ and are more sensitive to the high-mass meson contributions. 
This criterion is satisfied by ${\cal R}^c_{21}$ and ${\cal R}^c_{31}$ which one can see in Figs\,\ref{fig:r21} and \,\ref{fig:r31} where the contributions of the $\sigma_B,~\sigma'$ and $G_1,~G'_1$ have been subtracted. 
${\cal R}^c_{21}$ has a $(\tau,t_c)$ minimum at (0.18,12) in units of (GeV$^{-2}$, GeV$^2$) while ${\cal R}^c_{31}$ has the minimum at (0.24,10). 
%%%%%%%%%%%%%%%%%%%%%%%%%%%%%%%%%%%%%%%
%%%%%%%%%%%%%%%%%%%%%%%%%%%%%%%%%%%%%%%

%%%%%%%%%%%%%%%%%%%%%%%%%%%%%%%%%%%%%%%
%%%%%%%%%%%%%%%%%%%%%%%%%%%%%%%%%%%%%%%

 We obtain :
\bea
M_{G_2}
&=& 3079(15)_{t_c}(10)_\tau(63)_\Lambda(1)_{G^2}(2)_{G^3}(10)_{M_{\sigma_B}}(7)_{f_{\sigma_B}}
(20)_{M_{G_1}}(36)_{f_{G_1}} (6)_{M_{\sigma'}}(1)_{f_{\sigma'}}(193)_{M_{G'_1}}(207)_{f_{G'_1}}\nnb\\
&=& 3079(294)~{\rm MeV}~~~{\rm from}~~~{\cal R}^c_{21},
\eea
and:
\bea
M_{G_2}&=& 2879(9)_{t_c}(3)_\tau(64)_\Lambda(1)_{G^2}(2)_{G^3}(10)_{M_{\sigma_B}}(8)_{f_{\sigma_B}}
(25)_{M_{G_1}}(40)_{f_{G_1}} (6)_{M_{\sigma'}}(0)_{f_{\sigma'}}(226)_{M_{G'_1}}(233)_{f_{G'_1}}\nnb\\
&=& 2879(334)~{\rm MeV}~~~{\rm from}~~~{\cal R}^c_{31}~,
\eea
from which we deduce the mean value:
\beq
M_{G_2}=2992(221)~{\rm MeV}. 
\label{eq:mg2}
\eeq
This second radial excitation is far above the ground states and first radial excitations and may mix with the trigluonium scalar ground state\,\cite{PABAN,SNG}:
\beq
M_{3g}\simeq 3.1~{\rm GeV}.
\eeq
 %%%%%%%%%%%%%%%%%%%%%%%%%%%%%%%%%%%%%%%%%%%
\subsubsection*{\d The decay constant $f_{G_2}$}
%%%%%%%%%%%%%%%%%%%%%%%%%%%%%%%%%%%%%%%%%%%
%%%%%%%%%%%%%%%%%%%%%%%%%%%%%%%%%%%%%%%
\begin{figure}[H]
%\vspace*{-0.25cm}
\begin{center}
\includegraphics[width=10cm]{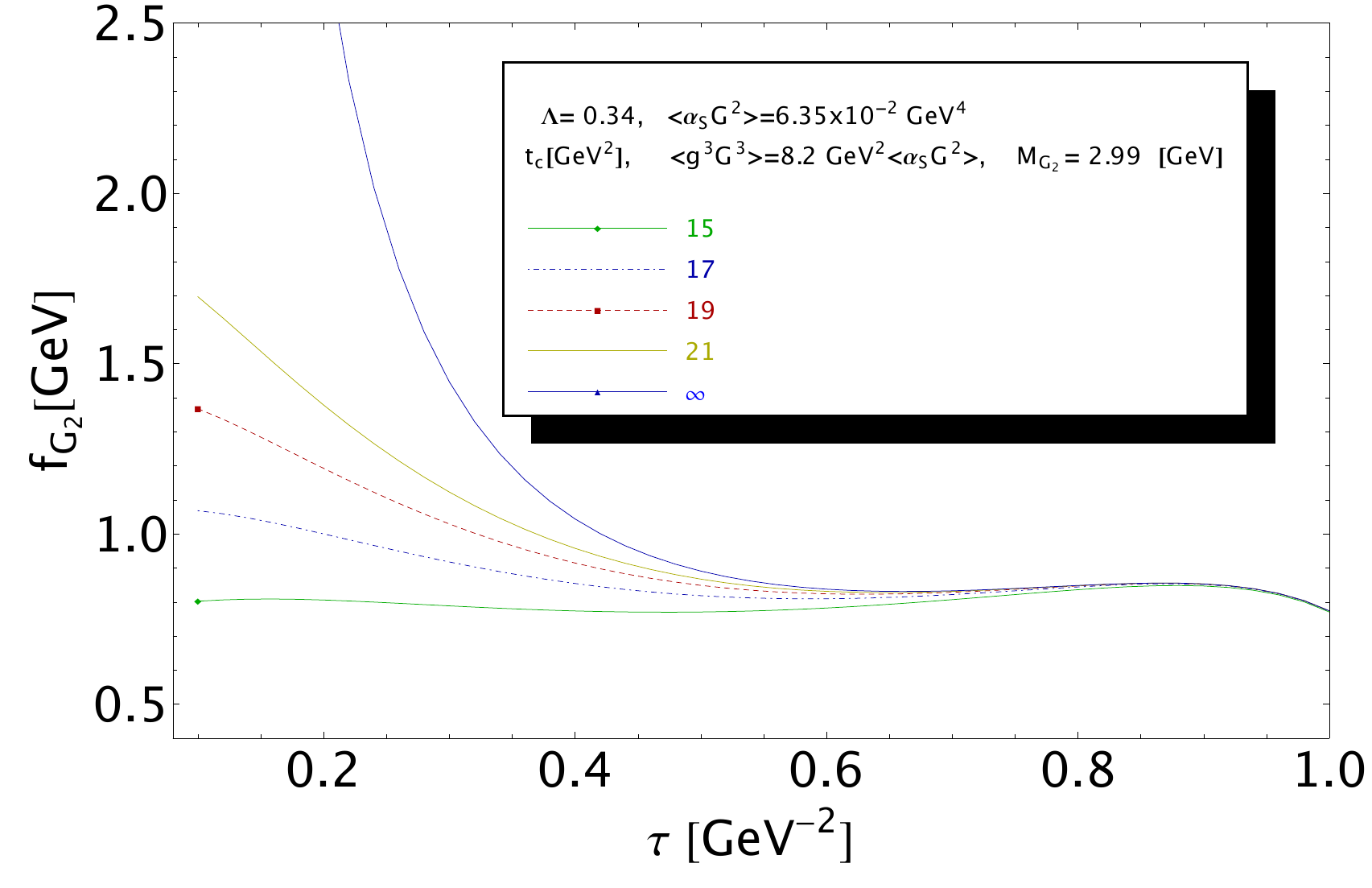}
\vspace*{-0.5cm}
\caption{\footnotesize  $f_{G_2}$   as a function of $\tau$ at N2LO for different values of $t_c$ for a four resonances parametrization
from  ${\cal L}^c_{3}$} 
\label{fig:fg2}
\end{center}
%\vspace*{-0.75cm}
\end{figure} 
%%%%%%%%%%%%%%%%%%%%%%%%%%%%%%%%%%%%%%%

We shall work with ${\cal L}^c_3$  (used to get $f_{G_1}$) for extracting the decay constant $f_{G_2}$. The analysis is shown in Fig.\,\ref{fig:fg2}  where one obtains:
\beq
f_{G_2}= 797(27)_{t_c}(57)_\Lambda(1)_{M_{\sigma'_B}}
(3){M_{G_1}}(3)_{f_{G_1}} (36)_{M_{G'_1}}(11)_{f_{G'_1}}(11)_{M_{G_2}}
=797(74)~{\rm MeV}.
\label{eq:fg2}
\eeq
The errors due to the other terms not quoted are zero. 
One should note that, like in the case of $f^{\rm eff}_{\sigma'}$ which is comparable in size, the previous coupling of the $G_2$ to the gluonic correlator can be interpreted as the sum of effective couplings of all higher states contributing to the spectral function. 

%%%%%%%%%%%%%%%%%%%%%%%%%%%%%%%%%%%%%
\section{Truncation of  the PT series, tachyonic gluon mass and the OPE}
%%%%%%%%%%%%%%%%%%%%%%%%%%%%%%%%%%%%%
\subsection*{\b Truncation of the PT series}
%%%%%%%%%%%%%%%%%%%%%%%%%%%%%%%%%%%%%
One can notice from the expression of the two-point correlator in Eq.\,\ref{eq:pert} that the radiative corrections to the unit operator are huge.  
%%%%%%%%%%%%%%%%%%%%%%%%%%%%%%%%%%%%%
\subsubsection*{\d $M_{\sigma_B}$ and $M_{G_1}$}
%%%%%%%%%%%%%%%%%%%%%%%%%%%%%%%%%%%%%

We compare explicitily 
the effect of PT radiative corrections for the predictions of $M_{\sigma_B}$ and $M_{G_1}$  and ${\cal R}^c_{42}$ in the case of  ``one resonance $\oplus$ QCD continuum" parametrization of the spectral function.
For e.g. $t_c=4$ GeV$^2$, one obtains:
\bea
M_{\sigma_B}&\simeq&768~({\rm LO})~~~~\rar~~~~1000~({\rm NLO})~~~~\rar~~~~ 1076~({\rm N2LO})~~{\rm from}~~~ {\cal R}^c_{20},\nnb\\
M_{G_1}&\simeq& 1492~({\rm LO})~~~~\rar~~~~1522~({\rm NLO})~~~~\rar~~~~ 1530~({\rm N2LO})~~{\rm from}~~~ {\cal R}^c_{42},
\eea
indicating that the $\alpha_s$  corrections increase the LO masses by about 30/\% for the $\sigma_B$ while the increase from NLO to N2LO is only about 8\%. For $M_{G_1}$ the effect of radiative corrections is much smaller. This relatively small effect is due to the fact that radiative PT corrections tend to compensate in the ratios of sum rules. 

%%%%%%%%%%%%%%%%%%%%%%%%%%%%%%%%%%%%%%
\subsubsection*{\d  The decay constant $f_{\sigma_B}$}
%%%%%%%%%%%%%%%%%%%%%%%%%%%%%%%%%%%%%%%%
Fixing e.g. $t_c\simeq 4$ GeV$^2$, we study the effect of the truncation of the PT series on the estimate of the coupling from ${\cal L}^c_{-1}$. Taking $\tau\simeq  0.5$ GeV$^{-2}$ at the inflexion point, we obtain the result in units of MeV for a ``one resonance" parametrization :  
\beq
f_{\sigma_B}= 202~({\rm LO})~~~~\rar~~~~406~({\rm NLO})~~~~\rar~~~~ 696~({\rm N2LO})~.
\eeq
The effect of radiative corrections is huge as expected. 
%%%%%%%%%%%%%%%%%%%%%%%%%%%%%%%%%%%%%%
\subsubsection*{\d  The decay constant $f_{G_1}$}
%%%%%%%%%%%%%%%%%%%%%%%%%%%%%%%%%%%%%%%%
Here we fix $t_c=3.75$ GeV$^2$ as obtained in the previous analysis . There is a large plateau in $\tau$ for different PT series. We obtain in units of MeV from ${\cal L}_3^c$ within a ``one resonance $\oplus$ QCD continuum":
\beq
f_{G_1}= 360~({\rm LO})~~~~\rar~~~~470~({\rm NLO})~~~~\rar~~~~ 580~({\rm N2LO})~,
\eeq
which shows relatively moderate PT corrections. 

%%%%%%%%%%%%%%%%%%%%%%%%%%%%%%%%%%%%%%%%%%%%
\subsection*{\b Gluon condensates and the OPE}
%%%%%%%%%%%%%%%%%%%%%%%%%%%%%%%%%%%%%%%%%%%%
Though the presence of the gluon condensates in the OPE are important in some moments for the stability points, their effects are relatively small ensuring the 
convergence of the OPE.

%%%%%%%%%%%%%%%%%%%%%%%%%%%%%%%%%%%%%%%%%%%%
\subsection*{\b The tachyonic gluon mass $\lambda^2$}
%%%%%%%%%%%%%%%%%%%%%%%%%%%%%%%%%%%%%%%%%%%%
As emphasized in\,\cite{CNZa} this effect is expected to explain the large scale of the scalar gluonium channel compared to the ordinary  $\rho$ meson channel. It is also expected\,\cite{CNZb} to give a phenomenological estimate of the uncalculated terms of the PT series. 

%%%%%%%%%%%%%%%%%%%%%%%%%%%%%%%%%%%%%
\subsubsection*{\d $M_{\sigma_B}$ and $M_{G_1}$}
%%%%%%%%%%%%%%%%%%%%%%%%%%%%%%%%%%%%%
Including the $\lambda^2$ contribution in the OPE, we check that its effect on the mass determination is negligible which can be due to the fact that the effects tend to compensate in the ratio of moments. 

%%%%%%%%%%%%%%%%%%%%%%%%%%%%%%%%%%%%%%
\subsubsection*{\d  The decay constants $f_{\sigma_B}$ and $f_{G_1}$}
%%%%%%%%%%%%%%%%%%%%%%%%%%%%%%%%%%%%%%%%
Within a two resonances parametrization, we find for a given value of $t_c$ and $\tau$ that the tachyonic gluon mass decreases the value of $f_{\sigma_B}$ by about 30-50 \% while it increases only slightly $f_{G_1}$ by about 5\%.

%\bea
%&&\lambda^2=0~~~~\rar~~~~~~~~\lambda^2=-0.065\,{\rm GeV}^2 \nnb\\
%f_{\sigma_B}&=&  533 ~~~~~~~~\rar~~~~~~~~ 272~~~~~~{\rm MeV}\nnb\\
%f_{G_1}&=&  569 ~~~~~~~~\rar ~~~~~~~~ 597~~~~~~{\rm MeV}~,
%\eea
%where the $\lambda^2$ effect is more important for $f_{\sigma_B}$ and acts in an opposite way. 

%%%%%%%%%%%%%%%%%%%%%%%%%%%%%%%%%%%%%%
\subsection*{\b Comparison with some other determinations}
%%%%%%%%%%%%%%%%%%%%%%%%%%%%%%%%%%%%%%
%%%%%%%%%%%%%%%%%%%%%%%%%%%%%%%%%%%%%%
\subsubsection*{\d The $\sigma_B$}
%%%%%%%%%%%%%%%%%%%%%%%%%%%%%%%%%%%%%%
The existence of the $\sigma_B$ with a mass about 1 GeV is compatible with the idea that it is the dilaton associated to 
conformal $U(1)_V$ invariance analogous to the case where the $\eta'$-meson is associated to the $U(1)_A$ symmetry. 
The dilaton has been extensively discussed in an effective Lagrangian approach\,(see e.g.\,\cite{CHANO,LANIK,JORA}) including the $U(1)_V$ symmetry. 

-- The mass in Eq.\,\ref{eq:msigmaf} is in a good agreement with the earlier sum rule result\,\cite{SNG0}:
\beq
M_{\sigma_B}\simeq (0.95\sim 1.10)~{\rm GeV},
\eeq
from a least square fit of the USR ratio of sum rules.  A  similar result has been obtained from the analysis\,\cite{STEELE2} based on Gaussian sum rules\,\cite{LAUNER} including instanton effects.% while a lower value around (300-600) MeV has been obtained in\,\cite{

--A full lattice calculation\,\cite{HART,MCNEILE} finds a relative suppressed mass of the lightest glueball which moves from  the quenched result about 1.6 GeV to the full QCD one around 1 GeV.  However, a more appropriate comparison with the lattice results requires a simulation beyond the one resonance contribution to the  scalar gluonium correlator.

-- A strong coupling analysis of the gluon propagator  %\,\cite{FRASCA} 
 finds a gluonium  pole  mass around $M_\sigma$ and $M_{f_0(980)}$ from the mass gap.\,\cite{FRASCA}. 

%%%%%%%%%%%%%%%%%%%%%%%%%%%%%%%%%%%%%%
\subsubsection*{\d The higher mass gluonium $G_1$}
%%%%%%%%%%%%%%%%%%%%%%%%%%%%%%%%%%%%%%
-- Its mass is given in Eqs.\,\ref{eq:mg11} and \,\ref{eq:mg1} in the case of {\it one or two resonances $\oplus$ QCD continuum} parametrization of the spectral function. These results agree with the earlier ones in Ref.\,\cite{SNG} from ratio of USR and can be compared to some other sum rules results\,\cite{KISSLINGER,HUANG,FORKEL}. We conclude from the paper of Ref.\,\cite{ZHU} that their result for $M_{G_1}\simeq 1.8$ GeV is obtained outside the
$\tau$ minima of their sum rule which leads to an overestimate. 

-- Our result for  $M_{G_1}$ is slightly lower than the one from quenched and full lattice calculations\,\cite{HART,MCNEILE,RAGO,MATHIEU}. However, a precise comparison of the results from the two approaches is delicate as  the lattice situation for the 1 GeV result remains unclear\,\cite{HART,MCNEILE}, while the lattice analysis is done by only retaining a single resonance contribution to the two-point correlator. 

%%%%%%%%%%%%%%%%%%%%%%%%%%%%%%%%%%%%%%%%%
\subsubsection*{\d The gluonium  radial excitations  $\sigma'$ and $G_2$ and }
%%%%%%%%%%%%%%%%%%%%%%%%%%%%%%%%%%%%%%%%%
We have also attempted to extract the masses and decay constants of the gluonium first radial excitations with the result in Eqs.
\,\ref{eq:msigmap}, \,\ref{eq:fsigmap}, \,\ref{eq:mg2} and\,\ref{eq:fg2}. In the next phenomenological analysis, we are tempted to identify the $\sigma'(1.32)$ with the $f_0(1.37)$.  The $G_2(2.5)$ is (within the large errors) in the range of the observed $f_0(2.02)$ and $f_0(2.2)$\,\cite{PDG}.

%%%%%%%%%%%%%%%%%%%%%%%%%%%%%%%%%%%%%%
\subsubsection*{\d Comments}
%%%%%%%%%%%%%%%%%%%%%%%%%%%%%%%%%%%%%%
-- In the example of the scalar gluonium channel, we have seen that (at least) NLO corrections are  important in the determination of the scalar gluonium masses  from the sum rules, while N2LO corrections are needed for a good determination of the decay constants.  Therefore, one should make a great care for quoting the LO results given in the literature.

-- The smallness of the tachyonic gluon mass contribution to the gluonium mass determinations may indicate that the PT series reaches its asymptotic form after the inclusion of the N2LO contribution which can be sufficient for extracting with a good accuracy these observables from the sum rules.  However, we have also noticed that the decay constant of the $\sigma_B$ is largely affected (-30 to -40\%) by the tachyonic gluon mass which tends to decouple the $\sigma_B$  from the two-point correlator. 

-- Some other analysis using ``two resonances $\oplus$ QCD continuum parametrization from sum rules with instantons \,\cite{STEELE2} and an inverse problem dispersive approach without instantons \,\cite{HSIANG} reproduce our results for $M_{\sigma_B}$ and $M_{G_1}$. The apparent discrepancies with the lattice results requests a clarification of the full QCD results and a two resonance parametrization of the two-point correlator. 

-- Some phenomenological implications of our results will be discussed in the next section.

%%%%%%%%%%%%%%%%%%%%%%%%%%%%%%%%%%%%%%
 %%%%%%%%%%%%%%%%%%%%%%%%%%%%%%%%%%%%%%%%%%%
\section{The conformal charge $\psi_G(0)$}
%%%%%%%%%%%%%%%%%%%%%%%%%%%%%%%%%%%%%%%%%%%
\subsection*{\b Test of the Low Energy Theorem (LET)}
 %%%%%%%%%%%%%%%%%%%%%%%%%%%%%%%%%%%%%%%%%%%
 LET suggests that the value of $\psi_G(0)$ is given by Eq.\,\ref{eq:let}. We test the accuracy of this relation by saturating the two-point function by  the two lowest ground state resonances $\sigma_B$ and $G_1$. In this way, we obtain :
 \beq
  \psi_G(0)= \sum_{i= \sigma,G,\cdots}2\,f_G ^2M_G^2\approx 1.95~{\rm GeV}^4~,
  \eeq
 which is slightly higher than  the LET estimate in Eq.\ref{eq:let} :
 \beq
  \psi_G(0)\vert_{\rm LET}= (1.46\pm 0.08)~{\rm GeV}^2~,
  \label{eq:let2}
  \eeq
% corresponds to the case of two resonances.  This agreement confirms the important role of the $\sigma$ meson for saturating LET  and shows  the self-consistency of
 without any appeal to some contributions beyond the OPE such as the  instantons ones.
 %%%%%%%%%%%%%%%%%%%%%%%%%%%%%%%%%%%%%%
 \subsection*{\b Sum rule extraction of conformal charge $\psi_G(0)$}
  %%%%%%%%%%%%%%%%%%%%%%%%%%%%%%%%%%%%%%
  Here, we shall, instead, determine $\psi_G(0)$ from the sum rule within the standard SVZ expansion. A similar analysis has been already done in\,\cite{SNG0} which we shall update here. 
  In so doing, we work with the combination of sum rules\,\footnote{We remind that this combination of sum rules has been firstly (and successfully)  introduced to quantify the deviation from kaon PCAC\,\cite{SNSU3} due to $SU(3)$ breakings and later to estimate the topological charge of the $U(1)_A$ channel\,\cite{SNG0} and its slope\,\cite{SNG0,SHORE} within the standard SVZ expansion without any additional instanton contributions.}:
  \beq
  \psi_G(0)\vert_{\rm LSR} =\int_0^{t_c} \frac{dt}{t}\,e^{-t\tau}\ga 1-\frac{t\tau}{2}\dr \frac{1}{\pi}\,{\rm Im}\psi_G(t)- \ga {\cal L}^c_{-1}-\frac{\tau}{2}{\cal L}^c_{0}\dr_{QCD}~,
  \eeq
  where the 1st part of the RHS is parametrized by the lowest resonances $\sigma_B,~G_1,~\sigma'_B$ and the 2nd part is the QCD expression of the moment sum rules ${\cal L}^c_n$. The other radial excitations $G'_1$ and $G_2$ having less certain values of the decay constants are included in the QCD continuum. 
    
  %%%%%%%%%%%%%%%%%%%%%%%%%%%%%%%%%%%%%%%
\begin{figure}[H]
%\vspace*{-0.25cm}
\begin{center}
%\centerline {\hspace*{-6.cm} \bf a) }
\vspace{0.25cm}
\includegraphics[width=10cm]{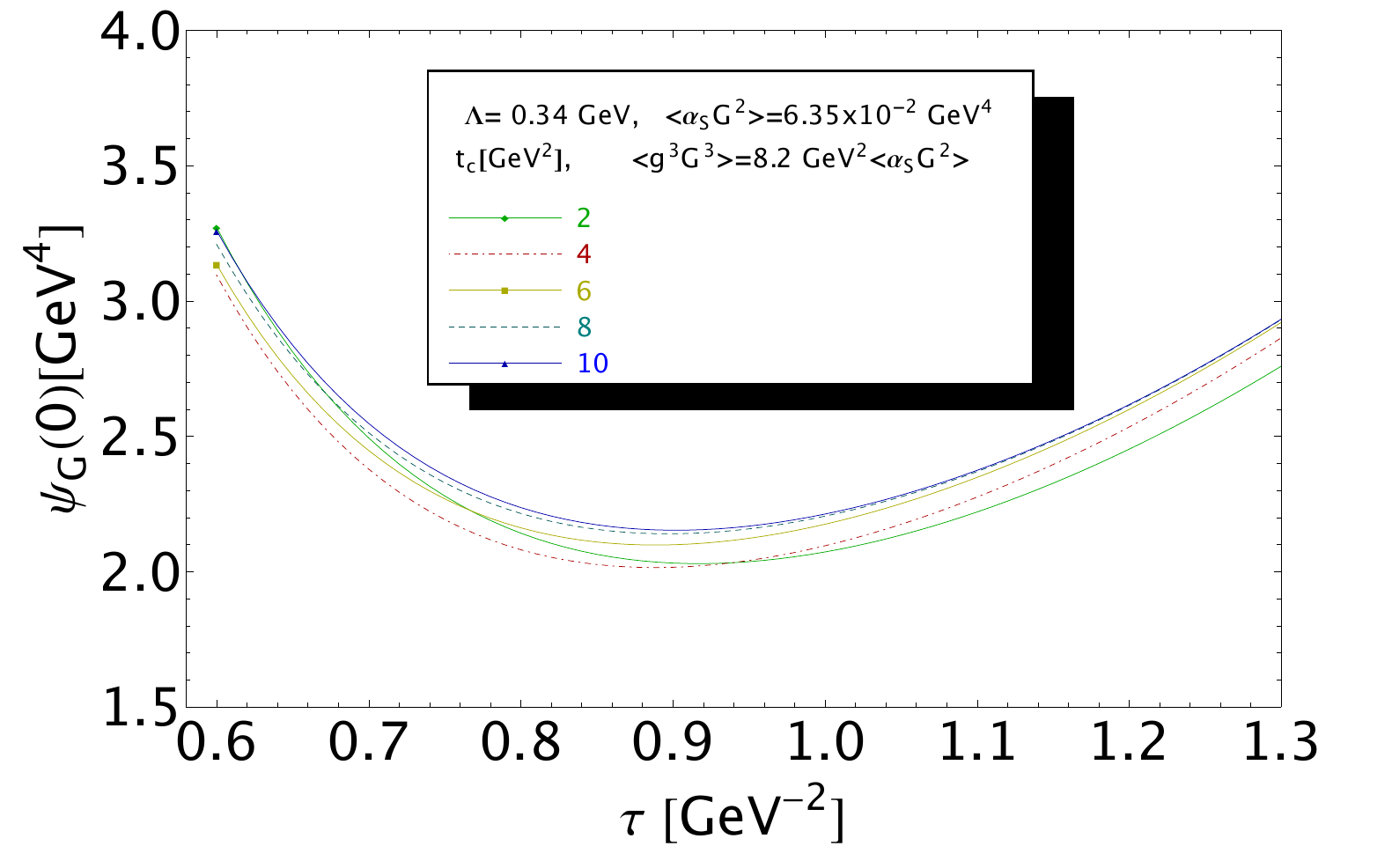} 
\vspace*{-0.5cm}
\caption{\footnotesize  $\psi(0)_G$  from LSR  as a function of $\tau$ for  different values of $t_c$.} 
\label{fig:psi0}
\end{center}
\end{figure} 
%%%%%%%%%%%%%%%%%%%%%%%%%%%%%%%%%%%%%%%

\b The result of the analysis is shown in Fig.\,\ref{fig:psi0}, where the curves present nice $\tau$ at $ (1.0\pm 0.04)$ GeV$^{-2}$. A $t_c$-minimum appears  at $(2\sim 4) $GeV$^2$ and a stability from 8 GeV$^2$. We obtain:
\bea
  \psi_G(0)\vert_{\rm LSR}& =& 2.09(2)_{t_c}(0)_\tau(18)_\Lambda(11)_{G^2}(10)_{G^3}(1)_{M_{\sigma_B}}(14)_{f_{\sigma_B}}(2)M_G(7)_{f_G}
  (0)_{M_{\sigma'_B}}(2)_{f_{\sigma'_B}} \nnb\\
  &=&2.09(29)~{\rm GeV^4}~,
  \label{eq:psi0}
  \eea
  where the central value is 1.4 times the LET estimate in Eq.\,\ref{eq:let} and the one 
  in\,\cite{SNG0} obtained within a one resonance and to LO. 
  
\d One can notice that adding the two radial excitation contributions changes this value to 2.23 $G_2$. The contribution of $\sigma_B$ is about 70\% which is crucial for recovering the LET result. 

\b We analyze the effect of this result in the estimate of $f_{\sigma_B}$ and $f_{\sigma'_B}$ obtained previously from the moment ${\cal L}_{-1}^c$ where $\psi_G(0)$ contributes. 

\d We find e.g. for $f_{\sigma_B}$ that the inflexion point in $\tau$ moves from $\tau=0.56$ GeV$^{-2}$ (see Fig.\,\ref{fig:fsigma}) to  0.46 for e.g. $t_c=3.75$ GeV$^2$  to which corresponds  an increase of about 4 MeV, which is negligible compared to the large error of $f_{\sigma_B}$. 

\d  For $f_{\sigma'_B}$, the LSR value of $\psi_G(0)$ increases the decay constant by about 100 MeV but this value is inside the range spanned by the large error of 342 MeV obtained in its determination. Therefore, we keep the result obtained in the previous section. 

 %%%%%%%%%%%%%%%%%%%%%%%%%%%%%%%%%%%%%%%%%%%
\section{The slope $\psi'_G(0)$ of the conformal charge }
%%%%%%%%%%%%%%%%%%%%%%%%%%%%%%%%%%%%%%%%%%%
The slope $\psi'_G(0)$ obeys the twice subtracted sum rule:
 \beq
  \psi'_G\vert(0)_{\rm LSR} =\int_0^{t_c} \frac{dt}{t^2}\,e^{-t\tau} \frac{1}{\pi}\,{\rm Im}\psi_G(t)-  {\cal L}^c_{-2}\vert_{QCD}+
 \tau\, \psi(0)_G~,
  \eeq
  where :
  \bea
{\cal L}^c_{-2}(\tau)&=&\beta^2(\alpha_s)\ga\frac{2}{\pi^2}\dr\tau^{-1}\sum_{n=0,2,\cdots}\hspace*{-0.25cm}D^{-2}_n~,
\eea
with:
\bea
D^{-2}_0&=&\Big{[}C_{00}-2C_{01}\gamma_E+C_{02}\big{[}\gamma_E-\frac{\pi^2}{6}\big{]}\Big{]}(1-\rho_0)\nnb\\
D^{-2}_2&=&-\tau\,C_{21}(1-\gamma_E)\lambda^2\,\nnb\\
D^{-2}_4&=&\tau^2\Big{[}C_{40}-C_{41}\big{[} \frac{1}{2} -\gamma_E\big{]}\Big{]}\gg,\nnb\\
%\eea
%\end{document}
D^{-2}_6&=&\frac{\tau^3}{2}\Big{[}C_{60}-C_{61}\ga\frac{1}{3}-\gamma_E\dr\Big{]}\ggg,\nnb\\
D^{-2}_8&=&\frac{\tau^4}{6}C_{80}(1.1\pm 0.5)\gg^2~.
\eea
  %%%%%%%%%%%%%%%%%%%%%%%%%%%%%%%%%%%%%%%
\begin{figure}[hbt]
%\vspace*{-0.25cm}
\begin{center}
%\centerline {\hspace*{-6.cm} \bf a) }
\vspace{0.25cm}
\includegraphics[width=10cm]{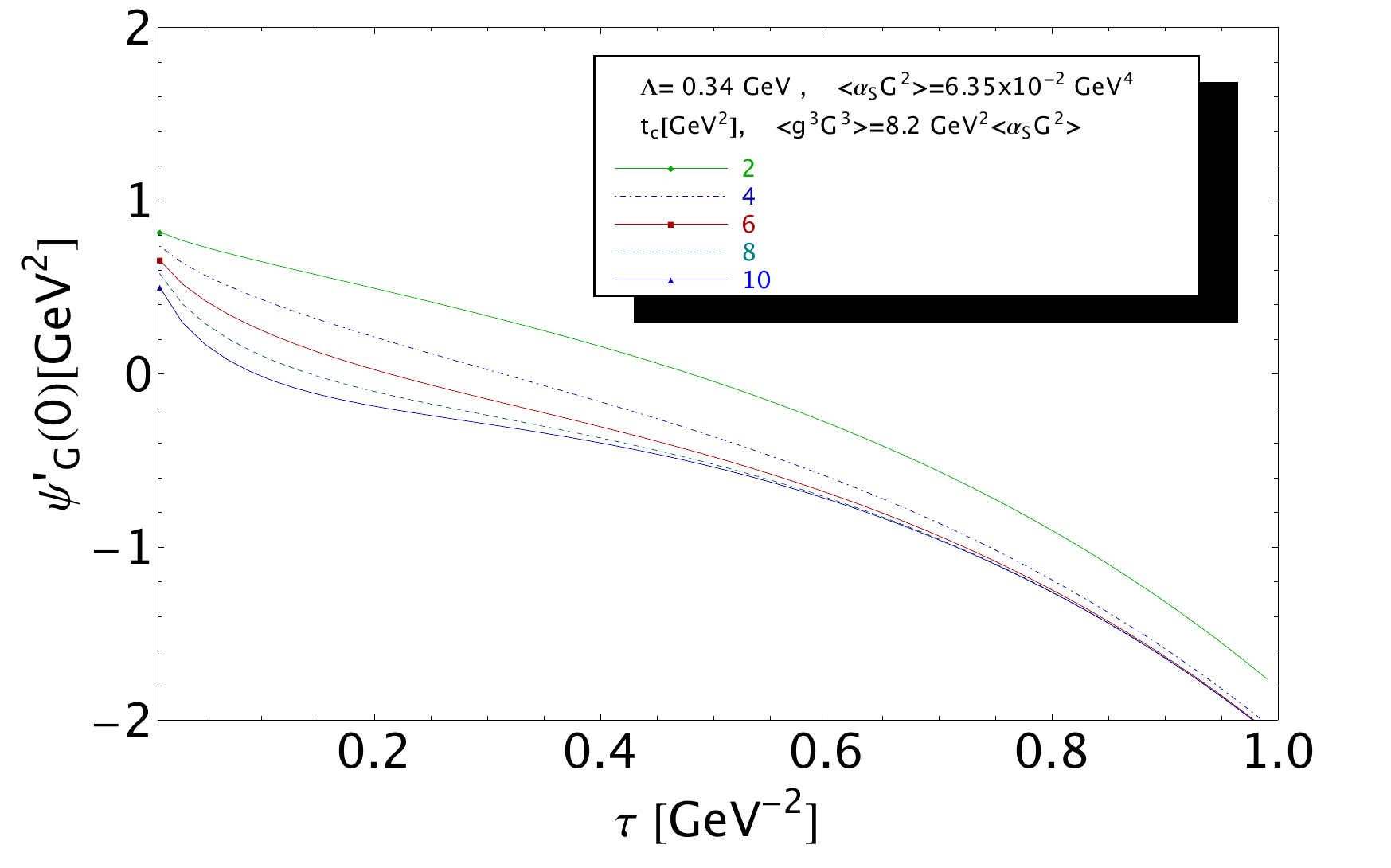} 
\vspace*{-0.5cm}
\caption{\footnotesize  $\psi'(0)_G$  from LSR  as a function of $\tau$ for  different values of $t_c$.} 
\label{fig:psip0}
\end{center}
\end{figure} 
%%%%%%%%%%%%%%%%%%%%%%%%%%%%%%%%%%%%%%%
We show the analysis in Fig.\,\ref{fig:psip0} from which we deduce from $t_c =(2\sim 8)$ GeV$^2$ at the $\tau$-minimum $\pm 0.04$ GeV$^{-2}$:
\bea
10^2\times \psi'_G(0)\vert_{\rm LSR}&=&-22(5)_{t_c}(4)_\tau(16)\Lambda(1)_{G^2}(0)_{G^3}(2)_{M_{\sigma_B}}(16)_{f_{\sigma_B}}\nnb\\
&&(1){M_{G_1}}(8)_{f_{G_1}} (1)_{M_{\sigma'}}(9)_{f_{\sigma'}}(11)_{\psi_G(0)}\nnb\\
&=&-22(29)~{\rm GeV^2},
\label{eq:psip0}
 \eea
 where the $\sigma_B$ contribution is important. The present result confirms and updates the earlier one in Ref.\,\cite{SNG0}. 
%\end{document}
  %%%%%%%%%%%%%%%%%%%%%%%%%%%%%%%%%%%%%%%%%%%
\section{Hadronic couplings of the $\sigma_B$ meson }
%%%%%%%%%%%%%%%%%%%%%%%%%%%%%%%%%%%%%%%%%%%
\subsection*{\b The $\sigma_B$ coupling to $\pi\pi$}
%%%%%%%%%%%%%%%%%%%%%%%%%%%%%%%%%%%%%%
%\subsubsection*{\d Hadronic width}
%%%%%%%%%%%%%%%%%%%%%%%%%%%%%%%%%%%%%%
%%%%%%%%%%%%%%%%%%%%%%%%%%%%%%%%%%
 One can use the previous result of the decay constant in Eq.\,\ref{eq:fsigmaf1} to predict the hadronic width of the $\sigma_B$. For this purpose, we consider the vertex (see e.g\,\cite{VENEZIA}):
\beq
V_\pi(q^2)=\la \pi_1\vert \theta_\mu^\mu \vert \pi_2\ra~~:~~~~~~~~~q=p_1-p_2~~~~~~~~~{\rm with}~~~~~~~~~~V(0)={\cal O}(m_\pi^2)~.
\eeq
In the chiral limit $(m_\pi^2= 0)$, one has :
\beq
V(q^2)=q^2\int_{4m^2_\pi}^\infty \frac{dt}{t(t-q^2-i\epsilon)}\frac{1}{\pi}\,{\rm Im} V(t)~.
\eeq
Using the fact that $V'(0)=1$\,\cite{NOVIKOV,NSVZ}, one can deduce the Low Energy Vertex sum rule  (LEV-SR)\,\cite{VENEZIA}:
\beq
\frac{1}{4}\sum_{G=\sigma_B,\cdots}g_{G\pi\pi}\frac{\sqrt{2}f_G}{M_G^2}=1~,
\label{eq:sigmapipi}
\eeq
where the contribution of the light $\sigma_B$ is enhanced. Then,  assuming (to a first approximation) that the $\sigma_B$ dominates the sum rule, one can deduce :
\beq
g_{\sigma_B\pi^+\pi^-}\approx (7.1\pm 2.5) ~{\rm GeV}~,
\eeq
to which corresponds the width:
\beq
\Gamma [\sigma_B\to \pi^+\pi^-+2\pi^0]=\frac{3}{2}\frac{\vert g_{\sigma_B\pi^+\pi^-} \vert^2}{16\pi M_{\sigma_B}}\ga 1-\frac{4m_\pi^2}{M^2_{\sigma_B}}\dr^{1/2} \approx (1.4\pm 0.8)~{\rm GeV}.
\label{eq:sigwidth} 
\eeq
To check the validity of this approximation, we use the average of the recent data from 2005\,\cite{PDG} which is :
\beq
\Gamma [f_0(1.37)\to \pi\pi] \simeq 215(14)~{\rm MeV}~~\lrar~~~g_{\sigma'_B\pi\pi}\simeq (3.2\pm 0.1)~{\rm GeV}~, 
\eeq
while the $\pi\pi$ width of the $f_0(1.5)$ of 39 MeV\,\cite{PDG} can be neglected. Including the $f_0(1.37)$ contribution, the previous sum rule gives:
\beq
g_{\sigma_B\pi^+\pi^-}\simeq 5.7 ~{\rm GeV}~~\lrar~~~\Gamma [\sigma_B\to \pi^+\pi^-+2\pi^0]\simeq 873~{\rm MeV},
\label{eq:sigmacoupl}
\eeq
in agreement with previous estimates\,\cite{VENEZIA,SNG}. 
%%%%%%%%%%%%%%%%%%%%%%%%%%%%%%%%%%%%%%%
\subsection*{\b The $\sigma_B$ coupling to $K^+K^-$}
%%%%%%%%%%%%%%%%%%%%%%%%%%%%%%%%%%%%%%
Up to $SU(3)$ breaking, the vertex sum rule in Eq.\ref{eq:sigmapipi} indicates that the $\sigma_B$ couples (almost) universally to Goldstone boson pairs, such that in this gluonium picture, one also expects a large coupling of the $\sigma_B$ to $K^+K^-$. This feature has been observed from the analysis of $\pi\pi\to K^+K^-$ scattering data\,\cite{ KMN,WANG1,WANG2}. 
%%%%%%%%%%%%%%%%%%%%%%%%%%%%%%%%%%%%%%%%%%%
\subsection*{\b $\sigma_B$ mass and width confronted to  the data}
%%%%%%%%%%%%%%%%%%%%%%%%%%%%%%%%%%%%%%%%%%%
It is informative to compare the previous mass and width to the one of the observed $\sigma$ from scatterings data analysis :

\d The values of the mass and width obtained in Eqs.\,\ref{eq:msigmaf} and \,\ref{eq:sigwidth} for $\sigma_B$ are comparable with the on-shell/Breit-Wigner mass and width obtained  in Eq.\,\ref{eq:sigdata}. 

\d These results together with the ones from data analysis reviewed in the first sections can indicate  the presence of an important gluon component  inside the observed $\sigma/f_0(500)$ meson wave function. 

%%%%%%%%%%%%%%%%%%%%%%%%%%%%%%%%%%%%%%%%%%%
\section{Hadronic couplings of the higher mass gluonium $G$}
%%%%%%%%%%%%%%%%%%%%%%%%%%%%%%%%%%%%%%%%%%%
\subsection*{\b $G_1$ coupling to $\eta\eta', \eta\eta$}
%%%%%%%%%%%%%%%%%%%%%%%%%%%%%%%%%%%%%%%%%%%
In order to compute the $g_{G_1\eta_1\eta_1}$ coupling of the $G_1$ to the singlet $\eta_1$, one starts from the \,\cite{VENEZIA,WITTEN}:
\beq
\tilde V_{\mu\nu}(q_1,q_2)=i\int d^4x_1d^4x_2\,e^{i(q_1x_1+q_2x_2)}\la 0\vert {\cal T} Q(x_1)Q(x_2) \theta_{\mu\nu}(0)\vert 0\ra~,
\eeq
where $\theta_{\mu\nu}$ is the energy-momentum tensor with 3 light quarks and :
\beq
Q(x)=\frac{\alpha_s}{16\pi} \epsilon^{\mu\nu\rho\sigma}G^a_{\mu\nu}G^a_{\rho\sigma}~,
\eeq
is the topological charge density, where one reminds that,  in the large $N_c$ limit (no quark loops), the solution to the $U(1)_A$ problem is\,\cite{VENEZIA,WITTEN}:
\beq
\Gamma_2(q)\equiv i\int d^4x\la 0\vert {\cal T }\,Q(x)Q(0)\vert 0\ra \rar  \Gamma^{\rm YM}_2(q\ll \Lambda)\rar  \Gamma^{\rm YM}_2(0)\simeq (180~{\rm MeV})^4.
\eeq
Using the expression of the topological charge  in the presence of quark loops\,\cite{DIVECCHIA} and the consistent one for the vertex $\tilde V_{\mu\nu}(q_1,q_2),$ one can deduce\,\cite{VENEZIA}
the constraint:
\beq
\la \eta_1\vert \theta_\mu^\mu\vert_{n_f=3} \vert \eta_1\ra\simeq \frac{9}{11}\frac{12}{f_\pi^2} \Gamma^{\rm YM}_2(0) = 2M_{\eta_1}^2=1.15~{\rm GeV}^2~,
\eeq
where the factor (9/11) comes from the ratio of the $\beta$-function $\beta_1$ for $n_f=3$ and $n_f=0$ and $M_{\eta_1}\simeq 0.76$ GeV\,\cite{VENEZIA,WITTEN}. Then, we deduce the LEV-SR \,\cite{VENEZIA}:
\beq
\frac{1}{4}\sum_{i=\sigma_B,G_1}g_{G\eta_1\eta_1}\sqrt 2 f_i\simeq 1.15~{\rm GeV}^2.
\eeq
Picking the singlet component of the physical $\eta,\eta'$ via the pseudoscalar mixing angle $\theta_P\simeq -(18\pm 2)^0$\,\cite{MONTANET} :  $\eta=\sin\theta_P \eta_1+\cdots~~,\eta'=\cos\theta_P \eta_1+\cdots, $
one deduces:
\beq
\sum_{G=\sigma_B,G_1}g_{G\eta\eta'} f_G\simeq \frac{4}{\sqrt 2} (\sin\theta_P)1.15\, {\rm GeV}^2\simeq -1.0\, {\rm GeV}^2~~~~{\rm and}~~~~g_{G_1\eta\eta}=(\sin\theta_P)  g_{G_1\eta\eta'}.
\label{eq:etavsr}
 \eeq
 Following Ref.\,\cite{VENEZIA} by using the constraint from $\eta'\to\eta\pi\pi$ (allowing a 1\% deviation from the precision of the data\,\cite{PDG}) , and a Gell-Mann, Sharp, Wagner-type model, for the intermediate  state $\sigma_B$, one can deduce the upper bound for $M_{\sigma_B}=1.08$ GeV :
 \beq 
\vert  g_{\sigma_B\eta\eta'}\vert \leq 0.34~{\rm GeV}~, 
 \eeq
 which is a relatively small contribution. Neglecting this contribution and the $f_0(1.37)$ one,   the LEV-SR in Eq.\,\ref{eq:etavsr} leads to :
 \beq
 g_{G_1\eta\eta'}\simeq  2.74(78)~{\rm GeV}~~~~~~~{\rm and}~~~~~~~ g_{G_1\eta\eta}\simeq  0.85(25)~{\rm GeV} ~.
  \eeq
 We deduce the width for ${\rm M}_{G_1}= 1.55~{\rm GeV}$, $p_{\eta\eta'}$ = 20 MeV and $p_{\eta\eta}$ = 517 MeV :
 \beq
 \Gamma (G_{1}\to\eta\eta')=
\frac{ \vert g_{G_1\eta'\eta}\vert^2}{8\pi}\frac{p_{\eta\eta'}}{M_{G_1}^2}\simeq (2.5\pm1.4) ~[(2.6\pm 0.9)\,{\rm Data}]~\,{\rm MeV},
\label{eq:gam-eta-etap}
\eeq
which implies:
\beq
 \frac{\Gamma (G\to\eta\eta)}{\Gamma (G\to\eta\eta')}= \sin^2\theta_P\frac{p_{\eta\eta}}{p_{\eta'\eta}}\simeq (2.3\pm 0.6)\,[(3.0)\,{\rm Data}]~. 
 \label{eq:gam-eta-eta}
 \eeq
The results agree perfectly within the errors with the data. 
 %%%%%%%%%%%%%%%%%%%%%%%%%%%%%%%%%%%%%%%%%%%
\subsection*{\b $\sigma',~G_1$ couplings to $\sigma_B\sigma_B$\label{sec:sigmap}}
%%%%%%%%%%%%%%%%%%%%%%%%%%%%%%%%%%%%%%%%%%%
One can also write a low-energy vertex sum rule (LEV-SR) for this coupling\,\cite{VENEZIA}:
\beq
\la \sigma_B\vert \theta_\mu^\mu\vert \sigma_B\ra = 2M_{\sigma_B}^2~.
\eeq
Using  the dispersion relation of the vertex, one can write the sum rule:
\beq
\frac{1}{4}\sum_{G=\sigma',G_1}g_{G\sigma_B\sigma_B}\sqrt 2 f_G\simeq 2M_{\sigma}^2, 
\label{eq:f0vertex}
\eeq
where here it is more appropriate to use the pole mass $M_\sigma\simeq$ 0.5 GeV for the virtual $\sigma_B$.

\d  Assuming that the $\sigma'$ dominates the sum rule which is also the only one which decays copiously into  $2(\pi\pi)_S$\,\cite{PDG}, one can deduce :
\beq
g_{\sigma'\sigma_B\sigma_B}\simeq (2.2\pm 0.73)~{\rm GeV}~,
\eeq
where we have used the previous estimate of the coupling in Eq.\,\ref{eq:fsigmap}. 

\d We confront this prediction  with the data of $f_0(1.37.1,5) \to 2(\pi\pi)$. 
%which corresponds to the momentum concerned in the decay.  
In so doing, we use the six most recent measurements of the $f_0(1.37)$ $4\pi$ width since 1995 compiled by PDG\,\cite{PDG} from which we deduce the mean:
\beq
\Gamma [f_0(1.37)\to 4\pi]\simeq 365(20)~{\rm MeV}.
\eeq
Using the ratio\,\cite{PDG} :  
\beq
\frac{\Gamma[f_0(1.37)\to 2(\pi\pi)_S]}{\Gamma [f_0(1.37)\to (4\pi )]}\simeq 0.51(9),
\eeq
we deduce:
\beq
\Gamma [f_0(1.37)\to 2(\pi\pi)_S] \simeq 186(35)~{\rm MeV}\lrar g_{f_0\sigma_B\sigma_B}\simeq (4.33\pm 0.39)~{\rm GeV}~.
\label{eq:f0pdata}
\eeq
where the estimate is off by about a factor two indicating that the sum rule may have overesetimated the decay constant of the $\sigma'$-meson which is (presumably) the sum of effective couplings of higher state resonances. 

\d One can use instead the experimental value of $ g_{f_0\sigma_B\sigma_B}$ into the LEV-SR. Neglecting the $G_1(1.55)$ contribution  which has a tiny $2(\pi\pi)S$ width of
14 MeV deduced from the data on $\Gamma_{\pi\pi}\simeq$ 39 MeV $\simeq$  34.5\%
of its total width,  $\Gamma_{4\pi}\simeq$ 48.9\,\% and $\Gamma_{2(\pi\pi)_S}/\Gamma_{4\pi}\simeq$ 26\,\%, we deduce from $ g_{f_0\sigma_B\sigma_B}$
and the LEV-SR in Eq.\,\ref{eq:f0vertex} the more accurate estimate :
\beq
f_{\sigma'}\simeq f_{\sigma'_B}= (329\pm 30)~{\rm MeV}.
\label{eq:fsigmap1}
\eeq
This value is in line with the intuitive expectations from ordinary $\rho$ and $\rho'$ mesons. 
%%%%%%%%%%%%%%%%%%%%%%%%%%%%%%%%%%%%%%%%%%%%%%
%%%%%%%%%%%%%%%%%%%%%%%%%%%%%%%%%%%%%%%%%%%
\section{Summary and conclusions}
%%%%%%%%%%%%%%%%%%%%%%%%%%%%%%%%%%%%%%%%%%%
%%%%%%%%%%%%%%%%%%%%%%%%%%%%%%%%%%%%%%%%%%%%%%%
The different results obtained in this paper are summarized in Table\,\ref{tab:res} which we shall briefly comment below. 
%%%%%%%%%%%%%%%%%%%%%%
\subsection*{\b The $\sigma/f_0(500)$ meson}
%%%%%%%%%%%%%%%%%%%%%%
 In this paper, we have started to review our present knowledge on the nature of the $\sigma$ meson from $\pi^+\pi^-, \gamma\gamma\to \pi^+\pi^-, K^+K^-$ data and from $J/\psi$ gluon rich channels $(\gamma\pi\pi, \omega\pi\pi, \phi\pi\pi)$ and $\phi$ radiative decays,$D_{(s)}$ semileptonic decays and $\bar pp, pp$ production processes. The data (a priori) favour / does not exclude  a large gluon component in the $\sigma$ wave function where the observed state may emerge from a maximal mixing with a $\bar qq$ state\,\cite{BN} rather than a pure four-quark $(\bar{uu})(dd)$  or a $\pi\pi$ molecule state. The four-quark or $\pi\pi$ molecule state  is not favoured by  the too small predicted direct $\gamma\gamma$ width of the $\sigma$\,\cite{MNO,WANG1,WANG2} and by the large $K^+K^-$ coupling found from scattering data\,\cite{KMN,WANG1}. The assumption that the $\sigma$ emerges from $\pi\pi$ rescattering is not also favoured from its large coupling to $K^+K^-$. 
%%%%%%%%%%%%%%%%%%%%%%%%%%%
\subsection*{\b The $f_0(980)$ meson from the data}
%%%%%%%%%%%%%%%%%%%%%%%%%%% 
\d  From the scattering data analysis, it is unlikely that the $f_0(980)$ is an $\bar ss$ state because its has a non negligible / large coupling to $\pipi$\,\cite{KMN,WANG1,WANG2}, while the strength of its $\gamma\gamma$ coupling is relatively small for a $\bar qq$ state\,\cite{SNA0,SN06}. Moreover, it is too light compared to the $\bar ss$ mass of about 1.4 GeV\,\cite{SN06}. 
%%%%%%%%%%%%%%%%%%%%%%%%%%%%%%%%%%%%%%%%%%%
\subsection*{\b Conclusions from the data }
%%%%%%%%%%%%%%%%%%%%%%%%%%%%%%%%%%%%%%%%%%%

The previous observations may favour the interpretation that the $\sigma$ and $f_0(980)$ emerge from a maximal mixing between a gluonium and a $(\bar uu+\bar dd)$ state like we have proposed in\,\cite{BN,SNG} and what had also been expected\,\cite{NSVZ} in their pioneering paper. 

%%%%%%%%%%%%%%%%%%%%%%%%%%%%%%%%%%%%%%%%%%%
\subsection*{\b The $\sigma_B$ meson from QCD spectral sum rules (QSSR)}
%%%%%%%%%%%%%%%%%%%%%%%%%%%%%%%%%%%%%%%%%%%
 \d A  QSSR analysis of the lowest ratios of sum rule ${\cal L}_{20}$  favours a $\sigma_B$ (the subindex $B$ refers to a pure unmixed gluonium state) having a mass $M_{\sigma_B}$=1.07 GeV (see Eq.\,\ref{eq:msigmaf1}) and a $\pi\pi$ width  Eq.\,(\ref{eq:sigmacoupl}) compatible with the ones from scattering data for a Breit-Wigner / On-shell $\sigma$-mass.
 
 \d Its decay constant can be deduced from ${\cal L}^c_{1}$ which is $f_{\sigma_B}=456(157)$ MeV (Eq.\,\ref{eq:fsigmaf1}). 
  
 \d One should also note that the set $(\Gamma_{\sigma_B}, M_{\sigma_B})$ obtained from the sum rule is fairly compatible with the on-shell mass and width from $\pi^+\pi^-\to \pi^+\pi^-,K^+K^-$ data . 
%%%%%%%%%%%%%%%%%%%%%%%%%%%%%%%%%%%%%%%%%%%
\subsection*{\b The ``scalar gluonium" $G_1$ from QSSR}
%%%%%%%%%%%%%%%%%%%%%%%%%%%%%%%%%%%%%%%%%%%

\d An analysis of the higher weight USR ${\cal L}_{21}$ and ${\cal L}_{31}$ leads to a gluonium mass:  $M_{G_1}=(1548 \pm 121)$ MeV (Eq.\,\ref{eq:mg1}) for a {\it ``3 resonances" $\oplus$ QCD continuum} parametrization of the spectral function while its becomes $(1515\pm 123)$ MeV (Eq.\ref{eq:mg11}) if one uses a ``one resonance"  $\oplus$ QCD continuum indicating that the presence of the $\sigma_B$ does not affect notably the value of $M_{G_1}$.  It is clear that higher moments are more appropriate in the QSSR approach to pick up the higher mass gluonium state due to the power mass suppression of the lowest mass $\sigma_B$ contribution in these sum rules.  

\d The corresponding gluonium decay constant is $f_{G_1}=(365\pm 110)$ MeV (Eq.\,\ref{eq:fgf1}) which is comparable with the one of $f_{\sigma_B}$. 

\d We have also seen that its small width into $2(\pi\pi)_S$ can be explained from the LET-V sum rule in Eq.\,\ref{eq:f0vertex} indicating that it is not a $\sigma-$like particle but instead decays into $\eta'\eta,~\eta\eta$ through the gluon vertex $U(1)$ anomaly. 

%%%%%%%%%%%%%%%%%%%%%%%%%%%%%%%%%%%%%%%%%%%
\subsection*{\b The radial excitations $\sigma',~G'_1$ and $G_2$ from QSSR}
%%%%%%%%%%%%%%%%%%%%%%%%%%%%%%%%%%%%%%%%%%%
We have attempted to extract the mass and coupling of the radial excitations  $\sigma'$ of  $\sigma_B$, $G'_1$ and $G_2$ of $G_1$. 

\d We found that $M_{\sigma'}\simeq (1121\pm 117)$ MeV is rather low compared to the radial excitation of ordinary mesons where we notice that a similar feature has been observed in the analysis of the $D^*D$ molecule described by high-dimension quark currents. 

\d $f_{\sigma'_B}\simeq 646(216)$ MeV can be large which we have interpreted as an effective sum of all radial excitations that the sum rule cannot disentangle. For a more phenomenological use, we identify the $\sigma'_B$ with the observed  $f_0(1.37)$ and extract the decay constant from its decay to $2(\pi\pi)_S\oplus$ LEV-SR analysis from which we obtain:  $f_{\sigma'}\simeq (329\pm 30)$ MeV (Eq.\,\ref{eq:f0vertex}). 

\d The mass of the radial excitation $G'_1$: $M_{G'_1}=1563(141)$ MeV (Eq.\,\ref{eq:mg1p}) of $G_1$ is also found to be almost degenerated with the ground state mass $M_{G_1}=1548(121)$ MeV (Eq.\,\ref{eq:mgf1}). Like in the case of $\sigma'_B$, we interpret the large coupling $f_{G'_1}=1000(230)$ MeV (Eq.\,\ref{eq:fg1p}) as a sum of the effective couplings of all higher radial excitations entering into the spectral function.    

\d The 2nd radial excitation gluonium $G_2$ is found to have a relatively high mass $M_{G_2}\simeq 2.99$ GeV.

%%%%%%%%%%%%%%%%%%%%%%%%%%%%%%%%%%%%%%%%%%%
\subsection*{\b Phenomenology}
%%%%%%%%%%%%%%%%%%%%%%%%%%%%%%%%%%%%%%%%%%%
\d The global picture which emerges from the approach indicates that the observed  gluonia candidates can de classified into two groups : 

%$\sigma/f_0(0.5)$, $f_0(1.37)$ and $f_0(1.5)$ have presumably a large gluon component in their wave functions. 

%\d Their couplings to the gluonium current have almost a similar strength within the errors.

\hspace*{0.5cm} -- The $\sigma$-like gluonia $[\sigma/f_0(0.5),f_0(1.37)]$ which emerge from $(\sigma_B,\sigma'_B)$ and which decay into $\pi\pi$ via OZI violating process while the $f_0(1.37)$ decays to $2(\pi\pi)_S$ via virtual $\sigma\sigma$.  

\hspace*{0.5cm} -- The $G$-like gluonia $[f_0(1.5),f_0(1.7)]$ which come from $(G_1,G'_1)$ and   which decay into $\eta'\eta$ and $\eta\eta$ through the gluon $U(1)$ anomaly vertex but not copiously to $\pi\pi$ and $2(\pi\pi)_S$. 
%The $f_0(1.5)$ width to $(2(\pi\pi)_S$ has been estimated to be about 49 MeV  (Eq.\,\ref{eq:gam-G-4pi} in good agreement with the data $(54\pm4)$ MeV\,\cite{PDG}. 

\hspace*{0.5cm} -- We may expect that high-mass 2nd radial  excitation $G_2(2.99)$ has similar properties and brings some gluon component to the gluonia candidates above 2 GeV 
which contributes to their $\eta'\eta$ and $\eta\eta$ decays.
%via a mixing with e.g the $G_1(1.58)$ or/and  with some quarkonium states.

\d  We remind that the masses of the $S_2(\bar uu+\bar dd)$, $S_3(\bar ss)$ quarkonia states and their radial excitations are predicted from the sum rules to be (in units of GeV)\,\cite{SNG} :
 \bea
 M_{S_2}&\simeq& 1~, ~~~~~~ M_{S'_2}\simeq (1.1\sim 1.4),\nnb\\
 M_{S_3}&\simeq& (1.47)~, ~~~~~~ M_{S'_3}\simeq (1.7\sim 2.4),
 \eea
which can mix with the previous gluonia states. 

\d A more complete analysis of the data can be done within some $\bar qq$ meson-gluonium mixing scheme. The case for the $\sigma(500)$-$f_0(980)$ has been studied in\,\cite{BN} but may/should be extended to the higher meson masses. Some attempts in this direction have been also done in\,\cite{SNG}\,\footnote{For a review prior 1996, see e.g.\,\cite{CLOSE}.} but needs to be updated.

\d We have not updated our analysis of the $\gamma\gamma$ and radiative widths obtained in \cite{VENEZIA,SNG}. We expect that these results are still valid within the errors. 

\d We emphasize that a more appropriate comparison of the results obtained in this work (e.g. with lattice calculations) requests an analysis beyond the "one resonance" contribution to the two-point correlator from these alternative approaches. Indeed, a simple one resonance for parametrizing the spectral function is  (obviously) insufficient for describing the complex and overpopulated spectra of the $I=o$ scalar mesons. 

%%%%%%%%%%%%%%%%%%%%%%%%%%%%%%%%%%%%%%%%%%%
\subsection*{\b The conformal charge  $\psi_G(0)$ from QSSR}
%%%%%%%%%%%%%%%%%%%%%%%%%%%%%%%%%%%%%%%%%%%
We have completed our analysis by the estimate of the conformal charge $\psi_G(0)$ and its slope $\psi'_G(0)$.

\d Using the previous values of the $\sigma$ and gluonia $G$ masses and couplings, we have obtained  $\psi^{LSR}_G(0)$=2.09 (29) GeV$^4$ (Eq.\,\ref{eq:psi0}) which can be compared with the LET prediction of about  1.46 GeV$^4$ (Eq.\ref{eq:let}). 

\d The result  indicates that the  $\sigma_B$ meson contributes predominantly in the spectral function for reproducing the LET prediction of $\psi_G(0)$. It also shows that the instanton effect in the estimate of $\psi_G(0)$ is (a priori) not necessary for a correct estimate of the conformal charge $\psi_G(0)$. A similar feature has been observed in the estimate of the topological charge and its slope in the $U(1)_A$ sector\,\cite{SHORE} . 

\d The value of the slope is  $\psi'_G(0)\vert_{LSR}$=  0.95(30) GeV$^2$. 
\subsection*{\b Prospects}
%%%%%%%%%%%%%%%%%%%%%%%%%%%%%%%%%%%%%%%%%%%
\d We plan to study in more details the mixing of the $\bar qq$ states with the higher gluonia masses $\sigma'(1.15)$, $G_1(1.53)$ and eventually with the $G_2(2.48)$.

\d We plan to extend the analysis to the case of (pseudo)scalar $C=\pm$ trigluonia channels which has been studied in\,\cite{PABAN} and \,\cite{PIMIKOV}. \\

%%%%%%%%%%%%%%%%%%%%%%%%%%%%%%%%%%%%%%%%
\vspace*{-0.25cm}
\begin{table}[H]
\setlength{\tabcolsep}{0.2pc}
% -----------------------------------------------------
% adapted from TeX book, p. 241
%\newlength{\digitwidth} \settowidth{\digitwidth}{\rm 0}
\catcode`?=\active \def?{\kern\digitwidth}
% -----------------------------------------------------
%\label{tab:res0}
\begin{center}
    {\footnotesize
  \begin{tabular}{lllll}
 % \begin{tabular*}{\textwidth}{@{}l@{\extracolsep{\fill}}|ccc}
  %lllllllllll}
 %{\begin{tabular}{@{}llll@{}} \toprule
%&\\
%\hline
\hline
\\
{\normalsize \bf  Observables\,}&\normalsize\bf Value&\normalsize\bf Eq.\#&\normalsize\bf Source&\normalsize \bf Comments\\
 \\
\hline
%\hline
\\
{\bf Masses\, [MeV]} \\
%\cline{0-0} 
%\\
{\it Ground states}\\
$M_{\sigma_B}$&1088(78)&\ref{eq:msigma0}& ${\cal R}^c_{20}$&1 resonance $\oplus$ QCD continuum\\
&1085(126)&\ref{eq:msigmaf}& ${\cal R}^c_{20}$&2 resonances $\oplus$ QCD continuum\\
&1070(126)&\ref{eq:msigmaf1}& ${\cal R}^c_{20}$&3 resonances $\oplus$ QCD continuum\\
$M_{G_1}$&1515(123)&\ref{eq:mg11}&${\cal R}^c_{42}$&1 resonance $\oplus$ QCD continuum\\
&1524(121)&\ref{eq:mg1}& ${\cal R}^c_{42}$&2 resonances $\oplus$ QCD continuum\\
&1548(121)&\ref{eq:mgf1}& ${\cal R}^c_{42}$&3 resonances $\oplus$ QCD continuum\\
{\it 1st Radial excitations} \\
$M_{\sigma'_B}$&1110(117)&\ref{eq:msigmap}&  ${\cal R}^c_{20}$ (after iteration)&3 resonances $\oplus$ QCD continuum\\
$M_{G'_1}$&1563(141)&\ref{eq:mg1p}&  ${\cal R}^c_{42}$&4 resonances $\oplus$ QCD continuum\\
$M_{G_2}$&2992(221)&\ref{eq:mg2}& mean from ${\cal R}^c_{21}$ and ${\cal R}^c_{31}$ & 5 resonances $\oplus$ QCD continuum
\\
%\\
{\bf Decay constants\, [MeV]} \\
%\cline{0-0} 
%\\
{\it Ground states}\\
$f_{\sigma_B}$&563(156)&\ref{eq:decay-const}& ${\cal L}^c_{-1}$&2 resonances $\oplus$ QCD continuum\\
&456(157)&\ref{eq:fsigmaf1}& ${\cal L}^c_{-1}$&3 resonances $\oplus$ QCD continuum\\
$f_{G_1}$&394103)&\ref{eq:decay-const}& ${\cal L}^c_{2}\oplus{\cal L}^c_{3}$&2 resonances $\oplus$ QCD continuum\\
&365(110)&\ref{eq:fgf1}& ${\cal L}^c_{2}\oplus{\cal L}^c_{3}$&3 resonances $\oplus$ QCD continuum\\
{\it Radial excitations}\\
$f_{\sigma'}$&329(30)&\ref{eq:fsigmap1}&LEV-SR $\oplus~f_0(1.37)\to 2(\pi\pi)_S$\,\cite{PDG} &$f_0(1.5)$ neglected\\
$f_{\sigma'}^{\rm eff}$&648(216)&\ref{eq:fsigmap} & ${\cal L}^c_{-1}$& Sum of higher states  effective couplings  \\
$f^{\rm eff}_{G'_1}$&1000(230)&\ref{eq:fg1p}&  ${\cal L}^c_{2}$ & Sum of higher states  effective couplings \\
$f^{\rm eff}_{G_2}$&797(74)&\ref{eq:fg2}& mean from ${\cal L}^c_{1}$ and ${\cal L}^c_{2}$ & Sum of higher states  effective couplings
\\
%\\
{\bf Decay widths\, [MeV]}  \\
%\cline{0-0} 
%\\
$\Gamma [\sigma_B\to\pi\pi]$&873 &\ref{eq:sigmacoupl}&LEV-SR& 700 : $\pi\pi$ scattering data \\
$\Gamma [\sigma'(1.37)\to 2(\pi\pi)_S]$& 186(35) &\ref{eq:fsigmap1}&data input\,\cite{PDG} $\oplus$ LEV-SR $\lrar$&$ f_{\sigma'}=329(15)$ \\
%$\Gamma [G_1(1.55)\to 2(\pi\pi)_S]$& 49&\ref{eq:gam-G-4pi}&LEV-SR&$(54\pm 4)$ (data) \cite{PDG} \\
$\Gamma [G_1(1.55)\to\eta\eta']$ & $(2.5\pm1.4)$&\ref{eq:gam-eta-etap}&-- & $(2.6\pm 0.9)$ (data) \cite{PDG}\\
$\frac{\Gamma [G_1(1.55)\to\eta\eta]}{\Gamma [G_1(1.55)\to\eta\eta']}$ &$(2.3\pm 0.6)$&\ref{eq:gam-eta-eta}&--& 3.0 (data) \cite{PDG}\\
\bf Conformal Anomaly  \\
%\\
 Charge : $ \psi_G(0)\vert_{\rm LSR}$  [GeV$^4$]&$2.09(29)$&\ref{eq:psi0}& ${\cal L}^c_{-1}\oplus {\cal L}^c_{0}$&  1.46(8)  [LET in Eq.\,\ref{eq:let}]\\
 Slope : $ 10^2\times\psi'_G(0)\vert_{\rm LSR}$  [GeV$^2$] &-22(29) &\ref{eq:psip0}& ${\cal L}^c_{-2} \oplus \psi_G(0)\vert_{\rm LSR}$ & \\
 \\
\hline
%\hline
\end{tabular}
%{\scriptsize
 \caption{Predictions from scalar di-gluonium LSR at N2LO and gluon condensates at NLO up to $D=8$. }
 \label{tab:res}
% \vspace*{0.5cm}
%}
}
\end{center}
\end{table}
%%%%%%%%%%%%%%%%%%%%%%%%%%%%%%%%%%%%%%%%%%%%
%\section*{Acknowledgements}
%%%%%%%%%%%%%%%%%%%%%%%%%%%%%%%%%%%%%%%%%%%
%We thank Claude Amsler, Ugo Gastaldi, Christoph Hanar and A. Rago for some exchanges. 
%%%%%%%%%%%%%%%%%%%%%%%%%%%%%%%%%%%%%%%%%%%

\vfill\eject
%%%%%%%%%%%%%%%%%%%%%%%%%%%%%%%%%%%%%%%%%%%
\input{bib_scalar.tex}

\end{document}

%% file: bib_scalar.tex
%%%%%%%%%%%%%%%